\documentclass[twocolumn]{aastex7}
\usepackage{amsmath}
\usepackage{subcaption}
\usepackage{float}
\usepackage{graphicx}
\graphicspath{{Final Figures/}}

\newcommand{\mgii}{Mg\,{\sc ii}}

\newcommand{\kms}{$\rm{km s^{-1}}$}

\begin{document}

\title{Cool CGM \mgii\ Absorption Across the Star-Forming, Green Valley, and 
Quiescent Transition}

\shorttitle{\mgii\ Across the SF-GV-Q Transition}

\author[orcid=0009-0002-1247-0096]{Simon Xinlin Wu}
\affiliation{ Department of Physics and Astronomy, North Carolina State University, Raleigh, NC 27695, USA}
\email[show]{xwu38@ncsu.edu}  
\author[orcid=0000-0002-3120-7173]{Rongmon Bordoloi}
\affiliation{ Department of Physics and Astronomy, North Carolina State University, Raleigh, NC 27695, USA}
\email[show]{rbordoloi@ncsu.edu}
\author[orcid=0000-0003-3769-9559]{Robert Simcoe}
\affiliation{MIT Kavli Institute for Astrophysics and Space Research, 77 Massachusetts Avenue, Cambridge, 02139, Massachusetts, USA}
\email{simcoe@space.mit.edu}

\author[orcid=0000-0003-0724-4115]{Andrew J. Fox}
\affiliation{AURA for ESA, Space Telescope Science Institute, 3700 San Martin Drive, Baltimore, MD 21218, USA}
\email{afox@stsci.edu}

 \author[orcid=0000-0002-0355-0134]{Jessica Werk}
 \affiliation{Department of Astronomy, University of Washington, Seattle, WA 98195, USA}
 \email{jwerk@uw.edu}

\author[orcid=0000-0002-7982-412X]{Jason Tumlinson}
\affiliation{Space Telescope Science Institute, 3700 San Martin Drive, Baltimore, MD 21218, USA}
\email{tumlinson@stsci.edu}

\author[orcid=0000-0002-7738-6875]{J. Xavier Prochaska}
\affiliation{Department of Astronomy \& Astrophysics, University of California Santa Cruz, 1156 High Street, Santa Cruz, CA 95064, USA}
 \email{xavier@ucolick.org}

\begin{abstract}
We investigate the distribution and kinematics of cool circumgalactic medium (CGM) gas  across the star-forming to quiescent transition using \mgii\ $\lambda\lambda$2796, 2803 absorption for 716 galaxies spanning $0.07<z<2.7$ (169 new and 547 archival galaxies). To compare galaxies uniformly across 10 billion years of cosmic time, we
introduce a star-formation offset metric ($\sigma_{\mathrm{SFO}}$), which
measures a galaxy's deviation from the star-forming main sequence of its
epoch. A key advantage of
$\sigma_{\mathrm{SFO}}$ is its ability to identify transitional green
valley galaxies as a distinct population across redshift, which would otherwise be obscured by binary
star-forming--passive classifications. We fit a virial-radius-normalized radial profile and find that the \mgii\ absorption strength declines with increasing projected distance from the host galaxy. The scatter around this mean profile correlates strongly with 
star-formation activity: radial profile residuals correlate positively with 
$\log_{10}(\mathrm{sSFR})$ and $\sigma_{\mathrm{SFO}}$, with star-forming 
galaxies showing excess absorption above the profile, green-valley galaxies 
showing intermediate residuals, and quiescent galaxies falling preferentially 
below it. The \mgii\ covering fraction in the inner CGM 
($R/R_{\mathrm{200}} < 0.25$) follows the same sequence, declining 
monotonically from star-forming through green-valley to quiescent systems. \mgii\ absorption kinematics are consistent with a predominantly bound 
cool CGM. Star-forming galaxies further exhibit a bimodal azimuthal dependence, with \mgii\ absorption enhanced along both the polar and disk directions, consistent with bipolar outflows and co-planar accretion. Together, these results indicate that the cool CGM tracks the 
quenching of star-formation in galaxies, with $\sigma_{\mathrm{SFO}}$ revealing 
a gradual decline in cool gas across the green valley rather than an 
abrupt star-forming--to--passive transition.

\end{abstract}


\section{Introduction} 

The circumgalactic medium (CGM) is broadly defined as the extended volume of diffuse gas surrounding galaxies, playing a central role in their evolution \citep{Tumlinson_2017}. Both observations and simulations have established that the 
dynamic, multiphase CGM acts as a central reservoir and regulator in the galaxy baryon 
cycle, where gas and energy are continuously exchanged between galaxies and their halos 
via supernova- and AGN-driven outflows and recycled inflows back onto the disk 
\citep{Bordoloi2011, Werk2014, Alcazar2017, Peroux_2020, Chen_2026, Faucher_2023, 
Nelson_2019, Borthakur_2015}. In this context, \mgii, which predominantly 
traces cool, $\sim10^4$\,K gas, provides a sensitive observational link between star 
formation and circumgalactic gas. Strong \mgii\ absorption can arise in metal-enriched material 
associated with supernovae \citep{Bordoloi2011, Bordoloi2014outflow} and in recycled material re-accreting onto galaxies to fuel future star formation 
\citep{Rubin2012}. Observationally, the \mgii\ $ \lambda\lambda$\,2796,\,2803 doublet, detected as absorption in the spectra of bright background sources, is 
redshifted into the wavelength range accessible to ground-based optical and 
near-infrared spectrographs over a broad redshift baseline of $z \approx 0.1$--$6$, 
making it a widely used tracer of CGM evolution across a large fraction of cosmic 
history \citep{Bergeron_1986,Bergeron_1991,Churchill2025, Chen2010, Chen2017, Dutta2020, Bordoloi2024}.

Systematic studies of \mgii\ absorption as a function of projected separation 
from diverse galaxy populations have yielded substantial progress in characterizing the 
statistical properties of the cool CGM. \mgii\ absorption strength is strongly 
anticorrelated with impact parameter $R$; the covering fraction $C_f$ approaches unity 
within $\sim$25\,kpc and rapidly declines beyond $\sim$100\,kpc 
\citep{Chen2010, Kacprzak2013, Bordoloi2024}. \mgii\ absorption strength
correlates strongly with galaxy star-formation activity, stellar mass \citep{Bordoloi2011, Lan2018, Das_2025}, and halo mass \citep{Churchill2013}, suggesting that strong \mgii\ absorption preferentially originates in star-formation-driven galactic outflows. This picture is reinforced by the observed azimuthal asymmetry of \mgii\ absorption. Strong systems aligned along the galaxy minor axis are associated with outflowing material 
\citep{Bordoloi2011, Bouche2012, Cherrey_2025}, while systems aligned along the major axis exhibit kinematics consistent with co-rotation, possibly tracing recycled accretion \citep{Ho2017, Martin2019}. Notably, significant cool \mgii\ absorption is also detected around passive, luminous red galaxies, suggesting that cool gas persists in quenched halos well after star formation has ceased \citep{ Huang2021, Zahedy_2018}. Galaxies residing in group environments 
show more extended \mgii\ radial absorption profiles than isolated systems, indicating that the group environment can significantly alter the cool CGM \citep{Chen2010, Bordoloi2011, Dutta2020}. These results are based primarily on statistical sampling of multiple independent halos using background quasars, and the large sightline-to-sightline scatter in \mgii\ radial profiles reflects genuine halo-to-halo (inter-halo) variations driven by the range of host galaxy properties described above. Recent gravitational lensing tomography studies that probe individual halos with multiple sightlines have demonstrated that the scatter in radial absorption profiles is substantially reduced within a single halo \citep{Lopez2018, Tejos2021, Shaban2025}, 
highlighting the importance of disentangling halo-to-halo variance from intra-halo (intrinsic) structure.

Despite this wealth of observational data, these studies collectively span a wide redshift range ($z \sim 0.1$--$6$), encompassing over 12\,Gyr of cosmic evolution, making it challenging to synthesize results into a coherent picture of CGM evolution. A fundamental difficulty arises from comparing galaxy populations across redshift: the 
star-forming main sequence (SFMS) itself evolves strongly with cosmic time, such that a galaxy on the main sequence at $z \sim 2$ would be classified as starburst at $z \sim 0.1$ \citep{Pearson2023, Whitaker2012, Speagle2014}. Meaningful comparisons across epochs therefore require a redshift-independent metric for characterizing 
star-formation activity. In this work, we adopt $\sigma_{\mathrm{SFO}}$, the offset of a galaxy's specific star-formation rate from the redshift-dependent SFMS expressed in units of the main-sequence scatter, as a redshift-independent metric for classifying star-formation activity. We apply this framework to one of the largest compiled samples ($N_{gal} \sim 716$) of galaxy-absorber pairs, 
combining new observations with archival data, to investigate how the cool CGM traced 
by \mgii\ evolves as a function of star-formation activity: from star-forming galaxies through the green valley to passive systems across 10 billion years  ($z \sim 0.07$--$2.7$) of galaxy evolution \citep{Salim2014, Schawinski2014}.

This paper is organized as follows. In Section~\ref{sec:Data}, we describe the galaxy and quasar spectral data used in this work. In Section~\ref{sec:Methods}, we outline our approach to measuring \mgii\ absorption and characterizing galaxy properties. In Section~\ref{sec:Results}, we present our results, and in 
Section~\ref{sec:Discussion}, we discuss their implications. Throughout, all calculations assume a flat $\Lambda$CDM cosmology with $H_0 = 67.7\ \mathrm{km\ s^{-1}\ Mpc^{-1}}$, $\Omega_M = 0.31$, and $\Omega_\Lambda = 0.69$ 
\citep{Planck2020}.

\begin{figure*}[ht!]
    \centering
    \includegraphics[height=6.5cm]{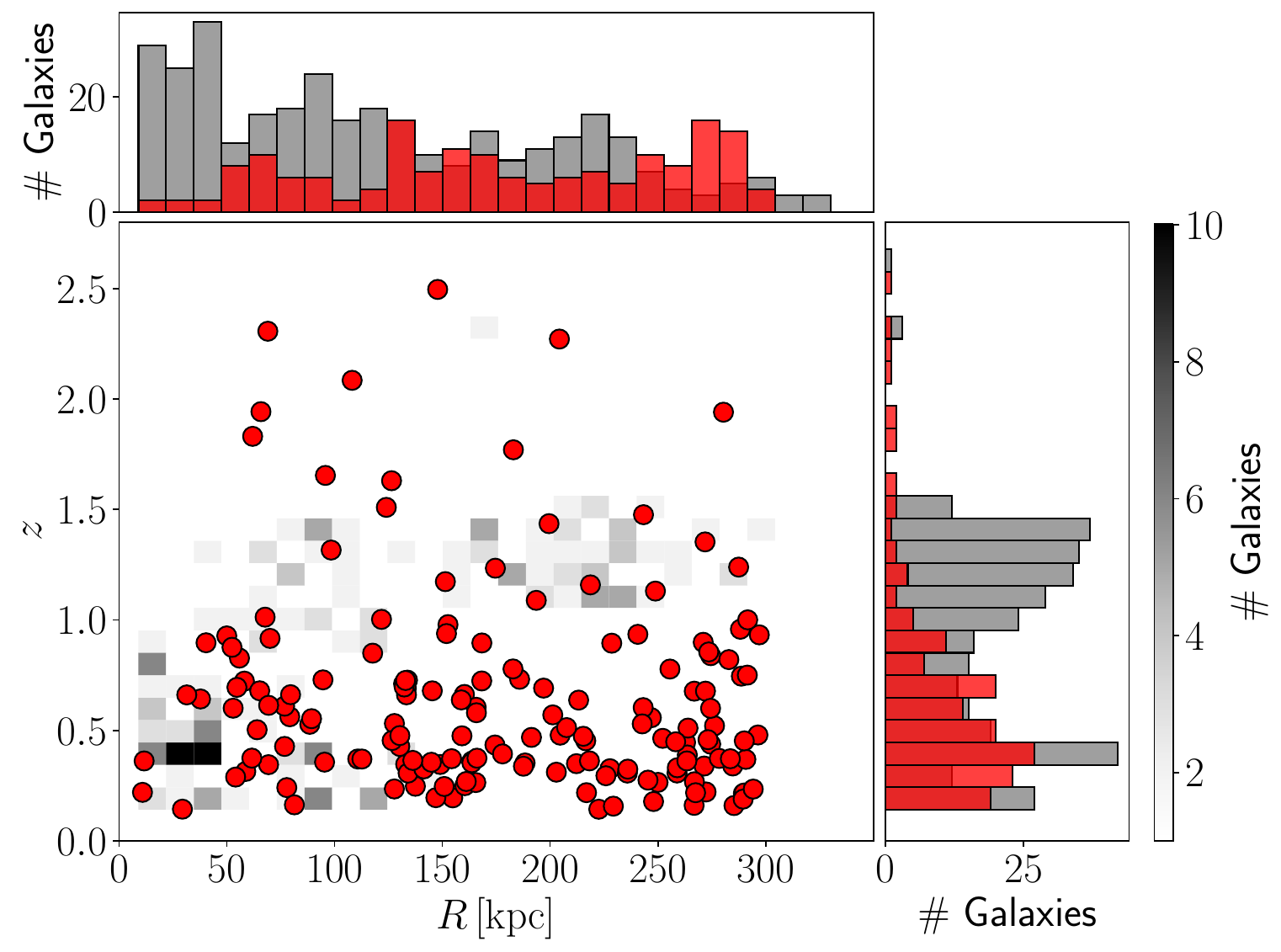} \hfill
    \includegraphics[height=6.5cm]{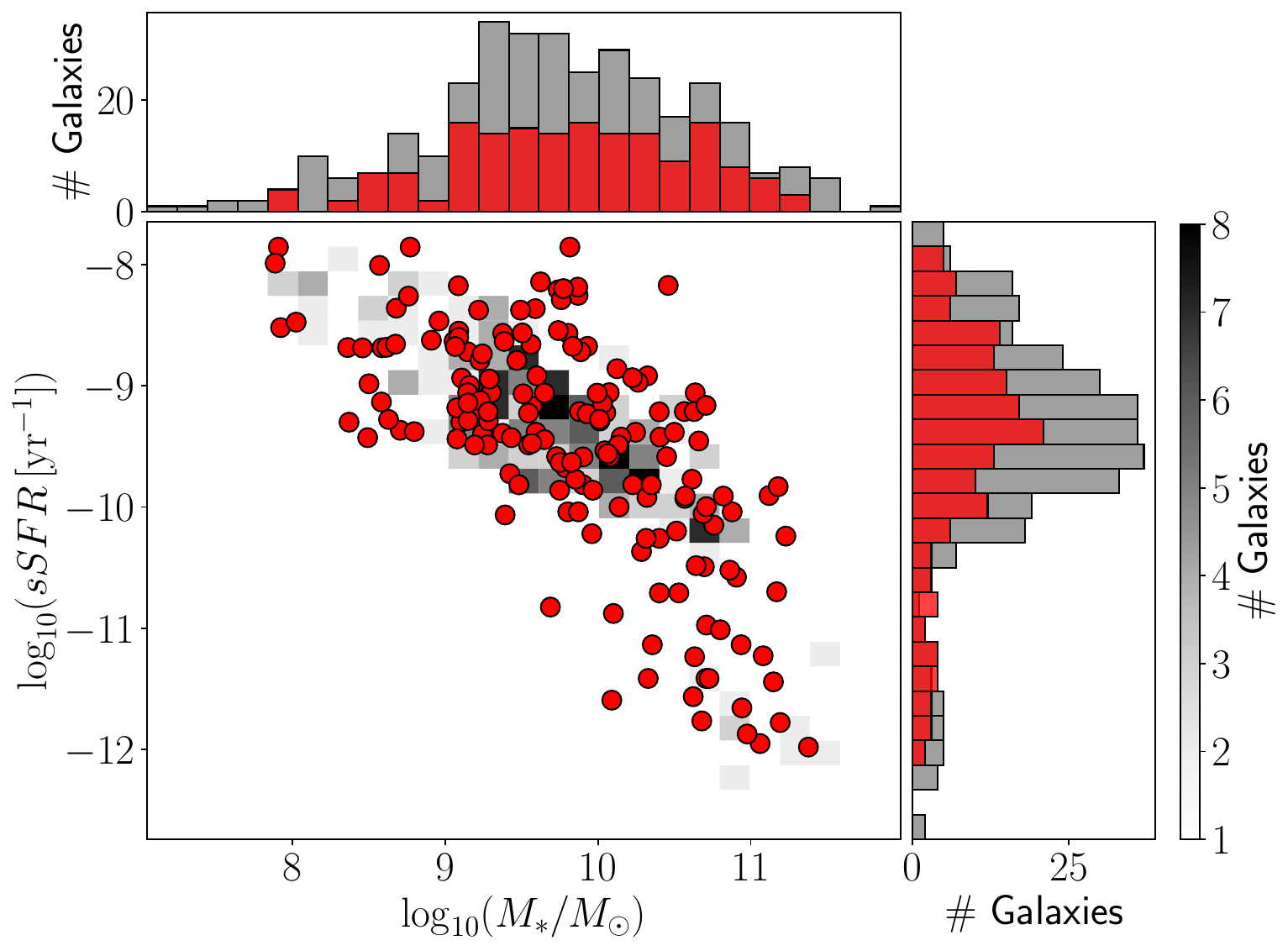}
    \caption{Properties of all galaxies used for analysis. New galaxies used in this work (red points) are shown overlaid on a 2D histogram of archival observations (gray, see Table \ref{tab:datasets} for details). \textit{Left panel:} Projected impact parameter ($R$) versus galaxy redshift ($z$). \textit{Right panel:} Specific star formation rate ($\log_{10} \mathrm{sSFR}$)  versus stellar mass ($\log_{10} M_*/M_{\odot} $). Marginal distributions for each variable are shown as 1D histograms above and to the right of each panel, using the same color scheme. Note that MAGIICAT galaxies are presented only on the left panel as they don't have published sSFR estimates.}
    \label{fig:2dhist_side_by_side}
\end{figure*}

\section{Data} \label{sec:Data}

Leveraging the large spectroscopic galaxy samples in the COSMOS field \citep{Lilly2009,Van_der_Wel_2021} and separately the CGM$^2$ survey \citep{Wilde_2021}, we select a sample of background QSOs at close projected separation from these galaxies. Below we describe these new observations (Section~\ref{subsec:newdata}) and the archival galaxy-absorber catalogs (Section~\ref{subsec:archival}) that together form 
our full sample.

\subsection{New Observations} \label{subsec:newdata}

Here we describe the new observations used for this work. The background quasars are selected from three tranches. We first describe how the background quasar samples are selected and then describe the corresponding galaxy samples. The observational details are presented in Table \ref{tab:qso}.

\textit{(i) COSMOS / MagE:}  Close QSO--galaxy pairs ($R<100$~kpc) are intrinsically rare, so we prioritize and select sightlines that pierce galaxy overdensities at close projected separations. We select 12 QSOs with at least one spectroscopically confirmed foreground galaxy at an impact parameter of $R<100$~kpc in the 
COSMOS field in the magnitude range $19 < r_{\mathrm{AB}} < 21.5$, ensuring sufficient continuum for absorption-line measurement, and observe them with the Magellan Echellette (MagE) on the Magellan Telescopes at Las Campanas Observatory. The spectra are reduced with the \texttt{MASE} pipeline \citep{Bochanski2009}. MagE's high blue sensitivity is essential for the redshifted \mgii\, $\lambda\lambda\;2796,2803$ doublet, which falls near $\sim$4700\,\AA\
at $z\sim0.7$, and a 3$\sigma$ detection threshold of  $W_r\approx30$\,m\AA\ ($N_{\mathrm{Mg\,II}}\sim10^{12}$\,cm$^{-2}$) for the highest SNR  sightlines. This tranche comes from a galaxy-selected survey; rather than identifying \mgii\ systems first and searching for host galaxies, we define a galaxy sample with bright background QSOs at close impact parameters that is selected without any prior knowledge of \mgii\ absorption.

\textit{(ii) COSMOS / VLT archival:} In the second tranche, we select 9  archival QSO spectra in the COSMOS field from the ESO archive. These QSOs are observed with the Very Large Telescope: 2 with UVES and 7 with X-shooter, obtained under ESO programme IDs 083.A-0401(A) and 086.A-0974 (PI: S. J. Lilly). We use the publicly available ESO Phase~3 reduced data products and co-add the individual spectrograph arms by inverse-variance weighting. The X-shooter spectra are shallower, reaching a 3$\sigma$ detection threshold of $W_r\approx$ 100\,m\AA. The UVES spectra reach a 3$\sigma$ detection threshold of $W_r\approx$ 40 $\mathrm{m\AA}$ and resolve individual \mgii\ components.

\textit{(iii) COS-Halos HIRES spectra, new CGM$^2$ galaxies:} In the third tranche, we reanalyze 11 Keck/HIRES QSO spectra from the COS-Halos survey \citep{Werk2013}, reduced as described therein, and cross-match them with galaxies from the CGM$^2$ survey \citep{Wilde_2021,Tchernyshyov_2022} that have not previously been analyzed for \mgii\ absorption. Because these counterparts are drawn from a galaxy catalog distinct from the original COS-Halos targets, the resulting pairs are independent of the archival COS-Halos sample used in Section~\ref{subsec:archival}. The original COS-Halos QSO sample was screened against known strong \mgii\ absorbers with $W_r>1$ \AA\ at $z>0.4$ \citep{Tumlinson2013}. However, only 11 of the 40 CGM$^2$ galaxies identified along these sightlines lie at $z>0.4$, so this selection is expected to introduce only modest bias in the overall galaxy sample. The HIRES spectra reach a 3$\sigma$ detection threshold of $W_r\approx$ 20 $\mathrm{m\AA}$.

For all QSO spectra, we impose a sensitivity floor of $\mathrm{SNR}_{200} \approx 9$, measured in a
200\kms\ window (matching our non-detection windows) at each galaxy redshift. This cutoff is set by the noisiest spectral slice in our sample that can still detect \mgii\  absorption with $3\sigma$ statistical significance. This removes portions of the spectrum with insufficient sensitivity and provides a uniform threshold across instruments of differing resolution. To exclude proximate, QSO-associated gas, we require galaxies to lie at least 2000 \kms\ blueward of the QSO systemic redshift ($\Delta v\leq-2000$ \kms).

\begin{deluxetable}{llrrr}
\tablecaption{Background quasar observations \label{tab:qso}}
\setlength{\tabcolsep}{2pt}
\tablehead{
\colhead{Telescope} & \colhead{Instrument} &
\colhead{$R$} &
\colhead{$\mathrm{SNR}_{200}$} & \colhead{$N_{\rm QSO}$} \\
\colhead{} & \colhead{} & \colhead{} &
 \colhead{} & \colhead{}
}
\startdata
Magellan & MagE      & 4100--4800        & $40$  & 12 \\
VLT  & UVES      & $\sim40{,}000$  & $90$  & 2  \\
VLT  & X-shooter & $\sim5400$       & $20$  & 7  \\
Keck & HIRES     & $\sim45{,}000$  & $150$ & 11 \\
\enddata
\tablecomments{Here, $\mathrm{SNR}_{200}$ is the median signal-to-noise ratio measured across each quasar sample over a 200~\kms\ window at $\sim$5000\,\AA.}
\end{deluxetable}

\paragraph{Galaxy Sample}
We now describe the galaxy sample used in this analysis. To identify galaxies at close projected separation of the background quasars, we search within 300~kpc of each QSO sightline for spectroscopically confirmed galaxies for both the COSMOS and CGM$^2$ fields. For the  sightlines in the COSMOS field, galaxies are drawn from a master catalog merged from three spectroscopic surveys: zCOSMOS-Bright ($0.1<z<1.2$), zCOSMOS-Deep ($1.4<z<3.0$), and LEGA-C ($0.6<z<1.0$) \citep{Lilly_2007,Van_der_Wel_2021}. We retain only robust zCOSMOS redshifts corresponding to confidence classes 2.5 and 3--5 \citep{Lilly2009,Lilly_2007} for zCOSMOS objects and require $\texttt{FLAG SPEC}=0$ for LEGA-C galaxies \citep{Van_der_Wel_2021}. For the HIRES sightlines, we cross-match the QSO fields against the CGM$^2$ galaxy catalog \citep{Wilde_2021,Tchernyshyov_2022} and require reliable redshifts with $\texttt{ZQ}>2$.

The spectroscopic catalogs have nonuniform targeting and redshift-completeness. zCOSMOS-Bright is an $I_{\rm AB}<22.5$ magnitude-limited survey and is  $48\%$ complete over the full survey area and $56\%$ complete in the central region \citep{Knobel_2012}. Among spectroscopically observed targets, the secure-redshift sample is 88\% complete overall and 95\% complete at $0.5<z<0.8$, with an estimated reliability of approximately 99\% \citep{Lilly2009}. zCOSMOS-Deep instead consists of a deeper, fainter sample in the central $1~\mathrm{deg}^{2}$ of the COSMOS field and has an effective spectroscopic sampling of approximately $55\%$ relative to its color-selected parent catalog \citep{Diener_2013,Lilly_2007}. In our compilation, LEGA-C supplements the zCOSMOS catalogs by providing additional high-quality spectroscopy. LEGA-C obtained spectroscopic redshifts for $35\%$ of its primary parent sample, with nearly all observed targets yielding redshifts \citep{Van_der_Wel_2021}. The CGM$^2$ survey is nearly spectroscopically complete for galaxies with $I_{AB}<22$ within $2'$ of the QSO sightline, with completeness decreasing to approximately 50\% at $I_{AB}<23$ \citep{Wilde_2021}. The typical spectroscopic-redshift uncertainties are approximately $110$ \kms\ for zCOSMOS-Bright and $50$--$100$ \kms\ for CGM$^2$ \citep{Lilly2009,Wilde_2021}.

After all quality cuts, the new sample comprises 169 galaxies within 300 kpc of background QSO sightlines.

\paragraph{Galaxy stellar mass and star-formation rate}
For galaxies in the COSMOS field, stellar masses and SFRs are obtained from the zCOSMOS survey \citep{Lilly2009}. Six galaxies lack these measurements in the zCOSMOS survey; for these we cross-match against the UltraVISTA $K_s$-selected catalog \citep{Muzzin_2013} and adopt
its stellar masses and SFRs. For the CGM$^2$ galaxies, stellar masses and SFRs are taken from the CGM$^2$ survey \citep{Wilde_2021}, where they are derived by fitting the
broadband photometric spectral energy distribution with CIGALE. From these stellar masses we then derive halo radius
$R_{200}$, defined as the radius within which the mean enclosed density is 200 times the critical density of the Universe at the galaxy redshift, halo mass, and circular and escape velocities for all galaxies following \citet{Bordoloi2024}, using
\textsc{UniverseMachine} abundance matching \citep{Behroozi2019} together with the
$c$--$M$ relation of \citet{Dutton_2014}.

\paragraph{Galaxy Environment}
We classify each galaxy's environment in two steps. First, we adopt the group memberships from the zCOSMOS-Bright group catalog \citep{Knobel_2012}. Second, to capture groups not flagged in this catalog, we perform an independent companion search within a cylindrical window of 600~kpc and $\pm600$ \kms\ in line-of-sight velocity, classifying any galaxy with more than two companions in this window as a group member. This procedure identifies 15 group galaxies from the catalog and 4 additional members from the companion search, for a total of 19 group galaxies in the new sample.

\begin{deluxetable}{lccccc}
\tablecaption{Galaxy--absorber datasets used in this work \label{tab:datasets}}
\tablehead{
  \colhead{Survey} & \colhead{$N_{\rm gal}$} & \colhead{$z$} & \colhead{$R$ (kpc)} & \colhead{Env.} &
  \colhead{$\sigma_{\mathrm{SFO}}$}
}
\startdata
This Work                         & 169 & 0.14--2.5   & $<300$  & iso+grp & yes \\
MAGG \citep{Dutta2020}            & 156 & 1--1.5      & $<330$  & iso+grp & yes \\
MEGAFLOW \citep{Cherrey_2025}     & 66  & 0.4--1.5    & $<150$  & iso     & yes \\
COS-Halos \citep{Werk2013}        & 40  & 0.1--0.3    & $<150$  & iso     & yes \\
EIGER \citep{Bordoloi2024}        & 4   & 2--2.7      & $<300$  & iso     & yes \\
Rubin \citep{Rubin_2018}          & 72  & 0.25--1.25  & $<50$   & iso     & yes \\
MAGIICAT \citep{Nielsen_2013,Nielsen_2018} & 209 & 0.07--1.1 & $<200$ & iso+grp & no \\
\enddata
\tablecomments{Summary of the new (this work) and archival galaxy--absorber datasets. The
$\sigma_{\mathrm{SFO}}$ column indicates whether the dataset provides SFRs suitable
for the SFMS offset analysis.}
\end{deluxetable}

\subsection{Archival Data} \label{subsec:archival}
To build a statistical sample spanning a wide range of galaxy properties, we combine the
new data with six archival \mgii\ galaxy--absorber surveys (Table~\ref{tab:datasets}). We apply the definitions used for group identification from Sections ~\ref{subsec:newdata} and define the offset from the star-formation main sequence ($\sigma_{SFO}$; see next section~\ref{subsec:sigma_SFO}) to archival datasets. The MAGG survey \citep{Dutta2020} contributes 156 isolated and group galaxies at $1<z<1.5$ with typical impact parameters of $R \simeq 137$--$262$ kpc
(16th--84th percentile) extending beyond the inner CGM. MEGAFLOW \citep{Cherrey_2025} adds an isolated sample of 66 galaxies with measured galaxy orientations. COS-Halos \citep{Werk2013} provides a low-redshift ($0.1<z<0.3$) sample of 40 galaxy--absorber pairs with associated velocities and Voigt-profile column densities $N$. The COS-Halos fields contain some probable group or cluster environments, but the available environmental information is not sufficient to classify all systems consistently; we therefore treat these galaxies as isolated for the $\sigma_{\mathrm{SFO}}$ analysis. The EIGER survey \citep{Bordoloi2024} contributes four high-redshift galaxy--absorber pairs at $2<z<2.7$ near cosmic noon, also with associated velocities and Voigt-profile column densities. These galaxies have not been found to have close companions, and are therefore classified as isolated, with a caveat that surveys are incomplete at these high redshifts. The survey from \citet{Rubin_2018} supplies a sample of 72 galaxies at small impact parameters ($R<50$~kpc), where the group environment is expected to be negligible, so we treat these galaxies as isolated. Finally, the isolated and group samples of \citet{Nielsen_2013,Nielsen_2018} provide a sample of 209 galaxies within 200~kpc of QSO sightlines, used only for the isolated and group
radial profile fits; because this survey doesn't provide SFR estimates and uses a $B-K$ color proxy rather than SFRs, it is excluded from the $\sigma_{\mathrm{SFO}}$ analysis.

After joining all datasets, the full sample contains 716 galaxies (93 in groups, 623
isolated). Of the isolated galaxies, 440 have SFR estimates and form the basis of our
main $\sigma_{\mathrm{SFO}}$ analysis. Figure~\ref{fig:2dhist_side_by_side} shows the
galaxy-property distributions for the new and archival samples.

\section{Methods} \label{sec:Methods} 

\subsection{Galactic Star Formation Classification}
\label{subsec:sigma_SFO}

\begin{figure*}[ht!]
    \centering

    \includegraphics[height=4.33cm]{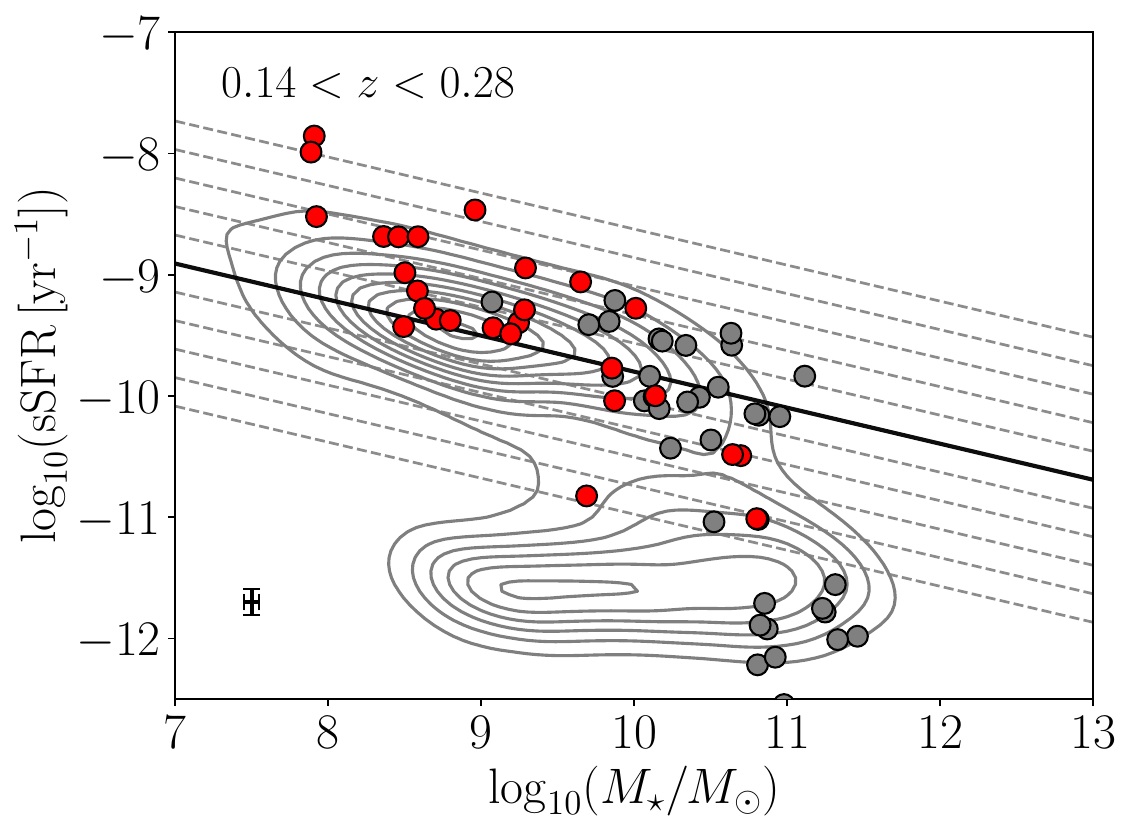} \hfill
    \includegraphics[height=4.33cm]{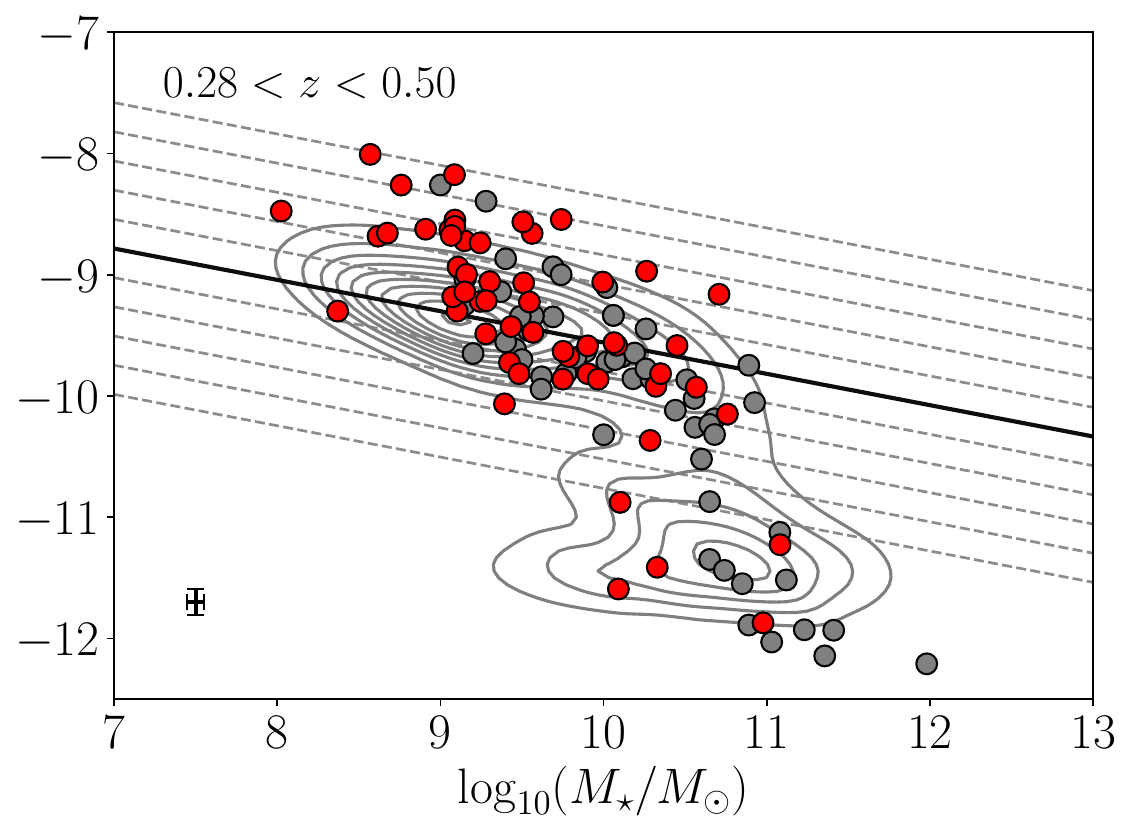} \hfill
    \includegraphics[height=4.33cm]{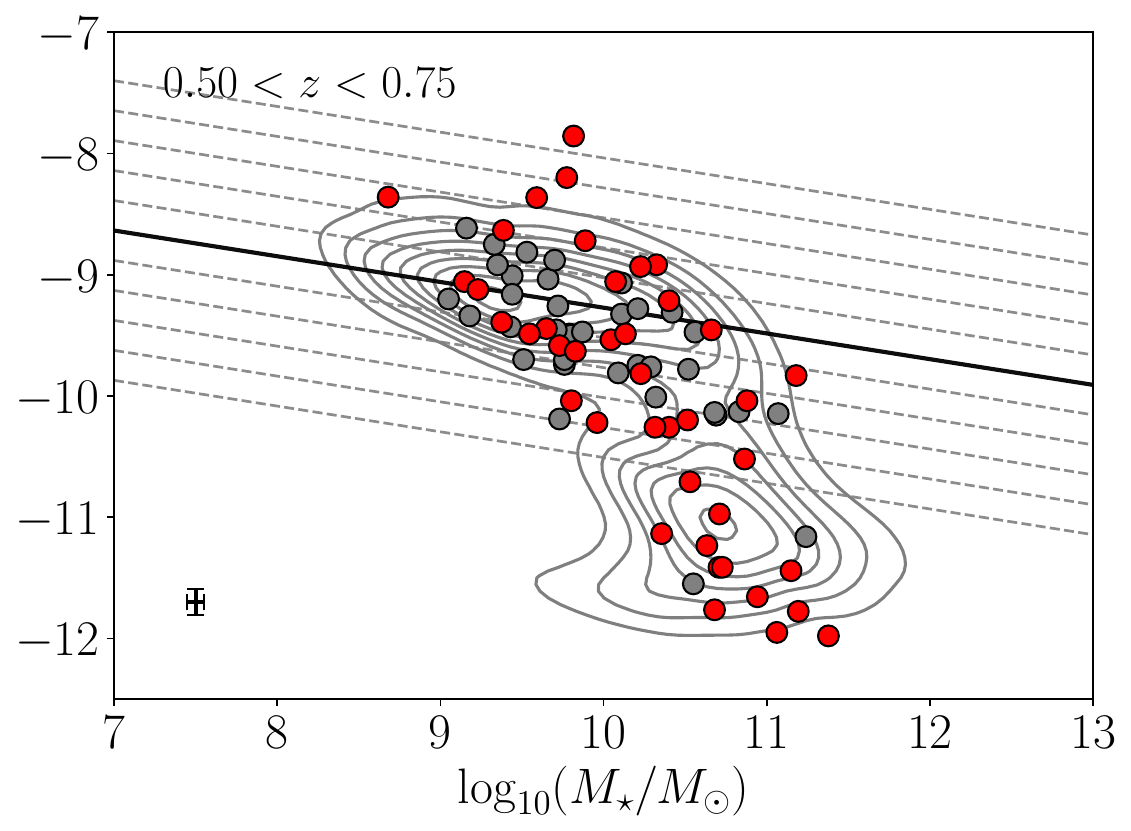} \\

    \includegraphics[height=4.33cm]{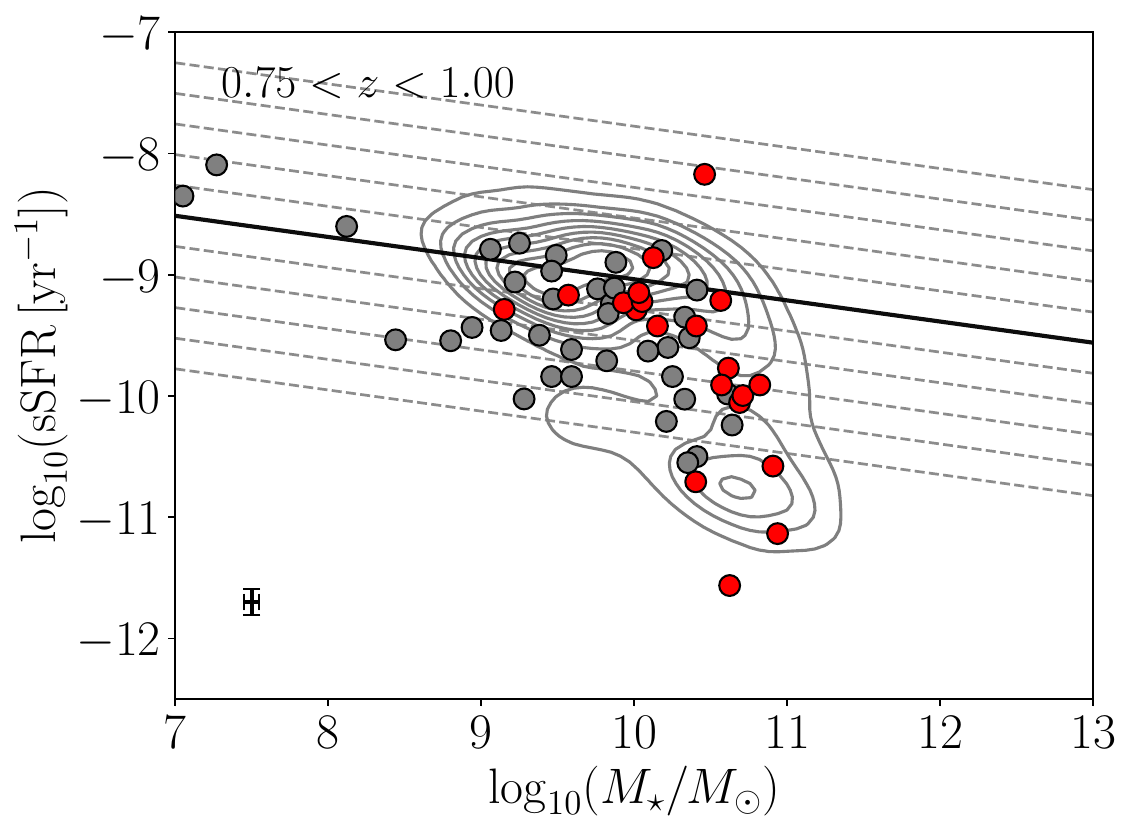} \hfill
    \includegraphics[height=4.33cm]{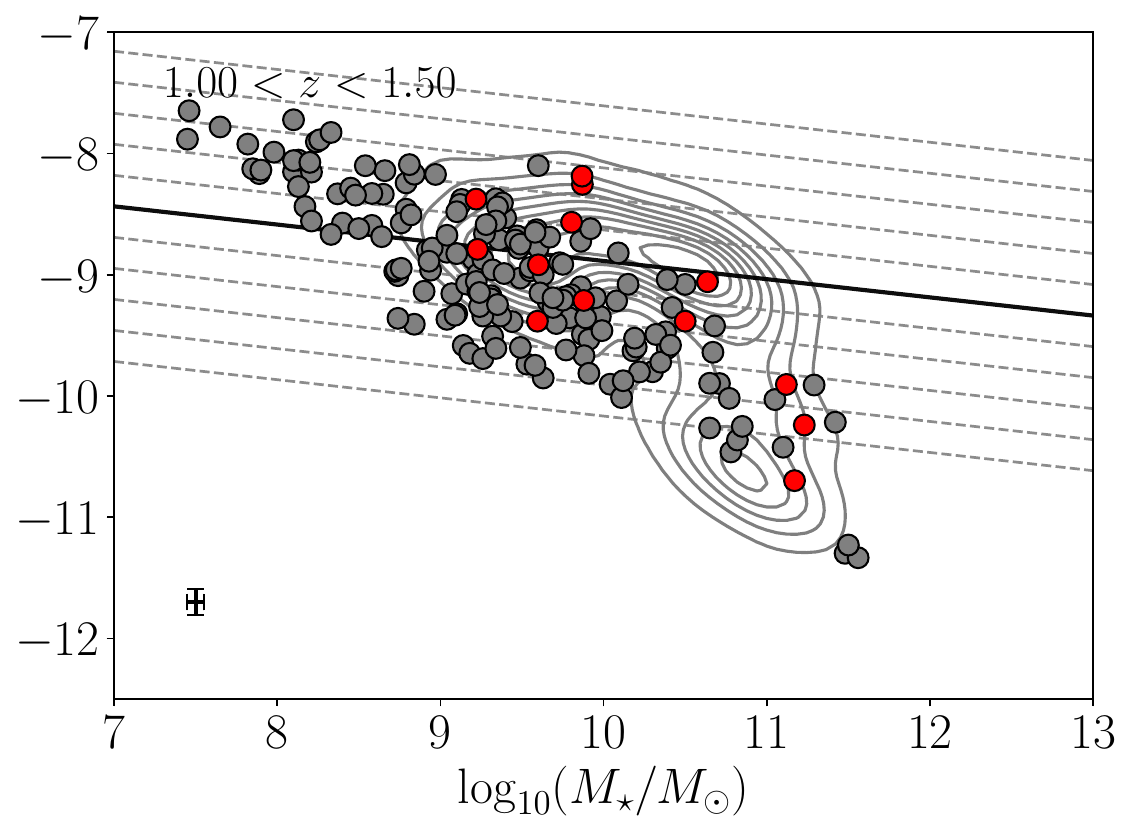} \hfill
    \includegraphics[height=4.33cm]{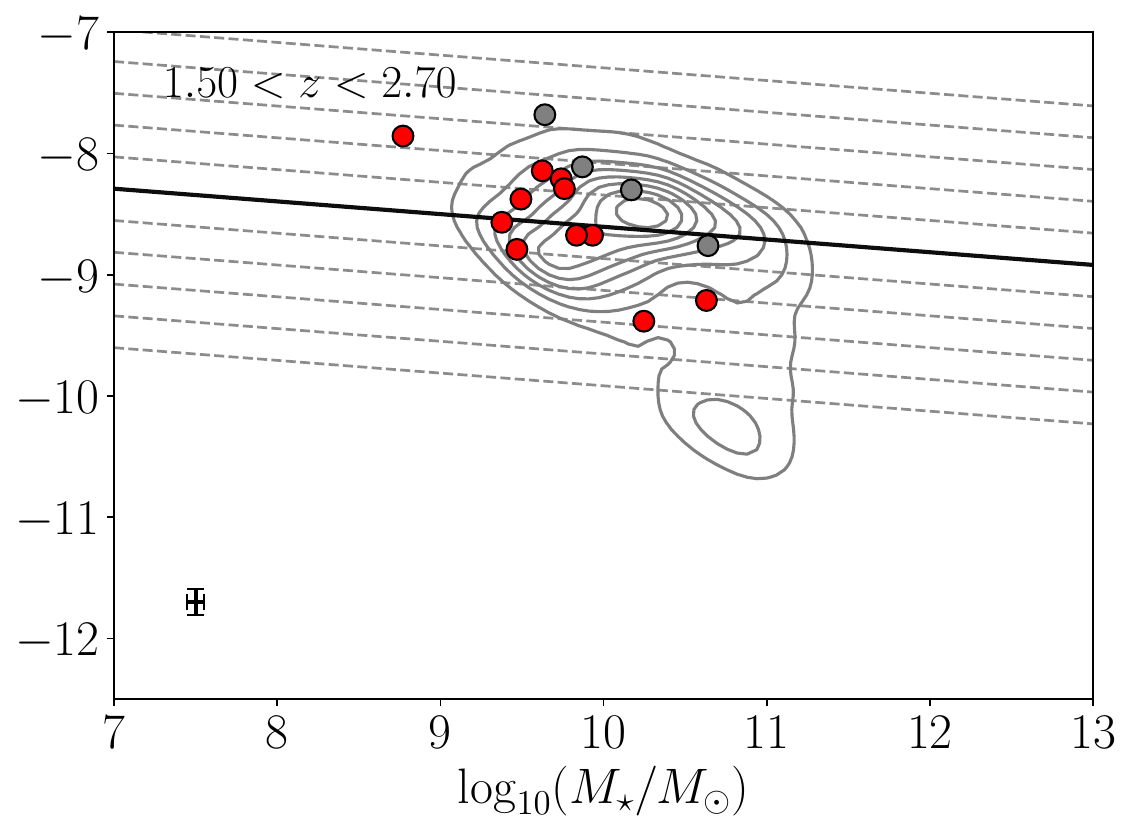}
    \caption{ $\log_{10}(\mathrm{sSFR})$ versus $\log_{10}(M_*/M_{\odot})$ across six example redshift bins showcasing galaxy distributions at different cosmic times, where the SFMS and $\sigma_{\mathrm{SFO}}$ are defined based on the central redshift of the bin. The background population is depicted with Gaussian kernel density estimation (KDE) contours overlaid with galaxies from this work (red circles) and archival datasets (gray circles). The black line in each panel defines the SFMS fit and gray dashed lines show $\sigma_{\mathrm{SFO}}$ from  -5 to 5 at increments of 1 $\sigma_\mathrm{SFO}$. A black crossbar at the bottom left of each panel shows typical zCOSMOS survey uncertainties of $\sim$0.05~dex and $\sim$0.11~dex for $\log_{10}(M_*/M_{\odot})$ and $\log_{10}(\mathrm{sSFR})$, respectively \citep{Darvish_2017}. Note: our definition of $\sigma_{SFO }$ also includes a mass cut as explained in Section ~\ref{subsec:sigma_SFO} , which is implemented for the SFMS fitting procedure but not in the display of the background population.}
    \label{fig:ssfr_mass_redshift_grid}
\end{figure*}

Galaxy populations follow a bimodal distribution in specific star-formation rate (sSFR)  vs stellar mass (M*) phase space with modes corresponding to star-forming and passive populations. Additionally, from $z=0$ to cosmic noon, both the global average sSFR and the star-forming main sequence (SFMS) shifts higher \citep{Ilbert}. Instead of dividing galaxies into ``blue'' (star-forming)  and ``red'' (passive) populations based purely on a constant sSFR cut across this large span of time, we reclassify galaxies based on the statistical offset ($\sigma_{\mathrm{SFO}}$) of a galaxy from the SFMS at a given redshift. With  $\sigma_{\mathrm{SFO}}$, we can quantify how the extent of quenching (negative $\sigma_{\mathrm{SFO}}$) affects the strength of \mgii\ absorption in the CGM while accounting for global galaxy evolution over time. Figure~\ref{fig:ssfr_mass_redshift_grid} shows the distribution of galaxies in our sample in front of the UltraVISTA galaxy background population \citep{Muzzin_2013} as well as the defined main sequence and star formation offsets across different redshift bins. We see in the figure that at lower redshifts, the main sequence slope of COSMOS field galaxies is negative and gradually becomes flatter at higher $z$. Concurrently, the normalization of the SFMS increases with redshift.

The overplotted archival samples are based on heterogeneous stellar-mass
and SFR measurements, which may contribute to differences in their locations within the sSFR--mass plane. For example, COS-Halos derives SFRs from dust-corrected H$\alpha$, H$\beta$, and
[O\,\textsc{ii}] emission-line measurements \citep{Werk2012}, while MEGAFLOW uses a stellar-mass-dependent [O\,\textsc{ii}] calibration \citep{Cherrey_2025}. Other surveys rely on broadband SED fitting \citep{Rubin_2018,Wilde_2021}, joint spectro-photometric fits \citep{Dutta2020,Bordoloi2024}, or a combination of UV+IR and SED-based estimates \citep{Kovac2014}. Such methodological differences can introduce survey-dependent offsets in $M_\star$ and SFR \citep{Conroy2013,Kennicutt_2012}, thereby adding scatter to the inferred sSFR and $\sigma_{\mathrm{SFO}}$ values and potentially moving some galaxies across the adopted population boundaries. We therefore interpret small offsets among individual datasets cautiously and focus primarily on trends that persist across the combined sample. Among all subsamples with $N>50$, our sample and that of \citet{Cherrey_2025} exhibit the closest alignment with the background sSFR--mass distribution, determined using a two-sample 2D K-S-test. 

Our implementation of $\sigma_{\mathrm{SFO}}$ for categorizing galactic star formation is described as follows. We start by defining a galaxy population for which the SFMS is estimated using photometric redshifts from the UltraVISTA $K$-selected catalog. These galaxies have robust photometric redshifts ($\Delta z/(1 + z) = 0.013$ \citep{Muzzin_2013}). We cross-match this catalog with the zCOSMOS galaxy catalog and use the spectroscopic redshifts for that subsample of galaxies present in both catalogs. 

We next divide this galaxy background sample into 21 redshift bins, applying a stellar-mass cut of $9.25 < \log_{10}(M_\star/M_\odot) < 11$ in each bin, since selection effects can bias galaxy surveys toward brighter systems \citep{Ilbert_2013}. After an initial sSFR cut in each bin to isolate the star-forming population mode, we perform a sigma-clipped fit to the star-forming main sequence (SFMS). We model the relation as

\begin{equation}
\log_{10}\!\left(\mathrm{sSFR}_{\mathrm{SFMS}}\right)
=
\alpha_{\mathrm{init}}
\left[
\log_{10}\!\left(\frac{M_\star}{M_\odot}\right) - M_{\mathrm{pivot}}
\right]
+
\gamma_{\mathrm{init}},
\end{equation}

where $M_{\mathrm{pivot}}=9.7$ dex is fixed for all redshift bins. In this parameterization,
$\alpha_{\mathrm{init}}$ is the slope of the SFMS, while $\gamma_{\mathrm{init}}$ gives the
normalization of the relation at the pivot mass. This yields, for each redshift bin, a first-pass
slope, normalization, and scatter  $(\alpha_{\mathrm{init}}, \gamma_{\mathrm{init}},
\sigma_{\mathrm{init}})$, which we associate with the central redshift $z_{\mathrm{center}}$
of that bin.

To obtain continuous functions for these main-sequence parameters across redshift, we fit the
sigma-clipped $\alpha_{\mathrm{init}}$, $\gamma_{\mathrm{init}}$, and $\sigma_{\mathrm{init}}$
values with linear functions of cosmic time:
\begin{align}
\alpha(t) &= m_{\alpha}t + b_{\alpha}, \\
\gamma(t) &= m_{\gamma}t + b_{\gamma}, \\
\sigma_{\mathrm{SFMS}}(t) &= m_{\sigma}t + b_{\sigma},
\end{align}

where $t$ is cosmic time in Gyr. The sigma clipping of the measured scatter values is important
because the highest-redshift bins can have inflated scatter due to larger measurement uncertainties,
smaller sample sizes, and stronger selection effects. After clipping these high-scatter outliers,
the fitted $\sigma_{\mathrm{SFMS}}(t)$ relation gives typical scatter values of
$\sim 0.2$--$0.3$ dex, comparable in magnitude to commonly reported intrinsic SFMS scatter values
\citep{Whitaker2012,Speagle2014}. This comparison verifies that the fitted scatter used to normalize
star-formation offsets remains physically reasonable and is not dominated by extreme high-redshift
outlier bins.

Using cosmic time in Gyr, the final smoothed SFMS parameter relations are

\begin{align}
\alpha(t) &= (-0.011 \pm 0.041) + (-0.0257 \pm 0.0056)t, \\
\gamma(t) &= (-8.021 \pm 0.025) + (-0.1519 \pm 0.0034)t, \\
\sigma_{\mathrm{SFMS}}(t) &= (0.275 \pm 0.015) + (-0.0036 \pm 0.0020)t
\end{align}

and the rest of the fit parameters can be found in Table ~\ref{tab:SFMS_parameters_pivot}. Star-formation offsets for each galaxy, given its $M_\star$, $z$, and sSFR, are then computed as

\begin{equation}
\sigma_{\mathrm{SFO}} =
\frac{
\log_{10}\left(\mathrm{sSFR}\right)
-
\log_{10}\!\left(\mathrm{sSFR}_{\mathrm{SFMS}}(M_\star,z)\right)
}{
\sigma_{\mathrm{SFMS}}(t(z))
},
\end{equation}

providing a measure of star-formation offset that can be compared uniformly across redshift.

As a test of how stellar-mass and SFR uncertainties affect $\sigma_{\mathrm{SFO}}$, we perform Monte Carlo perturbations using representative uncertainties for the zCOSMOS and CGM$^{2}$ samples. For each galaxy, we perturb $\log M_\star$ and $\log(\mathrm{SFR})$, recompute $\log(\mathrm{sSFR})=\log(\mathrm{SFR})-\log M_\star$, and then recalculate $\sigma_{\mathrm{SFO}}$. We take the standard deviation of the resulting $\sigma_{\mathrm{SFO}}$ distribution as the uncertainty for each galaxy. Adopting representative zCOSMOS uncertainties of $0.05$ dex in $\log M_\star$ and $0.10$ dex in $\log(\mathrm{SFR})$, based on \citet{Darvish_2017}, yields a median uncertainty of $\delta\sigma_{\mathrm{SFO}}\approx0.4$. This suggests that the classifications of most zCOSMOS galaxies are relatively well constrained, aside from objects near the adopted population boundaries. Repeating the test with uncertainties of $0.5$ dex in both $\log M_\star$ and $\log(\mathrm{SFR})$, representative of the CGM$^{2}$ sample, produces a characteristic uncertainty of $\delta\sigma_{\mathrm{SFO}}\approx2$. The $\sigma_{\mathrm{SFO}}$ values and classifications of some CGM$^{2}$ galaxies should therefore be interpreted with greater caution.

Based on the implemented statistical offset scheme, we divide galaxies into starburst ($\sigma_{\mathrm{SFO}}> 3$), main sequence ($-3 < \sigma_{\mathrm{SFO}} \le 3$), green valley ($-4.5 < \sigma_{\mathrm{SFO}} \le -3$), and quiescent ($\sigma_{\mathrm{SFO}} \le -4.5$) populations. These boundaries are chosen
to approximately align with the structure of the background 
Gaussian KDE distribution of galaxies, while
recognizing that the exact divisions between star-forming, transitioning,
and passive galaxy populations are not uniquely defined
\citep{Salim_2014,Renzini_2015}. We adopt this four-way split when quantifying the residual fraction $R_f$, (see details in Section~\ref{subsec: Radial Profile Residuals}) and inner CGM covering fraction $C_f$ (see details in the last paragraph of Section~\ref{subsec: Covering Fraction}), in order to explicitly trace changes in \mgii\ absorption across the star-forming--to--passive transition. For figures where we show $R_f$ or $C_f$ as a function of 
$\log_{10}(\mathrm{sSFR})$, we define the sSFR bin edges using the same
population boundaries adopted in $\sigma_{\mathrm{SFO}}$ space. To do this,
we evaluate the smoothed SFMS at the median redshift and median stellar mass
of the relevant galaxy sample, and then convert the $\sigma_{\mathrm{SFO}}$
boundaries into equivalent $\log_{10}(\mathrm{sSFR})$ boundaries. This ensures
that the sSFR bins correspond directly to the quiescent, green valley,
main-sequence, and starburst divisions used in $\sigma_{\mathrm{SFO}}$ space,
rather than being arbitrary fixed-width sSFR bins. Applying this four-way classification to the isolated subsample yields 21 starbursts, 327 main sequence galaxies, 35 green valley galaxies, and 57 quiescent galaxies. Because the starburst and green valley subsamples are inherently small, we also adopt a two-class division by combining green valley and quiescent galaxies into a single ``passive'' category ($\sigma_{\mathrm{SFO}}\le -3$)  when computing binned statistics that require higher signal-to-noise (e.g., $C_f(R)$ and $C_f(R/R_{\mathrm{200}})$ in Section ~\ref{subsec: Covering Fraction}). This two-class split is also used for analyses with limited ancillary data, including the velocity and azimuthal-angle comparisons in Section ~\ref{subsec: Velocities} and Section ~\ref{subsec: 2D Distribution}, respectively, where \citet{Werk2013} and \citet{Bordoloi2024} report velocities (used for comparison only) and \citet{Cherrey_2025} provides azimuthal-angle measurements (used as a combined dataset).

We therefore use two classification schemes throughout the paper: (i) a four-way split (starburst, star-forming, green valley, quiescent) when emphasizing trends across the green valley, and (ii) a two-way split (star-forming versus passive) for situations with limited statistics.

\subsection{\mgii\ Absorption-line Analysis }
\label{absorption line analysis}
Absorption line analysis is performed on the fully reduced and co-added quasar spectra. We perform the analysis at the systemic redshift of each galaxy. We first shift the quasar spectrum to the rest frame of the galaxy and create a spectral slice around a $\pm$1000 \kms\ velocity window centered on the \mgii\ $\lambda\lambda$ 2796, 2803 transition. This spectral slice is continuum normalized by performing a Legendre  polynomial fit. This continuum-normalized spectral slice is then visually inspected and a velocity window is defined over which \mgii\  rest frame equivalent width ($W_r$) is computed. The \mgii\ absorption is quantified as a detection if $W_r$ is significant at the $3\sigma$ level. We also inspect the doublet ratio and line profiles of the two \mgii\ transitions to verify the measurements. All the analysis is performed using the Python package \texttt{rbcodes} \citep{rbcodes}.

In 3 out of 28 detections, the weaker $\lambda2803$ line falls below $3\sigma$, in which case we additionally require these lines to be free from any visible contamination from common metal transitions at other redshifts and to be outside the range of the Ly$\alpha$ forest. We also cross check for presence of Fe II transitions to validate these lines.  Systems that do not meet the $3\sigma$ detection threshold are treated as non-detections, for which we calculate $2\sigma$ upper limits over a $\pm100$ \kms\  window centered on the galaxy rest frame similar to the treatment adopted in \citet{Bordoloi2024}.

For each detection, we perform Voigt profile-fits to measure accurate \mgii\  column density (N) measurements. We use the Python-based Bayesian Voigt profile fitter API \texttt{rbvfit} \citep{rbvfit} and fit for Doppler parameter b, N, and velocity v for each discernible cloud. For saturated components of the 2796 line, defined as those with normalized fluxes in three consecutive pixels below 0.05 or degenerate column density solutions, we treat the fitted column densities as lower limits. Consequently, the total column density $N_{tot}$ of the associated system is also treated as a lower limit. We mask simple  contamination from other intervening systems as needed and when more complex blends are present model these contaminating lines along with the \mgii\ absorption profiles. We also rescale archival non-detection limits to a common 2$\sigma$ threshold for our analysis. 

When more than one galaxy could be associated with an absorber, we use the galaxy at the smallest impact parameter within $\pm1000$ \kms\ as the primary galaxy for that absorber, a commonly adopted galaxy--absorber association method \citep{Schroetter_2016,Richter_2016,Dutta2020} noting that in group environments, others may instead prefer  methods that prioritize mass \citep{Dutta2020}. In a group environment, the galaxy closest to the QSO sightline should not automatically be assumed to be the sole source of the observed absorption, because the absorbing gas may arise from another group member, overlapping circumgalactic media, or a more extended intragroup medium \citep{Burchett_2013}. However, because the absorber-associated galaxy is defined identically (smallest impact parameter) in both isolated and group settings, differences in the observed absorption distributions between the two reflect physical effects of a group environment rather than association criteria.

Out of our sample of 169 galaxies, 28 are associated with detected \mgii\ absorption. All the detections along with the Voigt profile fits are presented in the appendix in Figures~\ref{fig:MgII_UVES}, \ref{fig:MgII_HIRES}, \ref{fig:MgII_MAGE_01}, and  \ref{fig:MgII_XSHOOTER}, respectively.

\subsection{\mgii\ Association with Galaxy Properties }
\label{Main analysis}
 Our primary goal in this paper is to understand the relationship between \mgii\ and the green valley transition using $\sigma_{\mathrm{SFO}}$. We account for the radial dependence of \mgii\ absorption by first fitting $W_r$ radial profiles using a Markov Chain Monte Carlo (MCMC)-based fitting algorithm. We model the radial profile as an exponential function:
\begin{equation}
\overline{W_r}(R) = A \, \exp\left(-\frac{R}{B}\right)
\end{equation}
This is mathematically equivalent to the log-linear models used in previous studies \citep{Nielsen_2013,Dutta2020,Cherrey_2025}, but returns a physically interpretable $W_r$ amplitude $A$ and cutoff radius $B$ \citep{Bordoloi2014outflow}. We first implement this model as a function of $R$ to compare isolated and group fits from this work and archival observations. We define our log-likelihood function as:
\begin{align}
\ln \mathcal{L} = 
& -\frac{1}{2} \sum_{i \in \mathcal{D}} 
  \left[ \frac{(W_i - \overline{W}_{i})^2}{\sigma_i^2} 
  + \ln (2\pi \sigma_i^2) \right] \notag\\
& + \sum_{i \in \mathcal{N}} 
  \ln \!\left[ \frac{1}{2} 
  \left( 1 + \mathrm{erf} 
  \!\left( \frac{W_{N,i} - \overline{W}_{i}}
  {\sqrt{2}\,\sigma_i} \right) \right) \right]
\label{eq:loglike}
\end{align}

where \( \mathcal{D} \) and \( \mathcal{N} \) denote the sets of detections and non-detections, respectively; \( W_i \) is the observed rest-frame equivalent width; \( \overline{W_i} \) is the model prediction at impact parameter \( R_i \); \( W_{N,i} \) is the upper limit threshold for non-detections ($2\sigma$); \( \sigma_i \) is the total uncertainty, given by
\begin{equation}
\sigma_i^2 = \sigma_{mi}^2 + \sigma_{\mathrm{sc}}^2,
\end{equation}
with \( \sigma_{mi} \) representing the measurement error and \( \sigma_{\mathrm{sc}} \) accounting for intrinsic scatter \citep{Chen2010,Dutta2020,Rubin_2018}. The free parameters of the model are \( A \), \( B \), and \( \sigma_{\mathrm{sc}} \).  We also derive our main radial profile fit for analysis as a function of normalized virial radius $R/R_{200}$, giving a radial profile of the form:

\begin{equation}
\overline{W_r}(R/R_{200}) = A \, \exp\left(-\frac{R/R_{200}}{B}\right)
\label{equation:radial_profile}
\end{equation}

Our main analysis comparing isolated galaxies across the green valley transition uses this $R/R_{\mathrm{200}}$ form, which offers a clearer one-to-one comparison of galaxies with different halo masses. From simulations, \citet{Alcazar2017} found that gas recycling occurs in a characteristic inner-halo zone that scales with halo size. Therefore, normalizing impact parameter by halo radius provides a physically motivated way to compare \mgii\ absorption around galaxies of different masses and redshifts \citep{guha2022,Churchill2013}.

After calculating the $R/R_{\mathrm{200}}$  radial profile fit (equation \ref{equation:radial_profile}), we compute log residuals, defined as the logarithmic difference between the observed data and the corresponding radial profile at a given $R/R_{\mathrm{200}}$

\begin{equation}
\Delta (R/R_{200})\equiv \log_{10} W_r (R/R_{200}) - \log_{10} \overline{W}_r(R/R_{200}),   
\label{equation:delta}
\end{equation}

 and compare how log residuals vary with galaxy properties $\log_{10}({M_*/M_{\odot})}$, $\sigma_{\mathrm{SFO}}$, and $\log_{10}(\mathrm{sSFR})$. This allows us to quantitatively study if the scatter seen in the  \mgii\ radial profile is correlated with galaxy properties.

In addition, we quantify the residual fraction $R_f$ to analyze these trends in discrete bins to compare continuous-fit results with our distinct definitions of star formation activity. $R_f$ represents the fraction of detections with equivalent widths above the best-fit model prediction, defined as:

\begin{equation}
R_f = \frac{N_{\text{det}} \left( W_r^{\text{det}} > \overline{W_r} \right)}{N_{\text{det}} +N_{\text{non}} \left( W_r^{\text{lim}} < \overline{W_r} \right)}
\label{equation:Rf}
\end{equation}

The numerator, $N_{\text{det}} \left( W_r^{\text{det}} > \overline{W_r} \right)$, is the number of detected absorbers with observed rest-frame equivalent widths $W_r^{\text{det}}$ exceeding the predicted value $\overline{W_r}$ from the model at the corresponding $R/R_{\mathrm{200}}$. The denominator is the total number of galaxies with associated detection absorption strength or 2$\sigma$ non-detection upper limits \( W_r^{\text{lim}} \) lower than the model prediction. For $R_f$, only detections are included in the numerator, as non-detection probability distributions above the fit encompass the area both above and below the fit.

To test if the enhancement of CGM \mgii\ in galaxies with higher star formation activity is evident at various physical distances,  we compute covering fractions \( C_f \), defined as the fraction of galaxies within a bin of data that show associated \mgii\ absorption stronger than a specified rest-frame equivalent width cut \( W_c \):
\begin{equation}
C_f = \frac{N_{\text{det}} \left( W_r^{\text{det}} > W_c \right)}{N_{\text{det}} + N_{\text{non}} \left( W_r^{\text{lim}} < W_c \right)},
\end{equation}

where the numerator is the number of detections required to have a \( W_r \) greater than or equal to the threshold \( W_c \) and the denominator is the total number of galaxies associated with detections or non-detection upper limits \( W_r^{\text{lim}} \) lower than $W_c$. All $C_f$ and $R_f$ binomial uncertainties for this paper are derived from Wilson score confidence intervals.

Next, we determine differences in absorption system kinematics between star-forming and passive halos by calculating individual cloud velocities and column densities using the aforementioned Voigt profile fitting method (section \ref{absorption line analysis}).

 Finally, our last aim is to quantify the 2D distribution of \mgii\ absorbers by determining the azimuthal angle between the galaxy major axis and the QSO sightline using archival HST/ACS F814W imaging and JWST/NIRCam F115W, F150W, F277W, and F444W imaging from COSMOS and COSMOS-Web \citep{IRSA_COSMOS_DOI,Casey_2023,Koekemoer_2007,Massey_2010}. We fit radial elliptical isophotes with \texttt{Photutils} and determine the semi-major axis from a contiguous set of intermediate-to-outer isophotes with stable orientations across radius. We retain only measurements with azimuthal angle uncertainties below $5^\circ$ and require the semi-major axis to extend at least 1.5 PSF FWHM from the galaxy center. We visually inspect the fitted isophotes to verify that the selected ellipses trace the galaxy-edge light distribution and are not dominated by neighboring sources or image artifacts.  For each acceptable set of isophotes, we adopt the inverse-variance-weighted mean azimuthal angle. The uncertainty is taken as the scatter in the azimuthal angles derived from the selected isophotes and for galaxies with acceptable measurements from multiple images, we retain the measurement with the smallest uncertainty. These criteria ensure that the orientations are spatially resolved, but may preferentially select galaxies with larger angular sizes and higher surface brightnesses. We restrict our analysis to galaxies with inclinations  $i>56^\circ$. We define $\theta=0^\circ$ along the projected major axis and $\theta=90^\circ$ along the projected minor axis.

\section{Results} \label{sec:Results}

\subsection{Radial distribution of \mgii\ absorption}
\label{subsec: Radial Profile}
\begin{figure*}[ht!]
    \centering
    \includegraphics[width=0.49\textwidth]{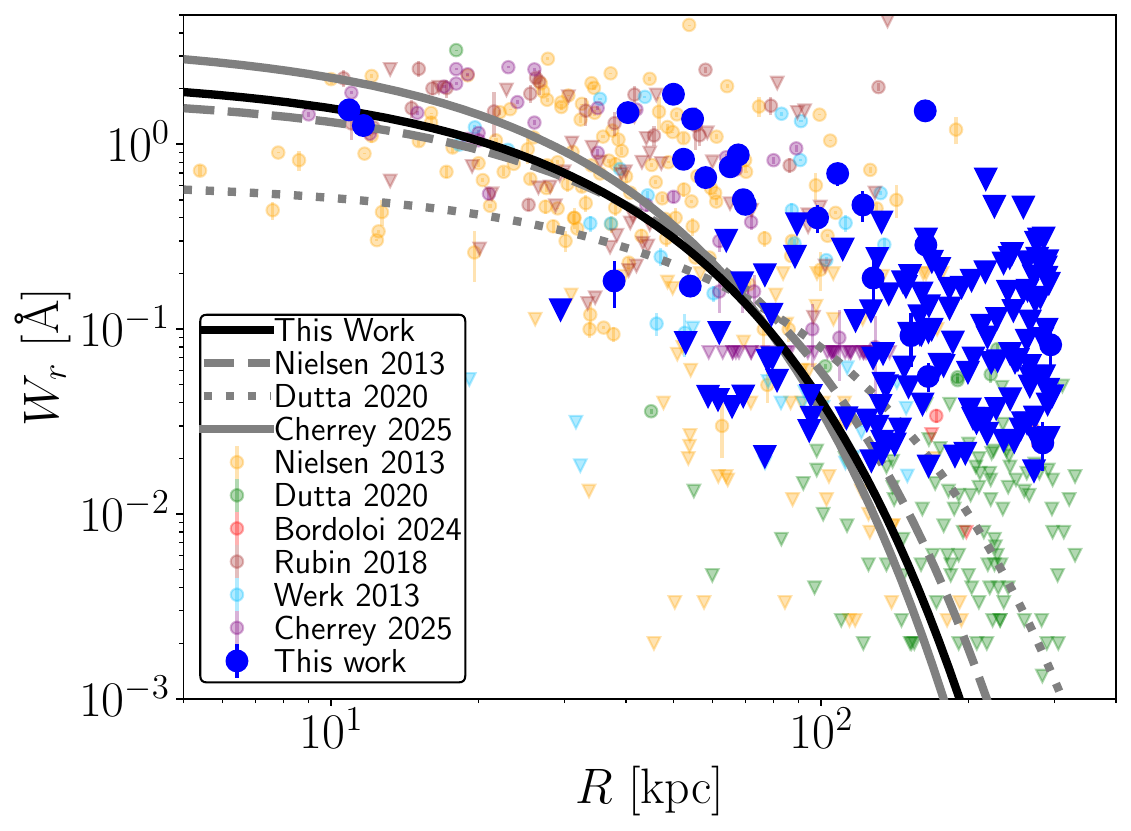} \hfill
    \includegraphics[width=0.49\textwidth]{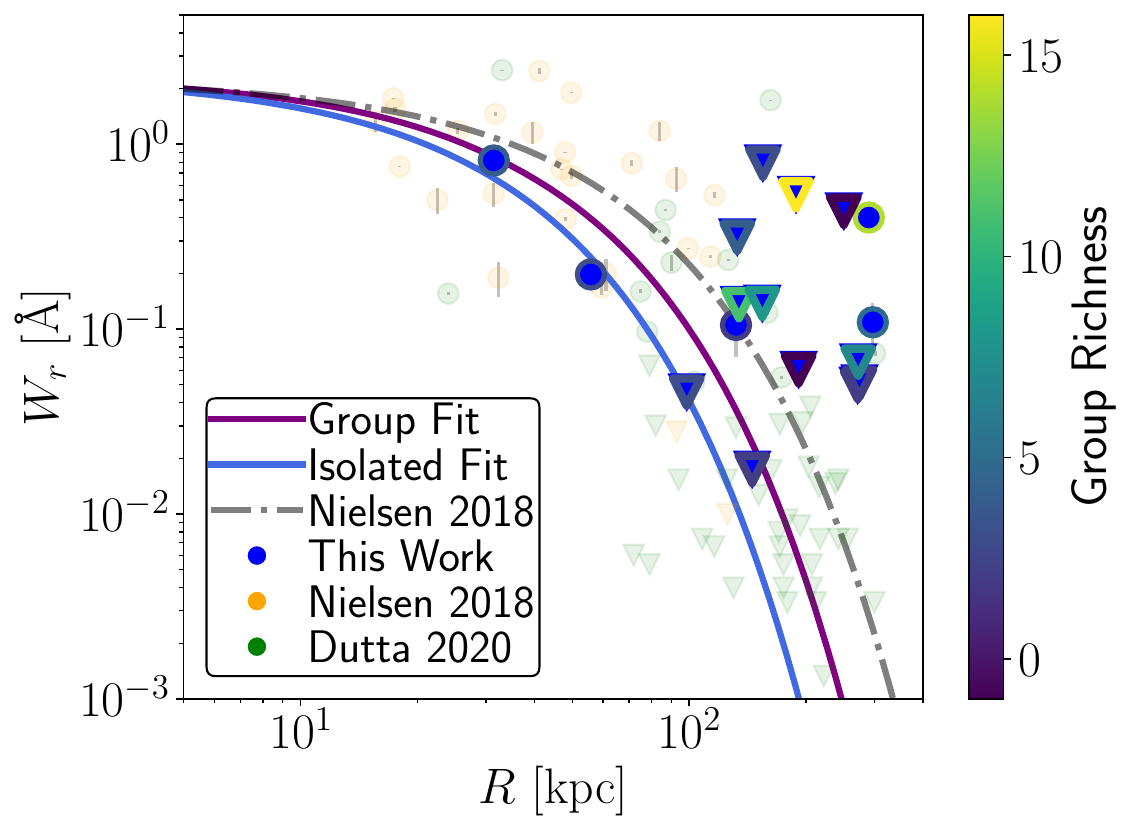}

\caption{\mgii\ radial absorption profiles with $W_r$ versus impact parameter $R$ for isolated (left) and group (right) galaxies. Circles and inverted triangles represent detections and upper limits, respectively. Data points are color-coded by literature source. In the left panel, the black line shows the fit from this work, gray lines show literature comparisons. In the right panel, data from this work has outlines color-coded by group richness. Nielsen 2013 and Nielsen 2018 refer to the isolated and group fits of their work, respectively.}
    
\label{fig:radial_profile_R}
\end{figure*}

\begin{figure}[tbh!]
\centering
\includegraphics[width=\columnwidth]{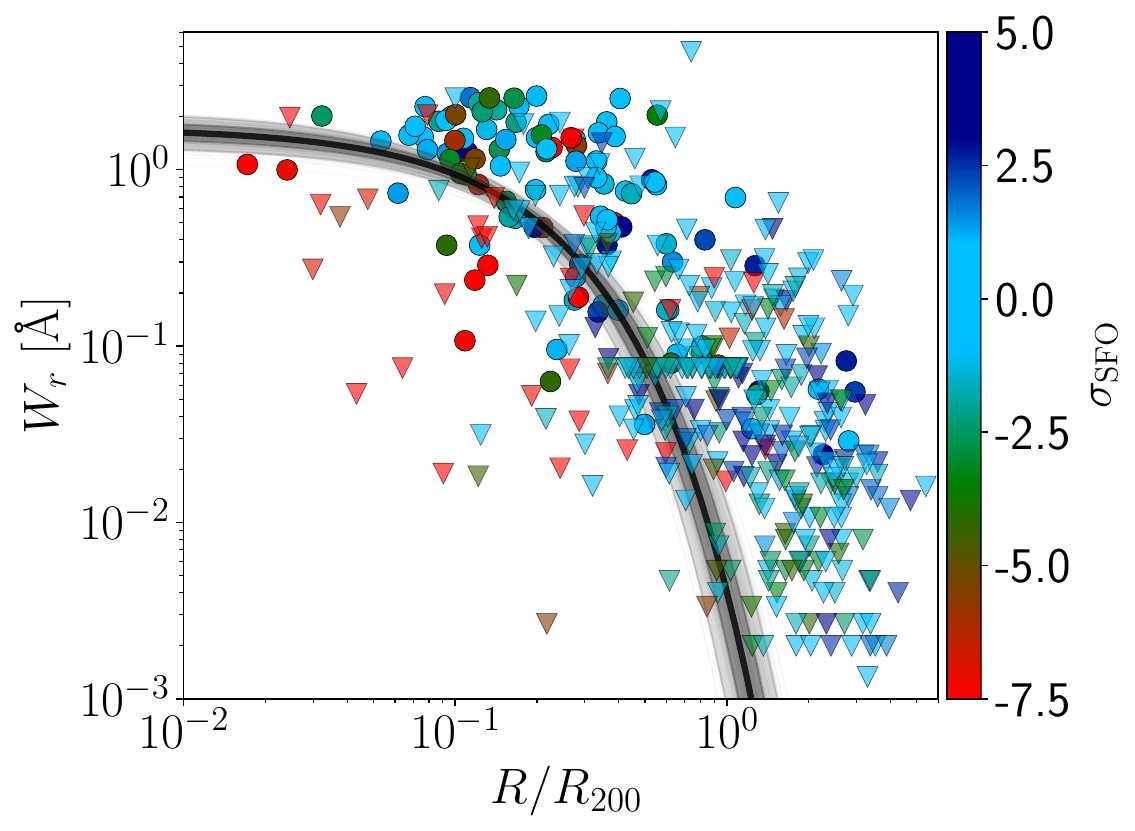}

\caption{\mgii\ radial absorption profile around isolated galaxies as a function of $R/R_{\mathrm{200}}$. The best-fit model is shown as a black line with 1$\sigma$ and 2$\sigma$ bootstrapped confidence intervals (gray shaded regions). Data points are color-coded by star formation offset $\sigma_{\mathrm{SFO}}$.}
\label{fig:radial_profile_Rvir}
\end{figure}

In this section we present the projected radial \mgii\ absorption profiles around isolated and group galaxies from this survey. Figure~\ref{fig:radial_profile_R} shows the \mgii\ radial absorption profiles  as a function of impact parameter $R$ for isolated (left) and group (right)  galaxies. For isolated galaxies, \mgii\ is regularly detected out to $\sim$100~kpc,  beyond which the detection rate drops sharply. Group galaxies exhibit a modestly  more extended profile, with significant detections reaching $\sim$200~kpc.  An exponential law with a cutoff provides an adequate description of  $\overline{W_r}(R)$ for both samples; best-fit posterior parameters are listed in 
Table~\ref{tab:fit_parameters}.

For isolated galaxies, the best-fit radial profile is comparable to 
\citet{Nielsen_2013} but less extended and steeper than \citet{Dutta2020}. 
The group profile has a larger exponential cutoff radius 
($B = 31.8^{+7.7}_{-9.3}$~kpc) than the isolated profile 
($B = 27.4^{+2.9}_{-3.2}$~kpc), consistent with enhanced cool CGM extent 
in denser environments.

Given the well-established correlation between CGM absorption strength 
and halo mass \citep[e.g.,][]{Chen2010, Churchill2013,Bordoloi2014,Bordoloi2018}, 
we additionally fit the $R/R_{\rm 200}$-normalized isolated profile. 
The best-fit exponential cutoff is at 
$B = 0.165^{+0.019}_{-0.021}~R_{\rm 200}$ 
(Table~\ref{tab:fit_parameters}), indicating that the bulk of cool CGM 
absorption is confined well within the halo radius. 
Figure~\ref{fig:radial_profile_Rvir} shows this fit with data points 
color-coded by $\sigma_{\rm SFO}$: the strongest absorbers are concentrated 
at small $R/R_{\rm 200}$, and a visual segregation by star-formation state 
is apparent. Quiescent systems ($\sigma_{\rm SFO} < -4.5$) scatter 
preferentially below the mean profile, while star-forming systems 
($-3 < \sigma_{\rm SFO} < 3$) extend to larger normalized radii. 
We quantify this segregation in Section~\ref{subsec: Radial Profile Residuals} 
using radial profile residuals and covering fractions.

\begin{table}[!tbh]
\centering
\small
\setlength{\tabcolsep}{2pt}
\caption{Radial profile fit posterior parameters}
\begin{tabular}{lccc}
\hline
Parameter & $R$: Group & $R$: Isolated & $R/R_{\mathrm{200}}$: Isolated \\
\hline
$A$ 
& $2.34^{+0.79}_{-0.57}$ 
& $2.28^{+0.27}_{-0.24}$ 
& $1.72^{+0.20}_{-0.19}$ \\

$B$ 
& $31.8^{+7.7}_{-9.3}$ 
& $27.4^{+2.9}_{-3.2}$ 
& $0.165^{+0.019}_{-0.021}$ \\

$\log_{10}(\sigma_{sc)}$ 
& $-0.17^{+0.05}_{-0.05}$ 
& $-0.11^{+0.02}_{-0.02}$ 
& $-0.09^{+0.02}_{-0.02}$ \\
\hline
\end{tabular}
\label{tab:fit_parameters}
\end{table}

\begin{figure*}[ht!]
    \centering
    \includegraphics[width=0.33\textwidth]{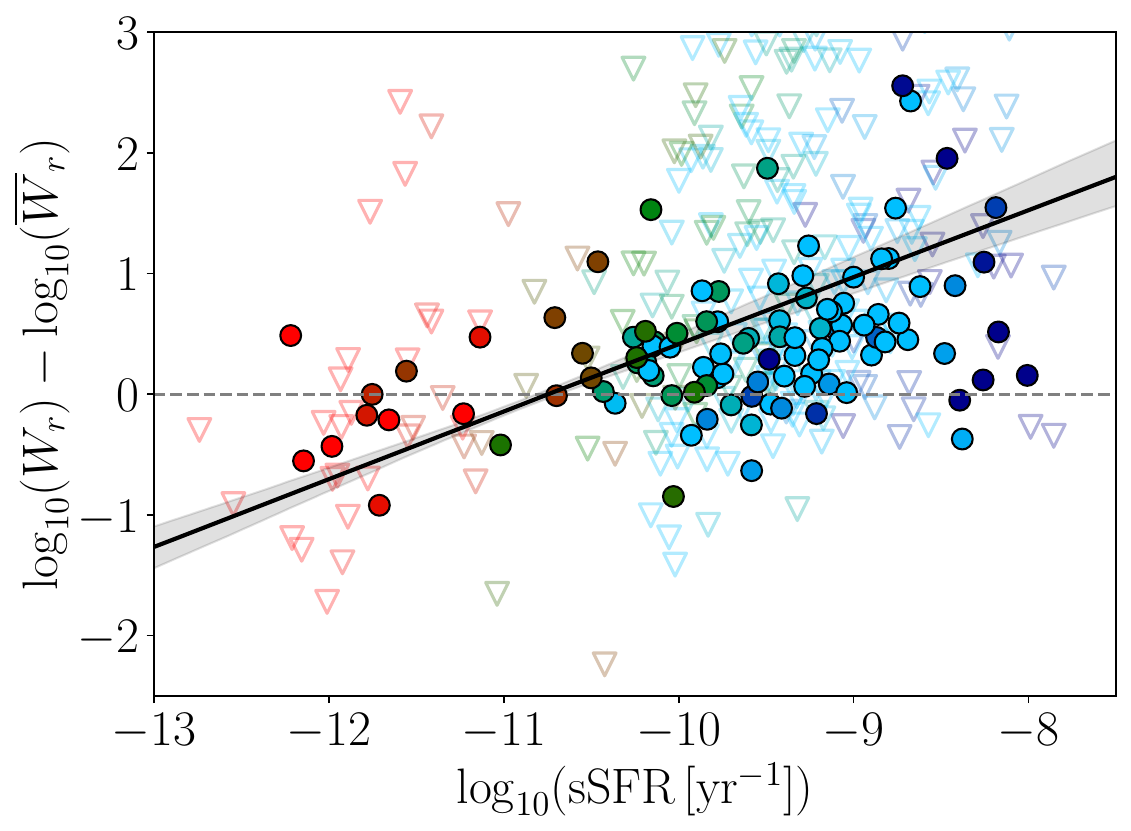} \hfill
    \includegraphics[width=0.325\textwidth]{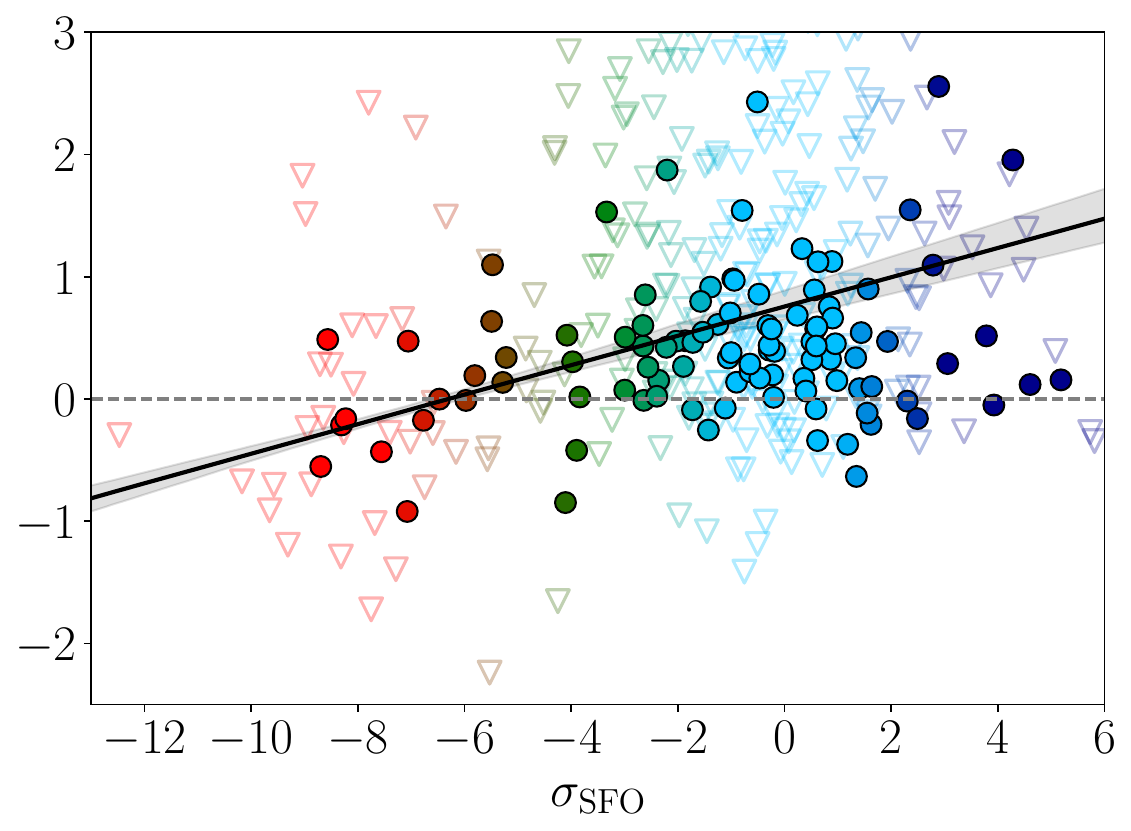} \hfill
    \includegraphics[width=0.325\textwidth]{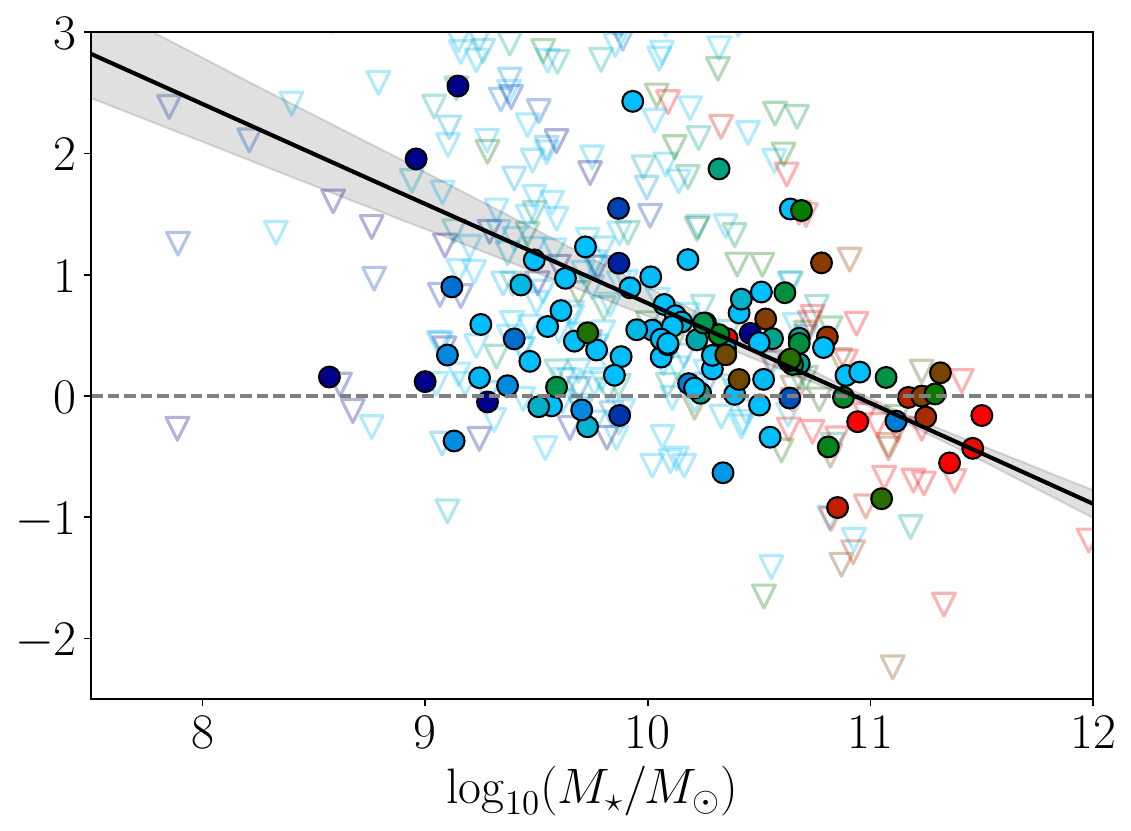}

    \includegraphics[width=0.65\textwidth]{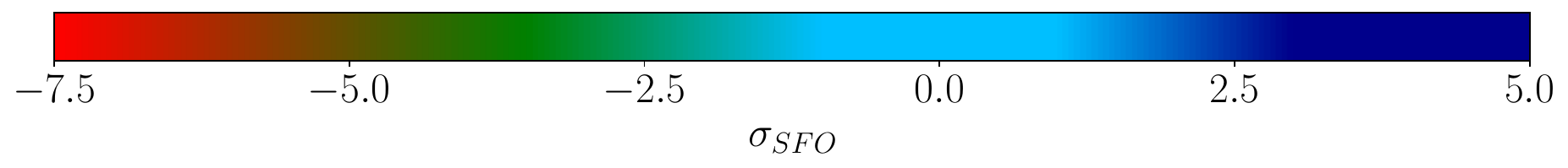}
    \vspace{0.5em}

 \caption{\mgii\ absorption $W_r$ radial profile log residuals versus 
galaxy properties. Data points are color-coded by $\sigma_{\rm SFO}$, 
with detections shown as filled circles and upper limits as transparent 
inverted triangles. The solid black line corresponds to the best fit of the residual distribution and the gray shaded regions show $1\sigma$ confidence intervals 
derived from the 16th and 84th percentiles of the MCMC radial profile fits.
The horizontal dashed line indicates zero residual. Residuals correlate 
positively with $\log_{10}(\mathrm{sSFR})$ and $\sigma_{\rm SFO}$ and 
negatively with $\log_{10}(M_\star/M_\odot)$, with the middle panel 
revealing a population of transitional green valley galaxies (missed by 
binary sSFR classifications) with a wide range of excess \mgii\ 
absorption indicative of a cool CGM in transition.}
    \label{fig:residuals_panel}
\end{figure*}

\subsection{\mgii\ absorption across the star-forming--to--quiescent transition}
\label{subsec: Radial Profile Residuals}

We quantify the dependence of \mgii\ absorption on host galaxy properties 
via log residuals with respect to the $R/R_{\rm 200}$ mean radial profile as defined in equation \ref{equation:delta}. By construction, $\Delta > 0$ ($\Delta < 0$) indicates absorption lying above (below) the mean radial trend.

Figure~\ref{fig:residuals_panel} shows $\Delta$ plotted against 
$\log_{10}(\mathrm{sSFR})$, $\sigma_{\rm SFO}$, and 
$\log_{10}(M_\star/M_\odot)$, respectively. Detections are shown as filled circles 
and upper limits as transparent inverted triangles. We first focus on 
detections to identify the general trends with galaxy properties. 
Both $\log_{10}(\mathrm{sSFR})$ and $\sigma_{\rm SFO}$ show a clear 
positive correlation with $\Delta$: galaxies with higher star-formation 
activity systematically show stronger \mgii\ absorption above the mean 
radial profile, while quiescent systems lie preferentially below it. 
Pearson correlation tests on detections confirm these trends: 
$r = 0.44$ ($p = 1.71\times10^{-6}$) for $\log_{10}(\mathrm{sSFR})$ 
and $r = 0.32$ ($p = 9.54\times10^{-4}$) for $\sigma_{\rm SFO}$, 
with a negative correlation of $r = -0.47$ 
($p = 2.78\times10^{-7}$) for $\log_{10}(M_\star/M_\odot)$; 
all results are summarized in Table~\ref{tab:residual_fit_results}. 
The upper limits are broadly consistent with these trends: 
non-detections at low $\sigma_{\rm SFO}$ and high $\log_{10}(M_\star/M_\odot)$ 
cluster preferentially on the expected sides of the correlation, 
reinforcing the picture that quiescent, massive galaxies have systematically 
weaker cool CGM absorption. Because the sample contains a substantial number of non-detections, we additionally evaluate each correlation using the generalized Kendall's $\tau$ statistic for censored data, treating the logarithmic residuals of non-detections as upper limits. Generalized Kendall's $\tau$ tests yield $\tau=0.185$ ($p=7.41\times10^{-9}$) for $\log_{10}(\mathrm{sSFR})$, $\tau=0.129$ ($p=5.37\times10^{-5}$) for $\sigma_{\mathrm{SFO}}$, and $\tau=-0.087$ ($p=6.28\times10^{-3}$) for $\log_{10}(M_\star/M_\odot)$, confirming correlations in the same directions. Thus, the correlations with both star-formation indicators remain highly significant when non-detections are included.

A key advantage of $\sigma_{\rm SFO}$ over a binary $\log_{10}(\mathrm{sSFR})$ division is its ability to identify transitional green valley galaxies as a distinct population. For example, \citet{Werk2013} divided the COS-Halos sample using a fixed threshold of $\log_{10}(\mathrm{sSFR}/\mathrm{yr}^{-1})=-11$ and therefore green valley galaxies in this work would be classified as star-forming, despite lying substantially below the SFMS. Although these approaches are useful for distinguishing broadly active and passive populations, they necessarily assign intermediate systems to one category or the other, potentially obscuring their intermediate CGM properties. This limitation is particularly relevant across the broad redshift range considered here $(0.07<z<2.7)$. Distinct from a time dependent sSFR cuts \citep{Damen_2009}, the fitted SFMS slope evolves relatively rapidly at low redshift in our model, such that the same sSFR can correspond to different offsets from the main sequence depending on both stellar mass and epoch rather than only time.

As seen in the left panel of Figure~\ref{fig:residuals_panel}, several detection-associated galaxies with intermediate $\Delta$ values ($N=5$ for $-0.1<\Delta<0.8$) that would be assigned to either the star-forming or passive category under a binary sSFR cut are identified by $\sigma_{\rm SFO}$ as transitional systems ($-4.5\lesssim\sigma_{\rm SFO}\lesssim-3$). The middle panel makes this distinction explicit: green valley galaxies span a non-negligible range of $\Delta$, with detections both above and below the mean radial profile, demonstrating that transitioning galaxies can retain significant cool-CGM absorption. Thus, $\sigma_{\rm SFO}$ provides a more finely resolved, redshift-dependent description of an intermediate population that would otherwise have been incorporated into one side of a binary division and represents direct observational evidence that population level \mgii\ CGM depletion occurs across the green valley rather than abruptly at quenching. Substantial sightline-to-sightline scatter persists throughout, consistent with additional contributions from CGM geometry and azimuthal structure \citep[e.g.,][]{Bordoloi2011,Cherrey_2025,Ho_2020}.

In the right panel (Figure \ref{fig:residuals_panel}), the negative $\Delta$--$\log_{10}(M_\star/M_\odot)$ correlation is not a direct mass effect. Because halo-radius normalization removes 
the leading-order halo-mass dependence on absorption strength, the residual trend instead reflects the fact that most of the low-mass galaxies are star-forming or starburst, whereas passive galaxies are predominantly found at the high mass end. This is evident from the $\sigma_{\rm SFO}$ color-coding in the right panel of Figure~\ref{fig:residuals_panel}: the negative slope at high $\log_{10}(M_\star/M_\odot)$ is driven by the increasing fraction of low-$\sigma_{\rm SFO}$ systems rather than any intrinsic mass dependence of the cool CGM.

\begin{figure*}[!tb]
    \centering

    \includegraphics[height=4.3cm]{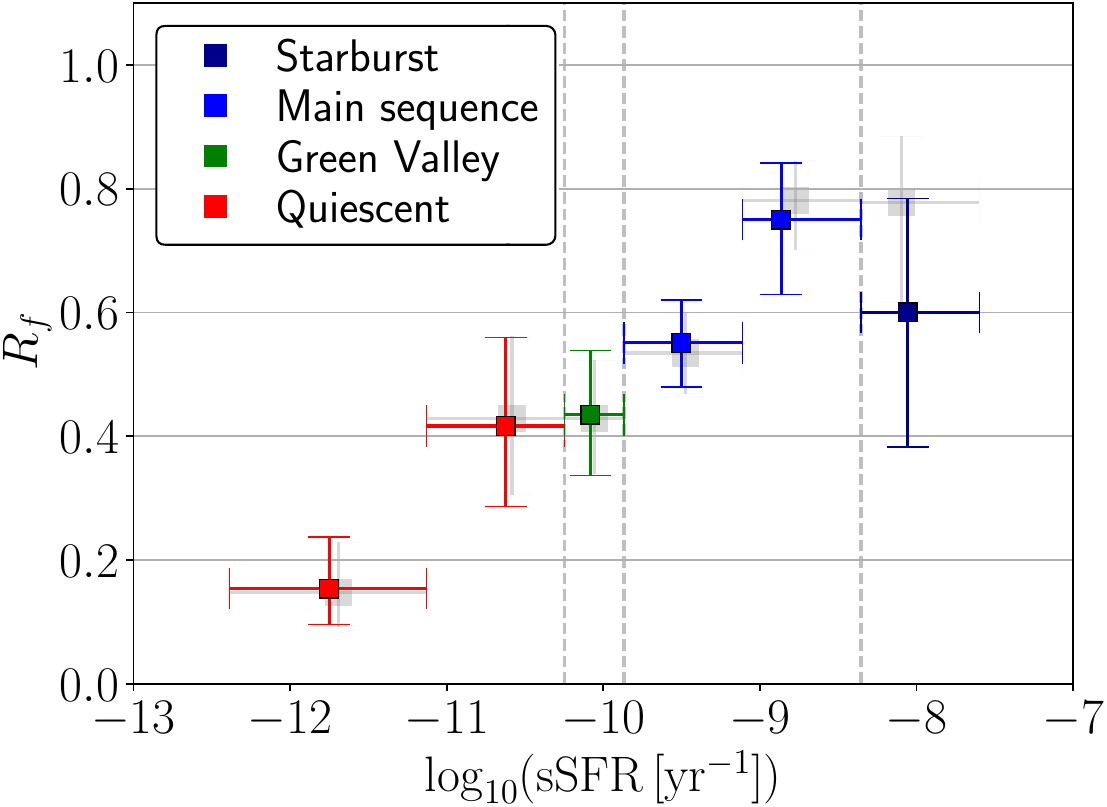}
    \includegraphics[height=4.3cm]{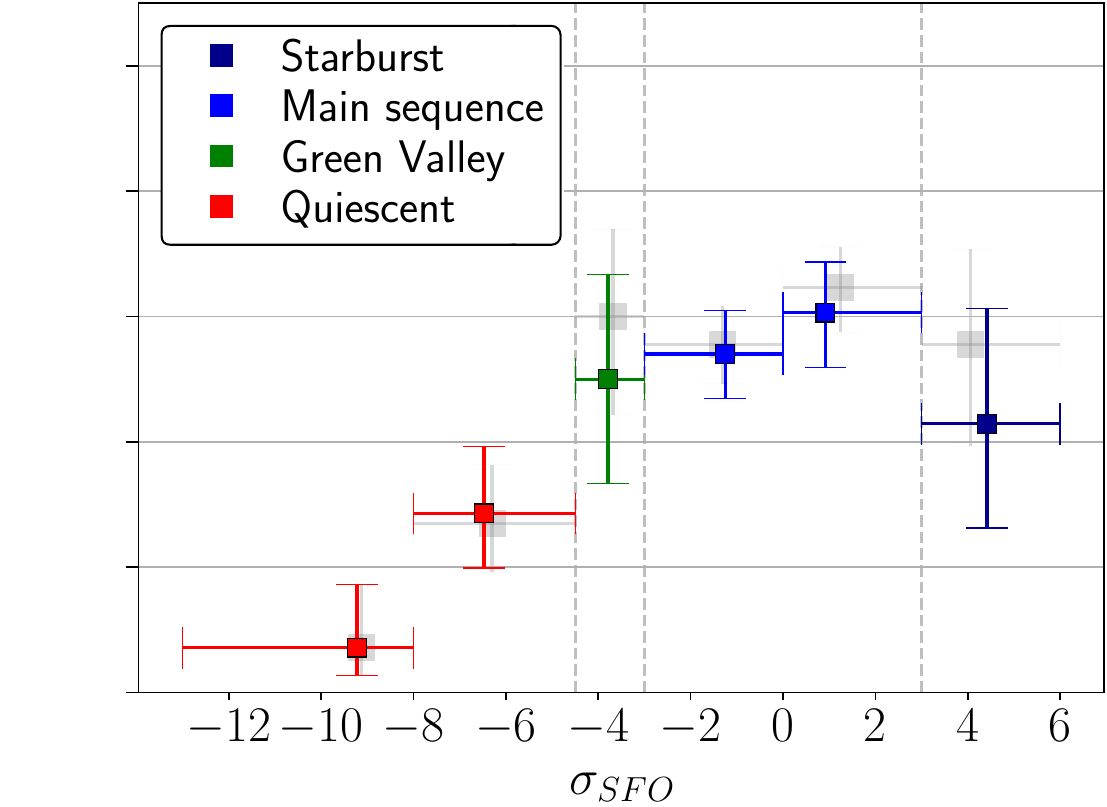}
   \caption{\mgii\ absorption residual fraction ($R_f$) as a function of galaxy properties. \textit{Left panel:} $R_f$ versus $\log_{10}(\mathrm{sSFR})$, with sSFR bins defined from the corresponding $\sigma_{\rm SFO}$ population divisions. \textit{Right panel:} $R_f$ versus star formation offset $\sigma_{\rm SFO}$. In both panels, the gray background points show the full sample, while the colored foreground points show galaxies within $0.5\,R_{\rm 200}$. Vertical dashed lines separate quiescent, green valley, main-sequence, and starburst populations from left to right. $R_f$ rises monotonically from quiescent to main-sequence systems in both parameterizations, demonstrating that cool CGM absorption strength tracks star-formation activity  continuously across the quenching sequence.}

    \label{fig:rf}
\end{figure*}

To further characterize these trends while accounting for non-detections, we compute the residual fraction $R_f$ in bins of $\sigma_{\rm SFO}$ and $\log_{10}(\mathrm{sSFR})$ (equation \ref{equation:Rf}). The sSFR bins are defined from the corresponding $\sigma_{\rm SFO}$ population divisions, ensuring a consistent comparison of quiescent, green valley, star-forming, and starburst galaxies across both parameterizations. Figure~\ref{fig:rf} shows $R_f$ for the full sample as gray background points and for galaxies within $0.5 R_{\rm 200}$ as colored foreground points. Within $0.5R_{\rm 200}$, $R_f$ increases monotonically from quiescent through green valley to star-forming galaxies. The separation between the quiescent and star-forming populations is statistically meaningful, with non-overlapping error bars, while green valley galaxies show intermediate $R_f$ values consistent with their intermediate $\sigma_{\rm SFO}$. Moreover, $R_f$ continues to decrease toward more negative $\sigma_{\rm SFO}$ within the passive population rather than reaching an immediate plateau at the green-valley--quiescent boundary. The starburst subsample contains only $N=7$ sightlines within $0.5R_{\rm 200}$, so its inferred $R_f$ should be interpreted with caution.

Taken together, these results show that \mgii\ absorption in 
the CGM is closely connected to the star-formation state of the host galaxy: \mgii\ absorption is enhanced during active star formation, on average decreases across the green valley, and reaches its lowest levels in the most quiescent systems. It is important to note that some green valley galaxies may be rapidly quenching or moving temporarily toward greater star-formation activity through rejuvenation \citep{Chauke2019,Angthopo2020}. Although individual galaxies may follow rapid or non-monotonic evolutionary pathways, the overall ordering of $\Delta$ and $R_f$ supports a picture of a population-level transition in which the dominant mode of quenching involves more gradual cool-CGM suppression accompanying the progression from star-forming through green valley to increasingly quiescent galaxies, rather than occurring primarily as an abrupt change at the onset of quiescence.

 \begin{table*}[ht!]
\centering
\caption{Linear fit, Pearson correlation, and Kendall's $\tau$ test parameters for 
$R/R_{\mathrm{200}}$ radial profile residuals}
\label{tab:residual_fit_results}

\resizebox{\textwidth}{!}{%
\begin{tabular}{lcccccc}
\hline
\textbf{Variable} 
& \textbf{Slope $m$} 
& \textbf{Intercept $b$} 
& \textbf{$r$} 
& \textbf{$p$-value} 
& \textbf{$\tau_{\mathrm{cens}}$} 
& \textbf{$p_{\mathrm{cens}}$} \\
\hline

$\log_{10}(\mathrm{sSFR/yr^{-1}})$ 
& $0.56^{+0.087}_{-0.073}$ 
& $5.98^{+0.95}_{-0.77}$ 
& $0.44$ 
& $1.71 \times 10^{-6}$ 
& $0.185$
& $7.41 \times 10^{-9}$ \\

$\log_{10}(M_*/M_{\odot})$ 
& $-0.82^{+0.10}_{-0.12}$ 
& $9.00^{+1.38}_{-1.13}$ 
& $-0.47$ 
& $2.78 \times 10^{-7}$ 
& $-0.087$
& $6.28 \times 10^{-3}$ \\

$\sigma_{\mathrm{SFO}}$ 
& $0.12^{+0.019}_{-0.016}$ 
& $0.78^{+0.13}_{-0.11}$ 
& $0.32$ 
& $9.54 \times 10^{-4}$ 
& $0.129$
& $5.37 \times 10^{-5}$ \\

\hline
\end{tabular}
}
\end{table*}

\subsection{Covering fraction}
\label{subsec: Covering Fraction}

This subsection quantifies the statistical incidence of \mgii\ absorbers as 
a function of distance using the covering fraction $C_f$. Figure~\ref{fig:covering_fraction_combined}  includes data from \citet{Nielsen_2013} in panels~(a)--(c). Since their sample lacks star-formation rate information, it is therefore excluded from the star-formation-split panels~(d)--(i).

In the inner CGM of isolated galaxies ($R \le 50$~kpc, or $R/R_{\rm 200} \le 0.25$), $C_f$ for absorbers with $W_r \ge 100$~m\AA\ approaches unity ($C_f \approx 0.9$--$1$). $C_f$ declines with increasing distance, approaching zero by $R \sim 150$~kpc, or $R/R_{\rm 200} \sim 0.75$. For strong absorbers ($W_r \ge 1000$~m\AA), the inner CGM shows $C_f \approx 0.4$, also declining to zero by $\sim$150~kpc, or $R/R_{\rm 200} \sim 0.75$.

Comparing group and isolated environments, Figure~\ref{fig:covering_fraction_combined}  shows similar radial extents for strong absorption, but a more extended profile of weak absorption ($100 \le W_r < 500$~m\AA) in group environments. The extended  group radial profile fit reported in Section~\ref{subsec: Radial Profile} therefore  originates from this weaker absorption component, which persists to at least 300~kpc, with $C_f(W_r > 100~\text{m\AA}) \approx 0.1$ between 200 and 300~kpc for groups, compared to 
$C_f(W_r > 100~\text{m\AA}) = 0$ for isolated galaxies over the same range. This extension of \mgii\ absorption beyond $\sim$150~kpc suggests that group environments dynamically redistribute lower-density \mgii\ gas throughout the CGM of group galaxies.

\begin{figure*}[ht!]
    \centering

    \begin{minipage}{0.32\textwidth}
        \centering
        (a) Isolated: $C_f$ vs $R$ \\
        \includegraphics[width=\linewidth]{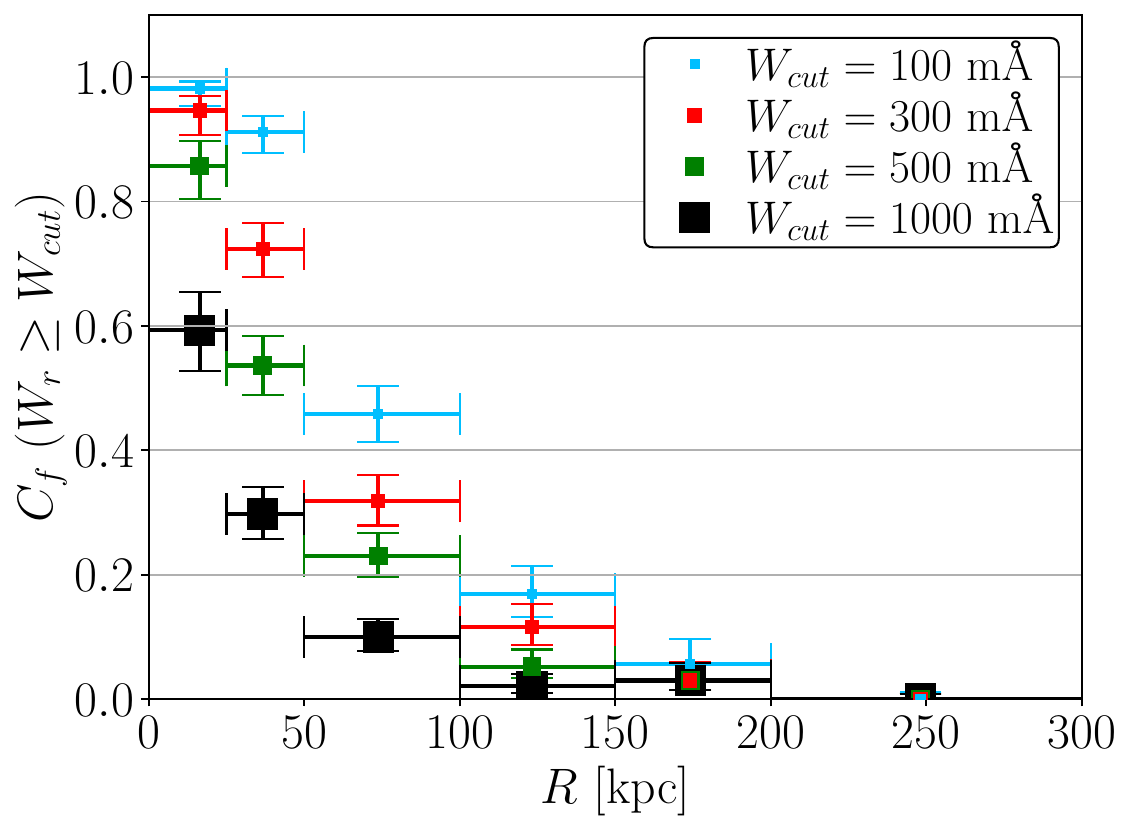}
    \end{minipage}
    \hfill
    \begin{minipage}{0.32\textwidth}
        \centering
        (b) Isolated: $C_f$ vs $R/R_{\rm 200}$ \\
        \includegraphics[width=\linewidth]{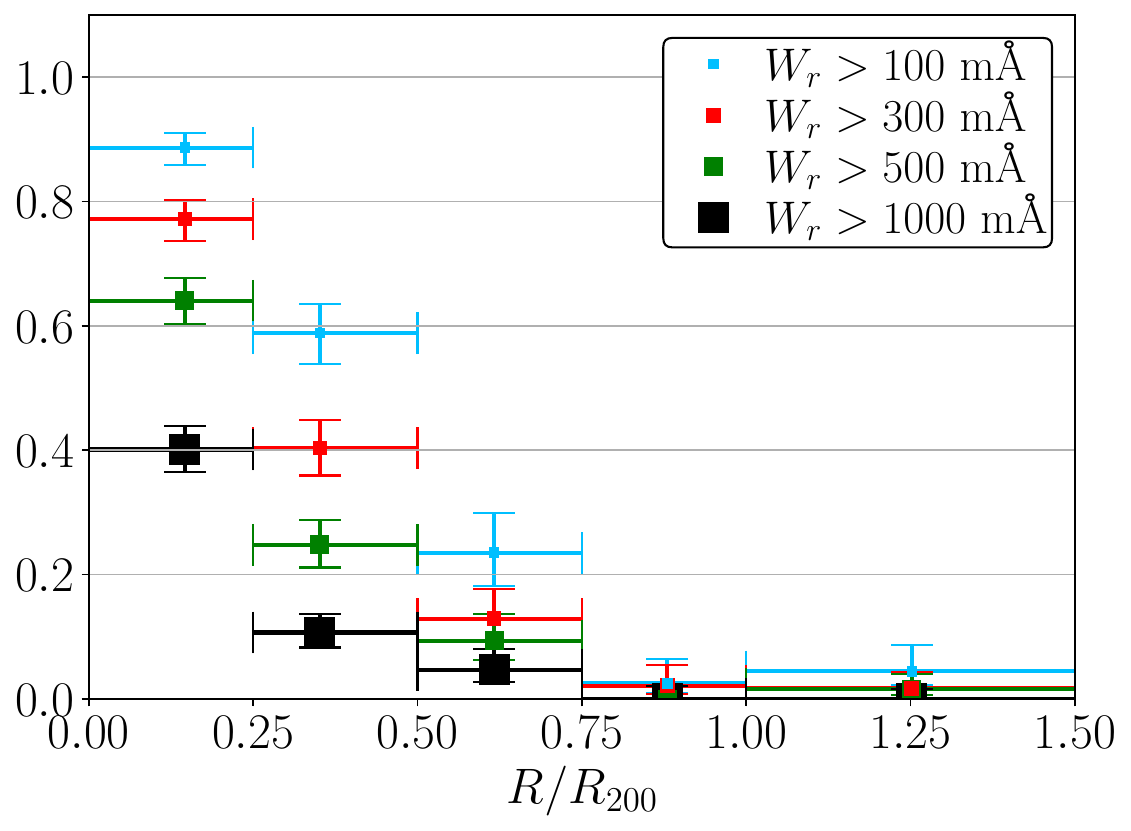}
    \end{minipage}
    \hfill
    \begin{minipage}{0.32\textwidth}
        \centering
        (c) Group: $C_f$ vs $R$ \\
        \includegraphics[width=\linewidth]{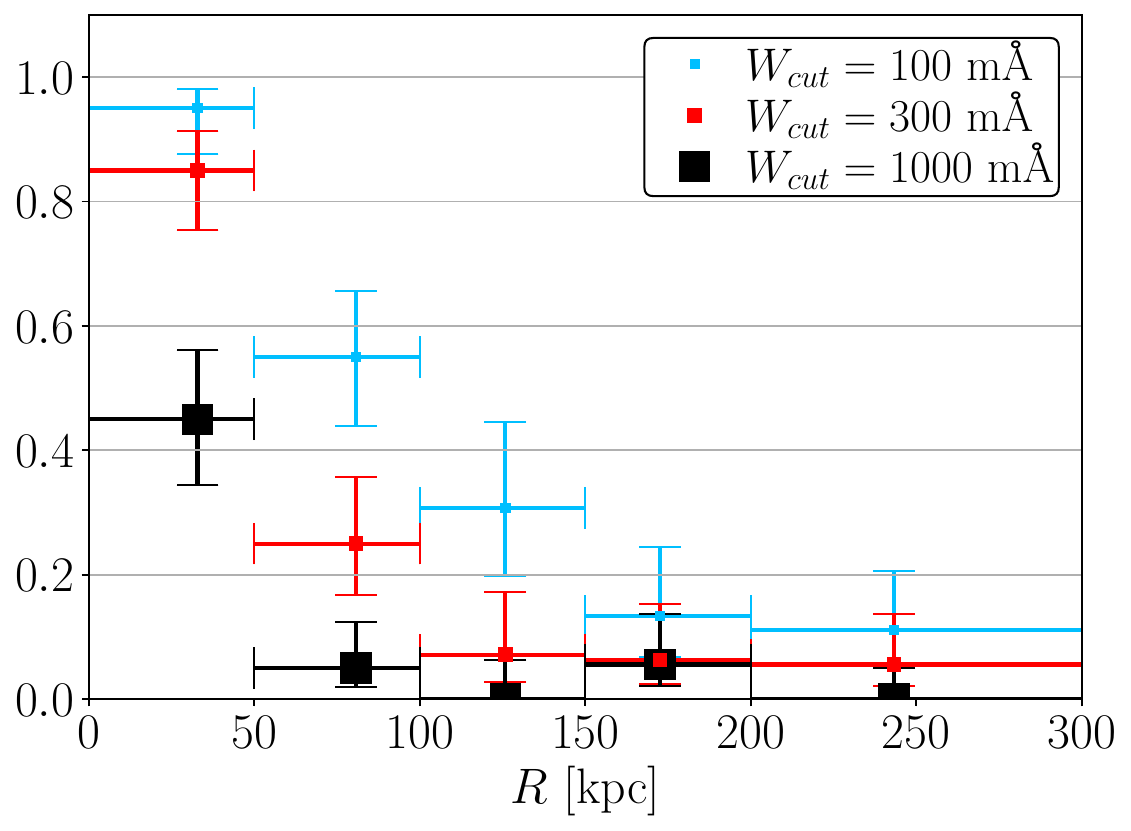}
    \end{minipage}

    \vspace{1em}

    \begin{minipage}{0.32\textwidth}
        \centering
        (d) Isolated: $C_f$ vs $R$ \\
        \includegraphics[width=\linewidth]{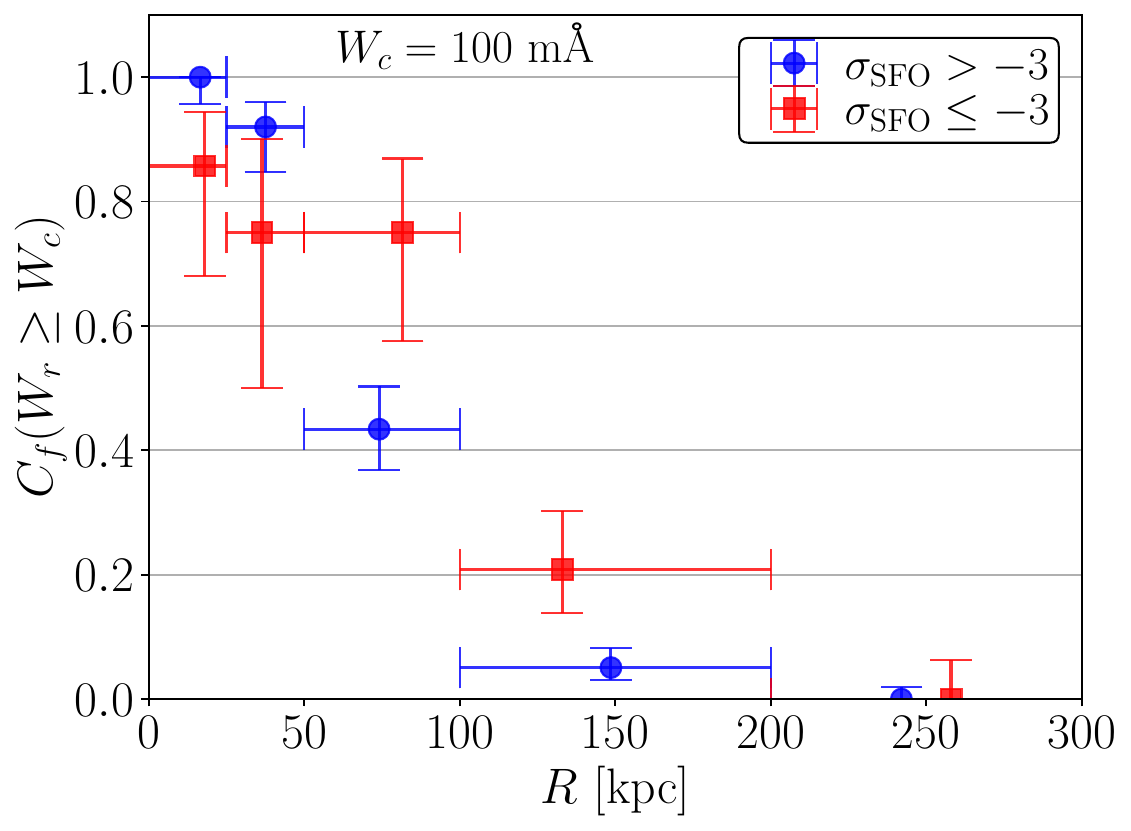}
    \end{minipage}
    \hfill
    \begin{minipage}{0.32\textwidth}
        \centering
        (e) Isolated: $C_f$ vs $R$ \\
        \includegraphics[width=\linewidth]{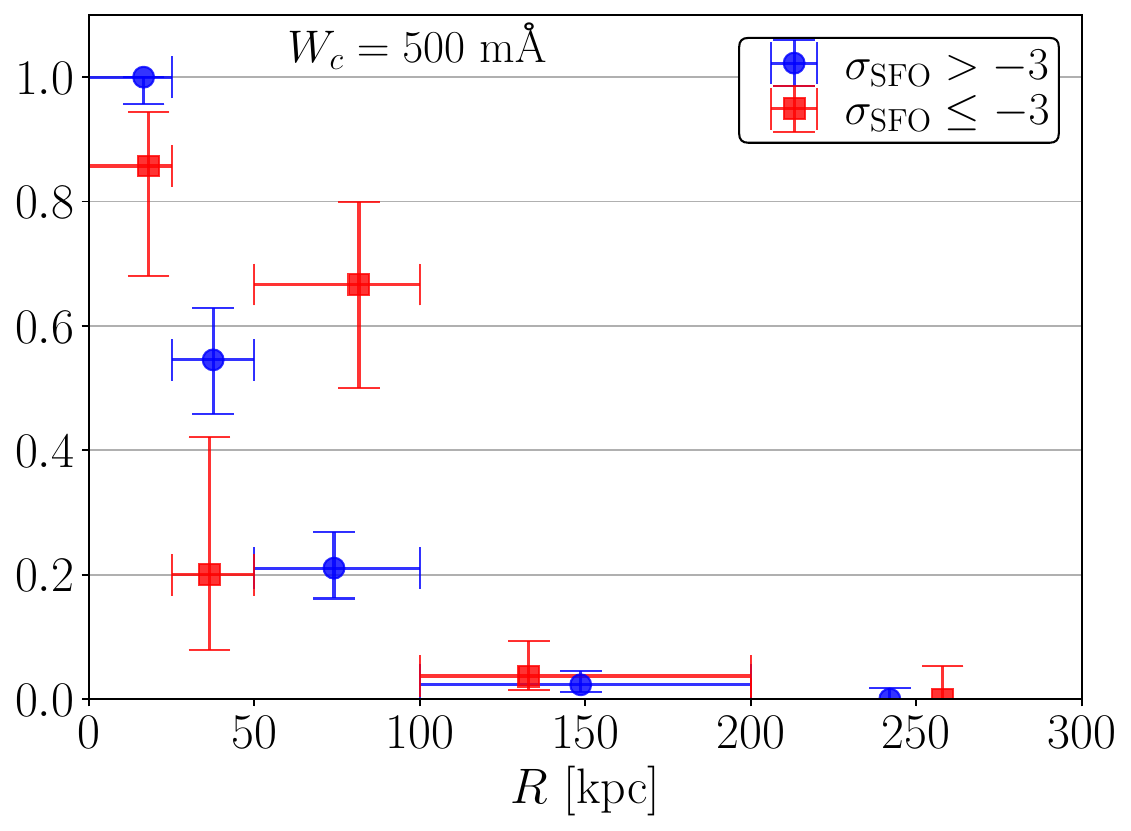}
    \end{minipage}
    \hfill
    \begin{minipage}{0.32\textwidth}
        \centering
        (f) Isolated: $C_f$ vs $R$\\
        \includegraphics[width=\linewidth]{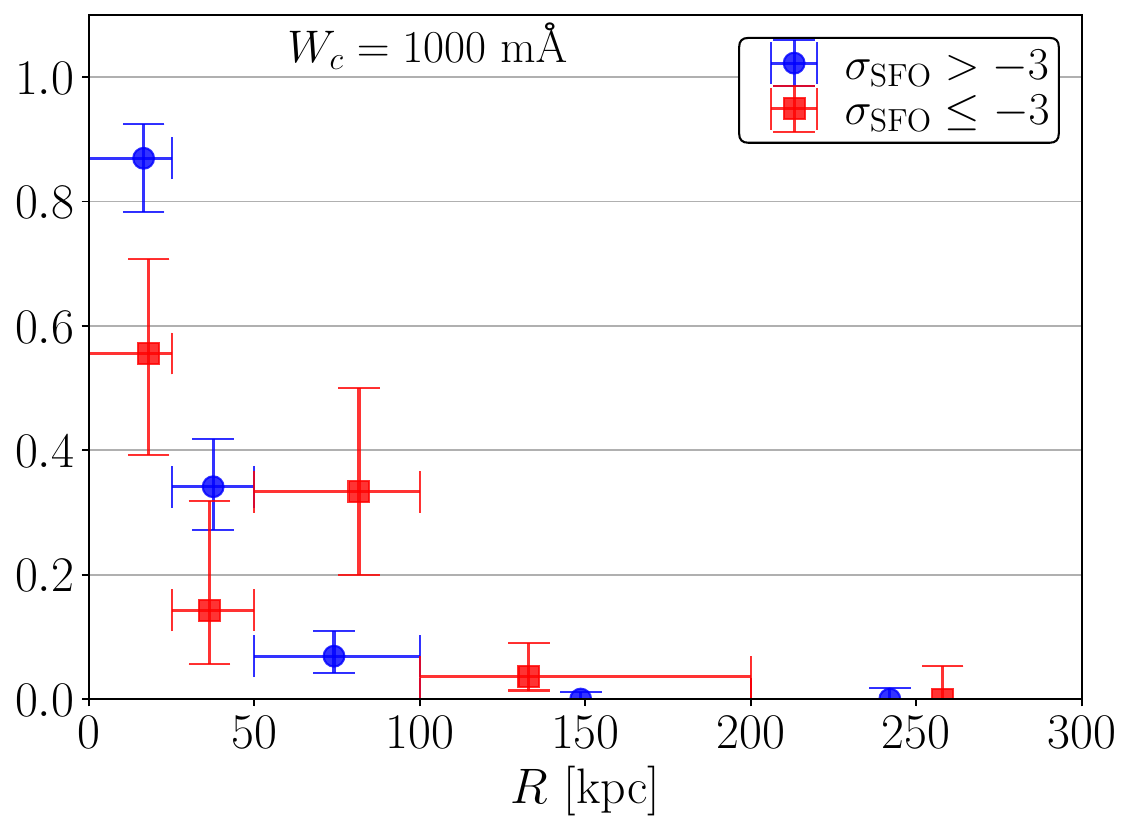}
    \end{minipage}

    \vspace{1em}

    \begin{minipage}{0.32\textwidth}
        \centering
        (g) Isolated: $C_f$ vs $R/R_{\rm 200}$\\
        \includegraphics[width=\linewidth]{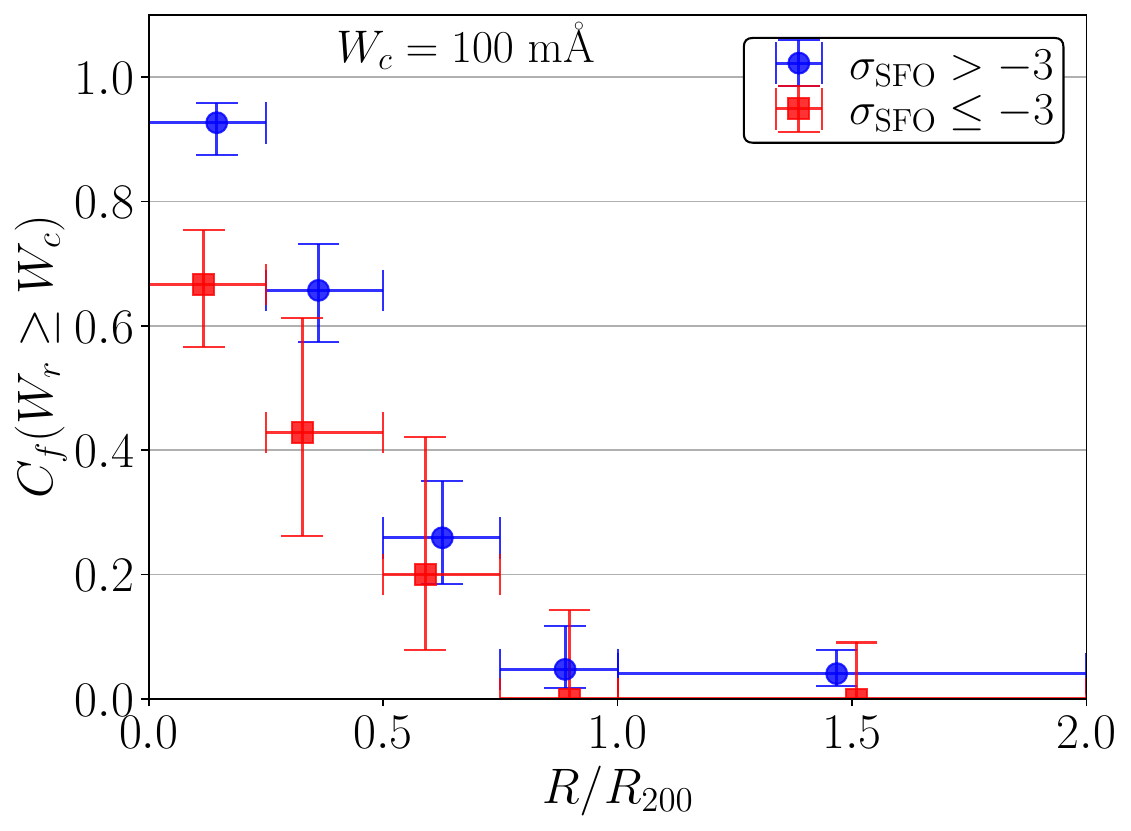}
    \end{minipage}
    \hfill
    \begin{minipage}{0.32\textwidth}
        \centering
        (h) Isolated: $C_f$ vs $R/R_{\rm 200}$ \\
        \includegraphics[width=\linewidth]{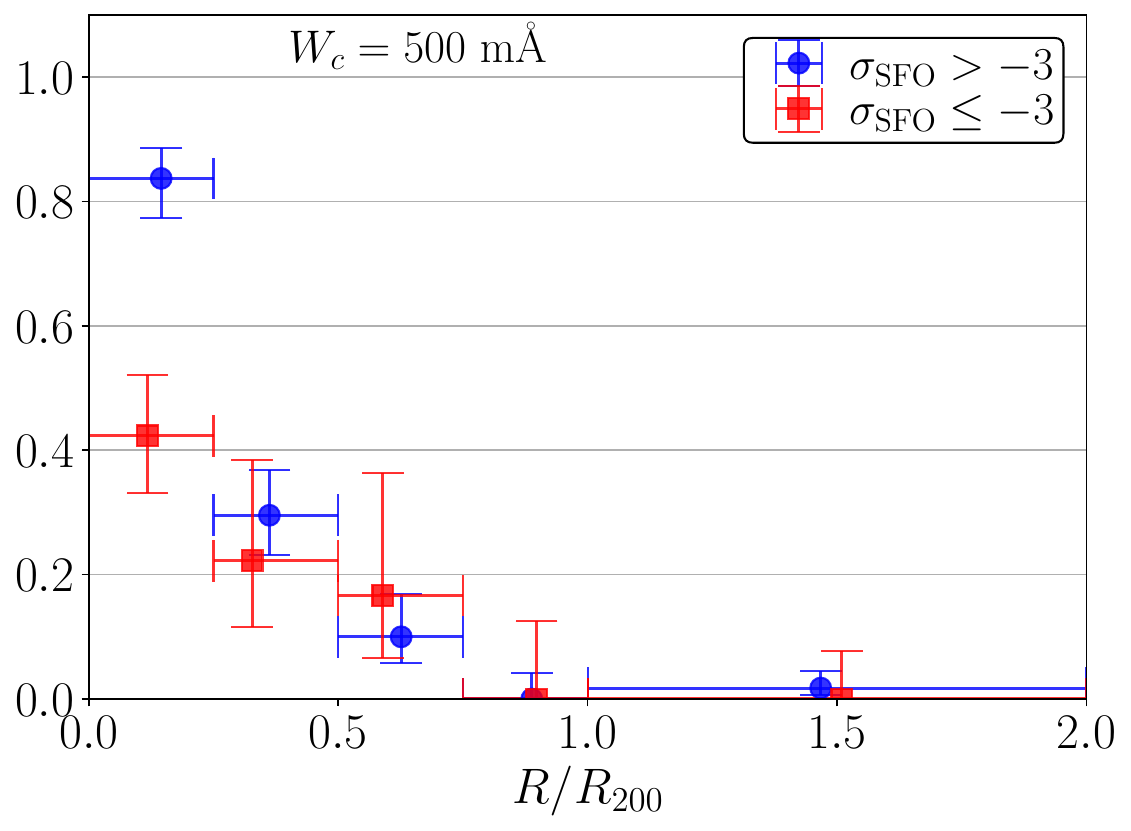}
    \end{minipage}
    \hfill
    \begin{minipage}{0.32\textwidth}
        \centering
        (i) Isolated: $C_f$ vs $R/R_{\rm 200}$ \\
        \includegraphics[width=\linewidth]{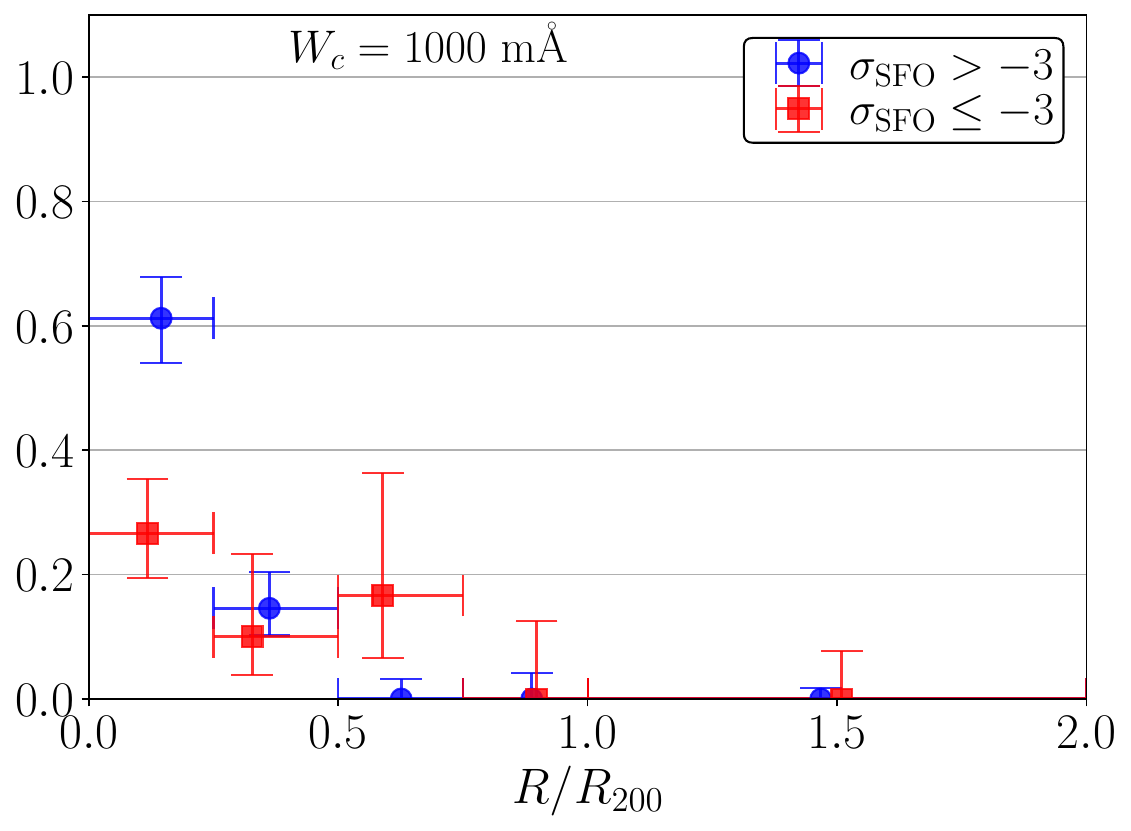}
    \end{minipage}
    \caption{\mgii\ absorption covering fraction ($C_f$) across isolated and group environments. Panels (a) and (b) show $C_f$ for the full isolated sample as a function of $R$ and  $R/R_{\mathrm{200}}$, respectively, at four equivalent width cutoffs ($W_{ c} = 100$, 300, 500, and 1000~m\AA). Panel (c) shows the corresponding group radial profile as a function of $R$. Panels (d)--(i) split the isolated sample into star-forming ($\sigma_{\rm SFO} > -3$) and passive ($\sigma_{\rm SFO} \le -3$) subsamples at three 
representative cutoffs, $W_c = 100$, 500, and 1000~m\AA\: panels (d)--(f) versus $R$ and panels (g)--(i) versus $R/R_{\mathrm{200}}$. The star-forming subsample shows systematically higher covering fractions in the inner CGM at all three cutoffs, with the contrast between star-forming and passive galaxies increasing toward higher $W_c$.
}
\label{fig:covering_fraction_combined}
    \label{fig:covering_fraction_combined}
\end{figure*}

We next divide the isolated sample into star-forming ($\sigma_{\rm SFO} > -3$) and passive ($\sigma_{\rm SFO} \le -3$) subsamples and compute $C_f(R)$ and $C_f(R/R_{\rm 200})$ for each. Panels~(d)--(f) of Figure~\ref{fig:covering_fraction_combined} show $C_f$ versus $R$, and 
panels~(g)--(i) show $C_f$ versus $R/R_{\rm 200}$, at $W_c = 100$, 500, and 1000~m\AA. Without $R_{\rm 200}$ normalization, passive galaxies show higher $C_f$ than star-forming galaxies at large $R$ and a less uniform radial decline. This is owing to the fact that passive galaxies typically possess larger halo masses and therefore larger halo radii. After normalizing 
by $R_{\rm 200}$, this trend reverses: the passive covering fraction declines more smoothly with radius and falls below the star-forming $C_f$ across most $W_c$ thresholds and $R/R_{\rm 200}$ bins. The contrast is strongest in the inner CGM and increases with $W_c$: the ratio $C_f^{\rm SF}/C_f^{\rm pass}$ rises from $\approx 1.4$ at $W_c = 100$~m\AA\ to 
$\approx 2.0$ at $W_c = 500$~m\AA\ and $\approx 2.3$ at $W_c = 1000$~m\AA. This indicates that the strongest \mgii\ absorbers preferentially arise around star-forming rather than passive galaxies.

To test whether this inner-CGM enhancement declines continuously during quenching, we compute $C_f$ in bins of $\log_{10}(\mathrm{sSFR})$ and $\sigma_{\rm SFO}$ for all detections and non-detections within $R/R_{\rm 200} < 0.25$, using a four-way $\sigma_{\rm SFO}$ split. Figure~\ref{fig:coveringfraction_inner} shows that $C_f$ increases 
with both $\log_{10}(\mathrm{sSFR})$ and $\sigma_{\rm SFO}$ across nearly all $W_c$. For strong absorbers ($W_c = 1000$~m\AA), $C_f \approx 0.22$ for the most passive bin, rising to $C_f \approx 0.65$ for the most star-forming bin --- a factor of $\sim$3 increase. Critically, the green valley bin shows an intermediate covering fraction ($C_f \approx 0.4$ at $W_c = 1000$~m\AA), demonstrating a clear, continuous transition through the green valley 
rather than a sharp threshold. This result extends the findings of \citet{Huang2021} and \citet{Chaudary_2025}, who reported a sensitive dependence of inner-CGM absorption on sSFR/SFR, by showing an explicit, monotonic decline in \mgii\ covering fraction as galaxies quench.

\begin{figure*}[ht!]
    \centering

    \begin{minipage}{0.37\textheight}
        \includegraphics[width=\linewidth]{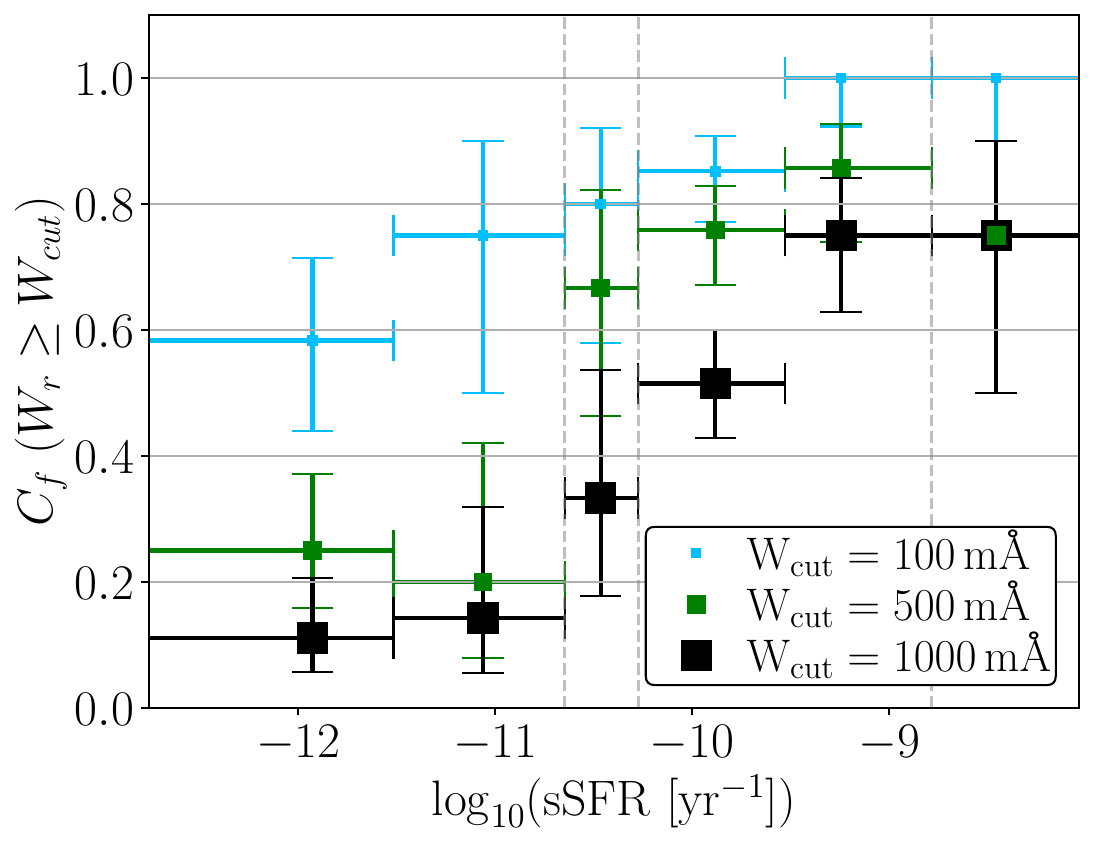}
    \end{minipage}
    \hfill
    \begin{minipage}{0.35\textheight}
        \includegraphics[width=\linewidth]{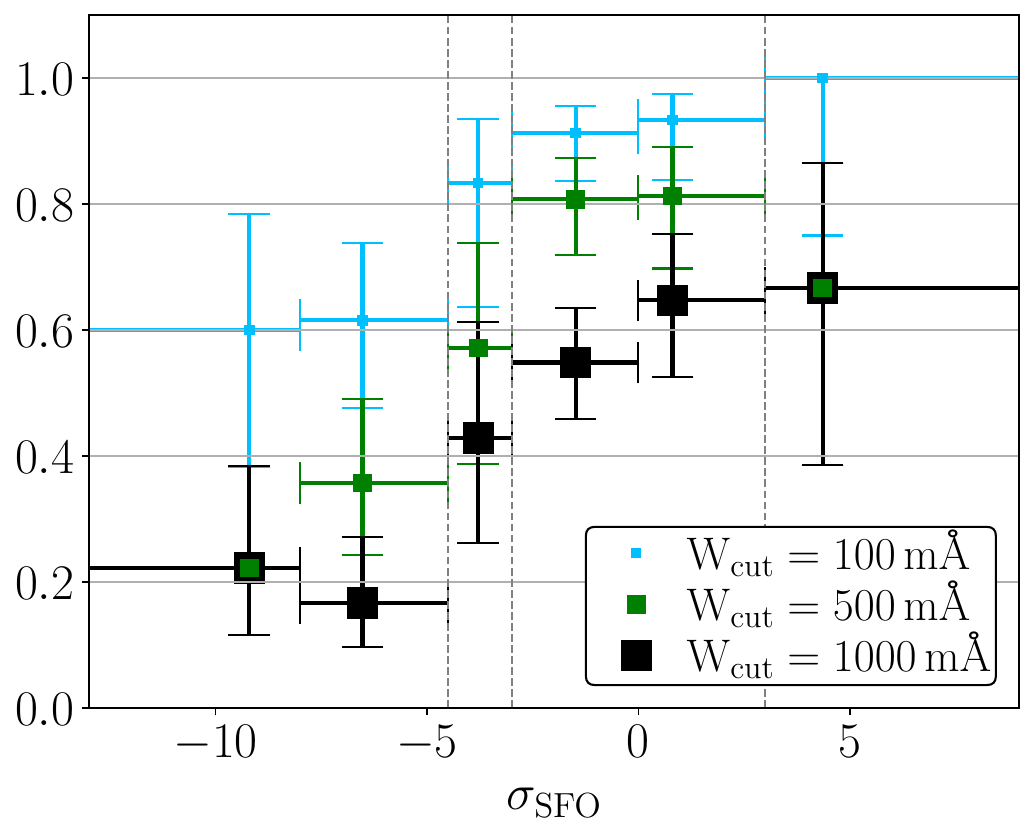}
    \end{minipage}

    \caption{\mgii\ absorption covering fraction ($C_f$) in the inner CGM ($R/R_{\mathrm{200}} < 0.25$) for isolated galaxies. \textit{Left panel:} $C_f$ versus $\log_{10}(\mathrm{sSFR})$. \textit{Right panel:} $C_f$ versus star formation offset $\sigma_{\rm SFO}$. The inner CGM range is chosen based on the enhanced \mgii\ absorption observed within $R/R_{\mathrm{200}} < 0.25$ in Figure~\ref{fig:covering_fraction_combined}, panels (g--i). Vertical dashed lines in the right panel show the four-way split cutoffs for star formation categories defined in Section~\ref{subsec:sigma_SFO}.
}
\label{fig:coveringfraction_inner}
    \label{fig:coveringfraction_inner}
\end{figure*}

\subsection{\mgii\ absorption kinematics}
\label{subsec: Velocities}

We next investigate the kinematic and column density distribution of absorption components in 
our detection sample. Escape velocities are computed from individual galaxy halos for isolated systems and from group halos for group systems.

\begin{figure*}[ht!]
    \centering

    \begin{minipage}[t]{0.44\textwidth}
        \centering
        \includegraphics[width=\textwidth]{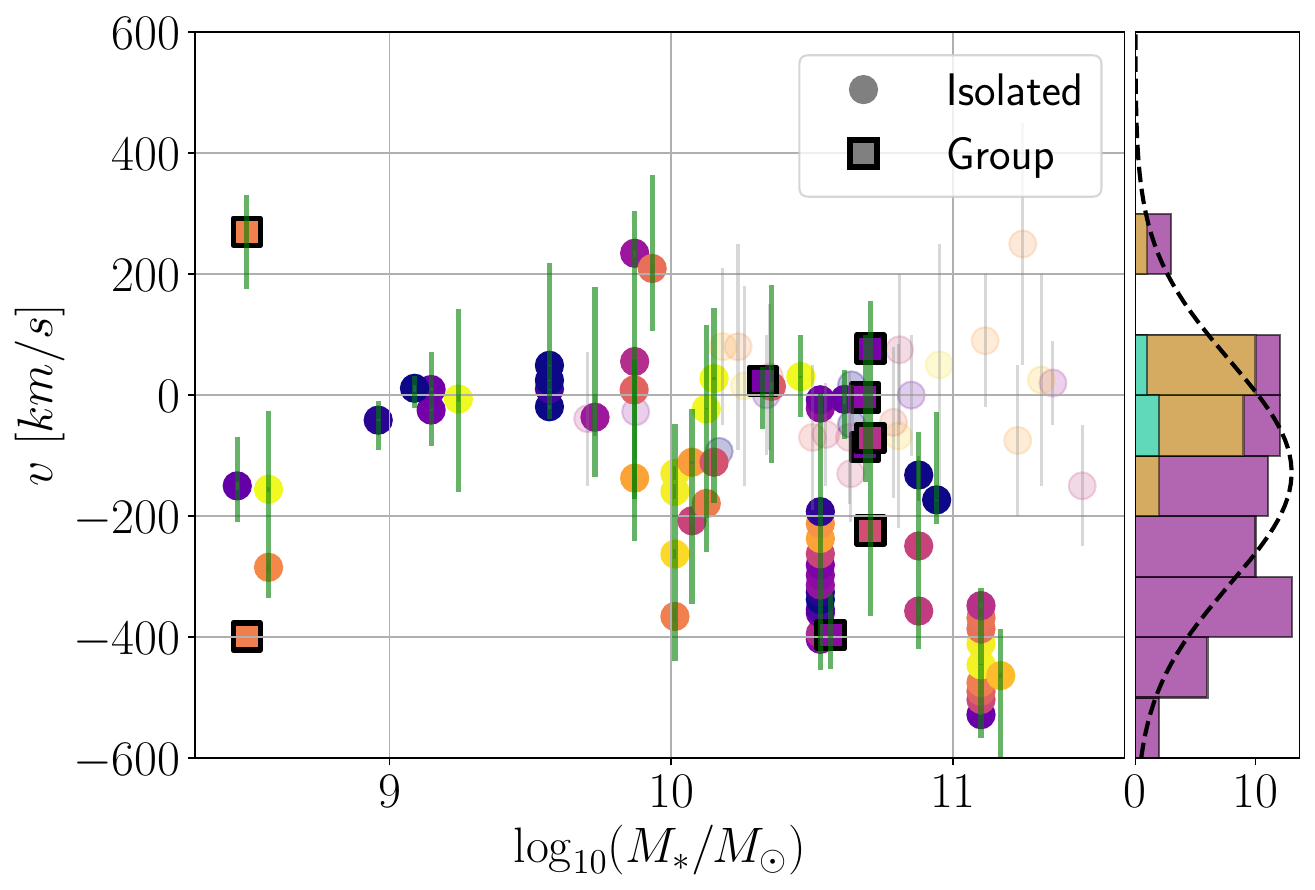}
    \end{minipage}
    \hfill
    \begin{minipage}[t]{0.42\textwidth}
        \centering
        \includegraphics[width=\textwidth]{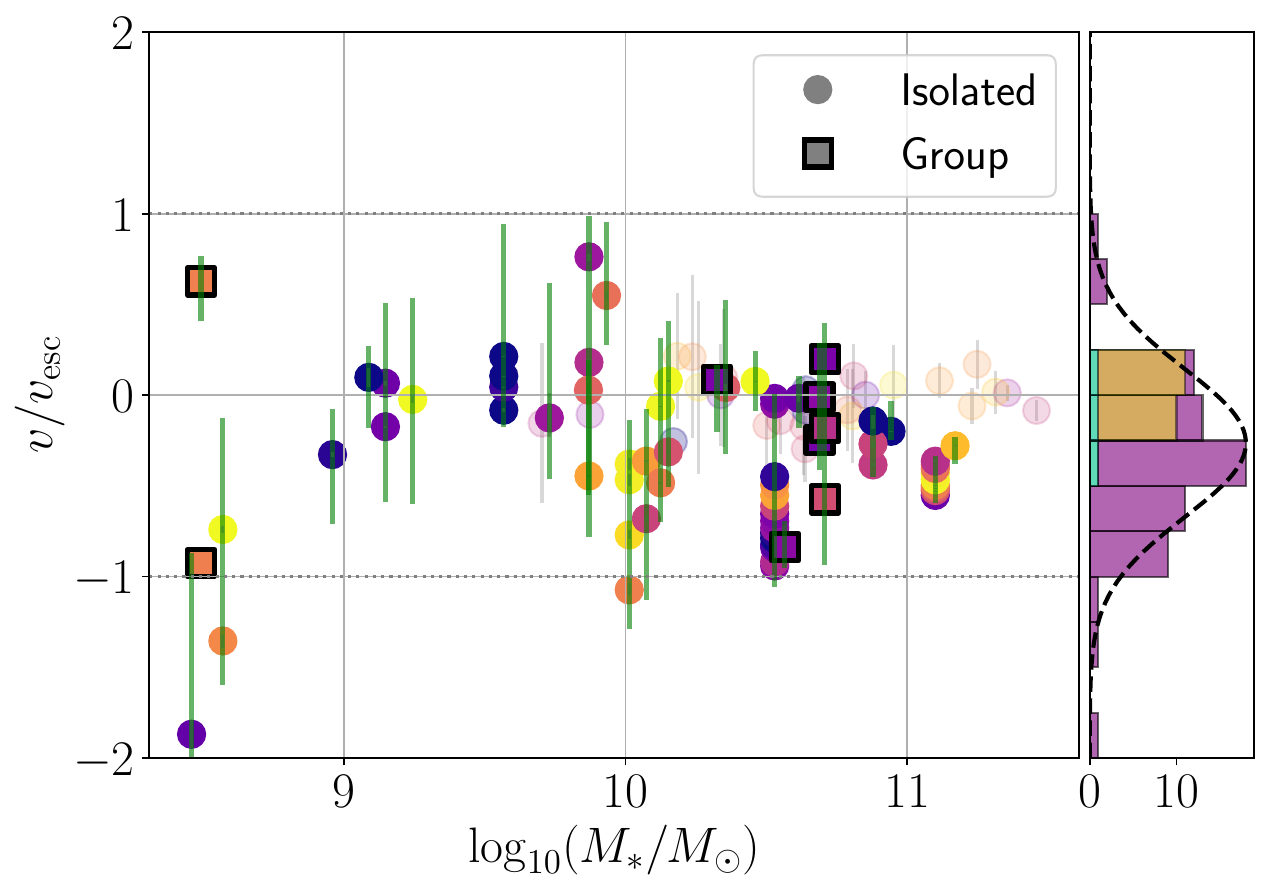}
    \end{minipage}
    \hfill
    \begin{minipage}[t]{0.095\textwidth}
        \centering
        \includegraphics[width=\textwidth]{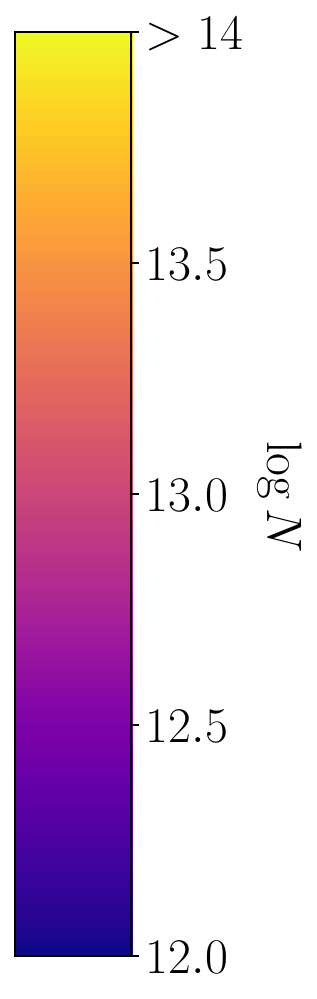}
    \end{minipage}

\caption{
Velocity distribution of \mgii\ absorbers. Circles and squares mark isolated and group detections, respectively, with absorption velocity ranges shown as green vertical lines. \textit{Left:} cloud velocity $v$ versus $\log_{10}(M_*/M_\odot)$. \textit{Right:} \mgii\ absorption component velocity normalized by escape velocity, $v/v_{\rm esc}$, versus $\log_{10}(M_*/M_\odot)$. Both panels share a common color scale for $\log_{10}(N)$. The side histograms show counts 
per 100\kms\ (left) or per 0.25 in $v/v_{\rm esc}$ (right), overlaid with Gaussian fits to the components. The COS-Halos galaxy absorber sample from \citet{Werk2013} and the $z \sim 2-3$ isolated galaxy absorber sample from \citet{Bordoloi2024} are shown as faint background points, color-coded gold and aqua in the histograms, respectively. Dashed horizontal lines in the right panel mark $|v/v_{\rm esc}| = 1$.}
\label{fig:v_vvesc_comparison}
\end{figure*}

We identify 69 individual absorption components within our 28 absorbers: 60 components in isolated environments and 9 components in group environments. The absorption component velocity distributions shown in Figure~\ref{fig:v_vvesc_comparison} are reasonably well described by a Gaussian in both $v$ and $v/v_{\rm esc}$, with the escape-velocity-normalized distribution providing a closer match to a normal distribution than the raw $v$ distribution. 
\mgii\ absorbers have a mean velocity of $-184.4 \pm 24.1$ \kms~ around isolated galaxies and 
have a mean velocity of $-90.0 \pm 73.1$ \kms~ around group galaxies. Absorption components around isolated galaxies have a mean absolute velocity of $207.2 \pm 20.7$ \kms, and 
around group galaxies, have a mean absolute velocity of $172.4 \pm 51.6$ \kms\, showing no 
significant difference in velocity magnitude between environments, though the 
group component sample remains small ($N = 9$).

For each absorber system, we normalize the column densities of its individual 
components to that system's maximum component column density. We then test whether this 
relative column density depends on velocity offset from the galaxy's systemic 
redshift, using a Spearman correlation across all absorption components in the sample. We find 
a significant anticorrelation ($\rho = -0.339$, $p = 1.115\times10^{-2}$): \mgii\ absorption components with larger velocity offsets tend to have lower relative column densities. In other words, within a given absorber system, the strongest absorption components usually are found  closest to the galaxy's systemic velocity, while the weaker components are typically found further away in velocity space.

We next assess whether the observed component velocities are dynamically consistent 
with halo rotation. The mean circular velocity of the host galaxies, evaluated 
at each absorber's projected impact parameter $R$, is $181.7 \pm 102.3$~\kms, 
comparable in both mean and scatter to the observed line-of-sight cloud 
velocities. As shown in Figure~\ref{fig:v_vvesc_comparison}, nearly all absorption components are consistent with being  
gravitationally bound ($|v/v_{\rm esc}| < 1$); only components around the 
lowest-mass dwarf hosts approach $v \sim v_{\rm esc}$.

We further divide the absorption around the isolated galaxy sample into star-forming (30 components) and 
passive (30 components) subsamples using a binary cut at $\sigma_{\rm SFO} = -3$. 
Absorption components around isolated star-forming galaxies have a mean velocity of $-58.8 \pm 24.4$ \kms, 
and around passive galaxies the absorption components have a mean velocity of $-310.7 \pm 26.2$ \kms, respectively. 
Around passive galaxies, absorption components also show a higher mean absolute velocity ($311.6 \pm 25.8$ \kms) 
than star-forming clouds ($102.8 \pm 18.2$ \kms). Because passive galaxies in 
our sample tend to reside in more massive halos, when normalized by 
escape velocity, \mgii\ absorption around star-forming galaxies show $|v/v_{\rm esc}| = 0.40 \pm 0.08$ and 
around passive galaxies show $|v/v_{\rm esc}| = 0.49 \pm 0.05$, respectively. The velocity dispersions tell 
a different story: around passive galaxies absorption components show a larger velocity dispersion 
($\sigma_{|v|} = 141.3 \pm 18.6$~\kms) than star-forming galaxies 
($\sigma_{|v|} = 99.4 \pm 13.1$~\kms), but after normalizing by escape velocity 
the trend reverses, with star-forming galaxies showing larger dispersion 
($\sigma_{|v/v_{\rm esc}|} = 0.44 \pm 0.06$) than passive galaxies 
($\sigma_{|v/v_{\rm esc}|} = 0.26 \pm 0.03$). This reversal suggests a more dynamically active CGM in star-forming galaxies, consistent with outflow-driven kinematics, once one accounts for the differences in halo mass.

\subsection{2D Distribution of Absorbers}
\label{subsec: 2D Distribution}

This subsection examines the azimuthal dependence of CGM \mgii\ absorption. 
Combining our COSMOS field targets with the MEGAFLOW sample \citep{Cherrey_2025}, we assemble a total of 89 galaxies within $R/R_{\rm 200} < 1$. Using a binary cut at $\sigma_{\rm SFO} = -3$, 63 galaxies are classified as star-forming and 26 as passive. The top panels of Figure~\ref{fig:az_angle_combo}, show the resulting 2D distribution of \mgii\ absorption for star-forming (top left) and passive (top right) galaxies,  respectively. QSO sightlines passing near the galaxy's major axis (disk) are presented at 0$^\circ$ azimuthal angle  and sightlines aligned along the minor axis (pole) are presented at 90$^\circ$ azimuthal angle. Detection strength and frequency decline rapidly between $R/R_{\rm 200} = 0.25$ and $0.75$. For star-forming galaxies, the azimuthal dependence becomes more pronounced with increasing absorption strength (Figure~\ref{fig:az_angle_combo}, bottom panels). At $W_c=100~\mathrm{m\AA}$, the covering fractions in the disk-oriented
($\theta<30^\circ$) and intermediate-angle
($30^\circ<\theta<60^\circ$) and consistent within their uncertainties while the minor axis oriented
($\theta>60^\circ$) sightlines show nominally higher covering fraction (median outside of uncertainty of other bins),
indicating weaker angular dependence for weak  \mgii\
absorption. In contrast, at $W_c=1000~\mathrm{m\AA}$, the
sightlines probing the minor axis of star-forming galaxies show a clear covering-fraction enhancement,
consistent with strong \mgii\ absorption preferentially tracing
bipolar outflows along the projected minor axis. The disk-oriented sightlines ($\theta<30^\circ$) show a nominally higher covering fraction than the intermediate-angle bin ($30^\circ<\theta<60^\circ$) for $W_c=500~\mathrm{m\AA}$ and $W_c=1000~\mathrm{m\AA}$, but the two are consistent within their uncertainties. Thus, the data securely establish a polar enhancement for strong absorption, while providing tentative evidence for an additional moderate disk-aligned enhancement of strong absorption that is consistent with bimodal azimuthal dependence observed in previous studies  \citep{Bordoloi2011,Bouche2012,Bordoloi2014_mgii,Martin2019}.

We find no significant azimuthal dependence around passive galaxies 
(Figure~\ref{fig:az_angle_combo}, bottom right), though the passive sample's 
sparse sampling and correspondingly large error bars limit our ability to rule 
out a weak asymmetry. The absence of a detectable signal in the passive sample, 
contrasted with a tentative bimodal trend in the star-forming sample, is 
consistent with the azimuthal asymmetry being linked to ongoing star-formation 
activity. This supports an interpretation in which galactic outflows and/or 
inflows associated with active star formation drive the observed angular 
dependence, reflecting a dynamic baryon cycle that regulates galactic feedback \citep{Bordoloi2014_mgii}.

\section{Discussion and Conclusion}
\label{sec:Discussion}

\textbf{Green valley galaxies as a CGM transition population.} A central result 
of this work is that green valley galaxies are not simply an intermediate 
classification bin, but trace a genuine, continuous transition in cool CGM 
content. Three independent diagnostics converge on this picture. First, the 
halo-radius-normalized residuals $\Delta$ (Section~\ref{subsec: Radial 
Profile Residuals}) show that green valley galaxies span a broad range of 
excess \mgii\ absorption, with detections both above and below the mean 
radial profile. This behavior is distinct from the more tightly clustered 
star-forming and quiescent populations. Second, the residual fraction $R_f$ 
is intermediate for green valley systems, falling between the star-forming 
and quiescent values. Third, the inner-CGM covering fraction $C_f$ for 
green valley galaxies is likewise intermediate at all $W_c$ thresholds we 
test (Section~\ref{subsec: Covering Fraction}). Together, these independent 
measures all place green valley galaxies between the star-forming and 
quiescent endpoints, indicating that the cool CGM does not vanish abruptly 
when star formation shuts off, but declines gradually as galaxies migrate 
across the green valley.

\begin{figure*}[ht!]
    \centering

    \includegraphics[width=.9\textwidth]{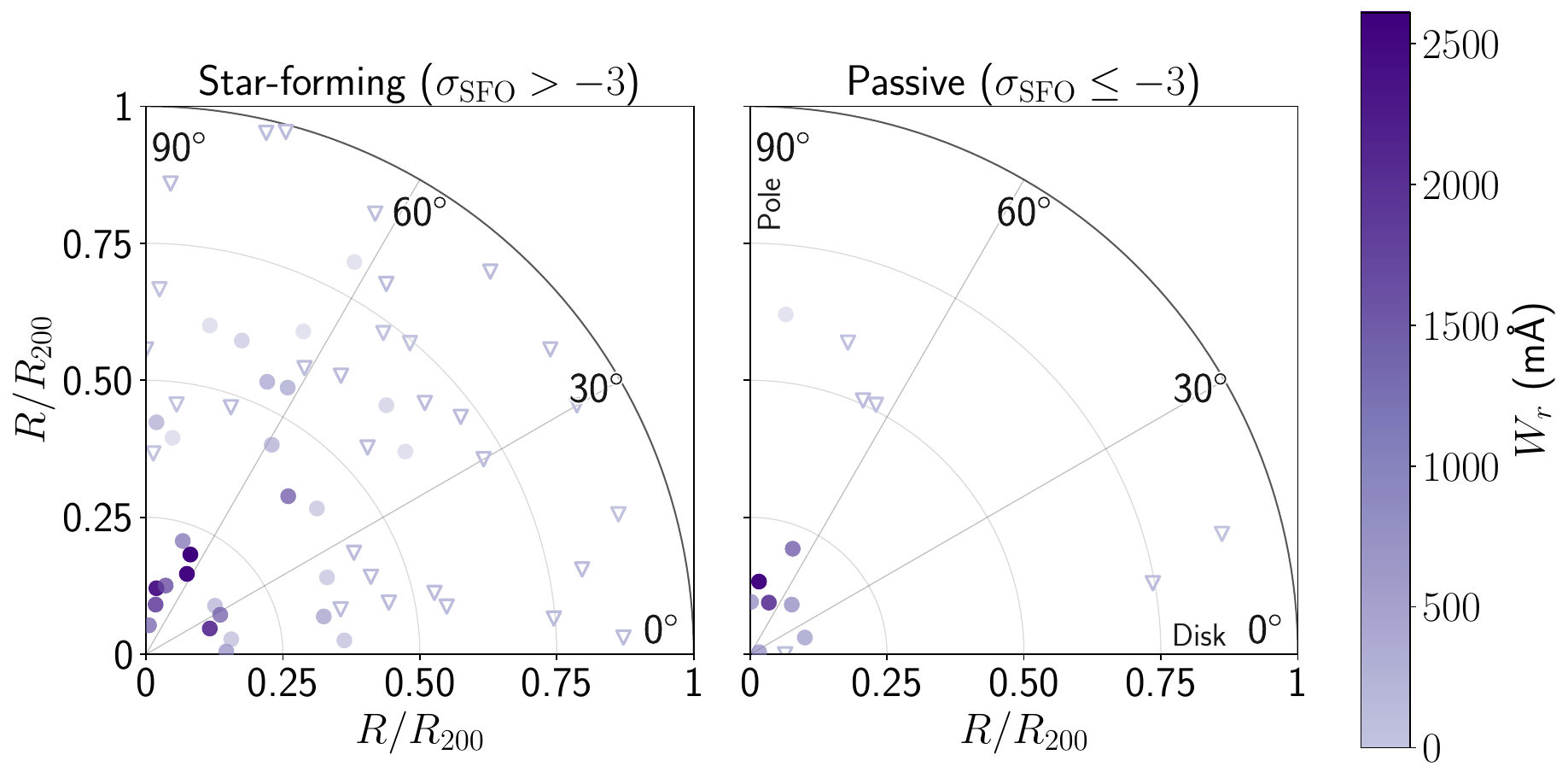}
    
    \vspace{0.1cm}

    \includegraphics[width=.9\textwidth]{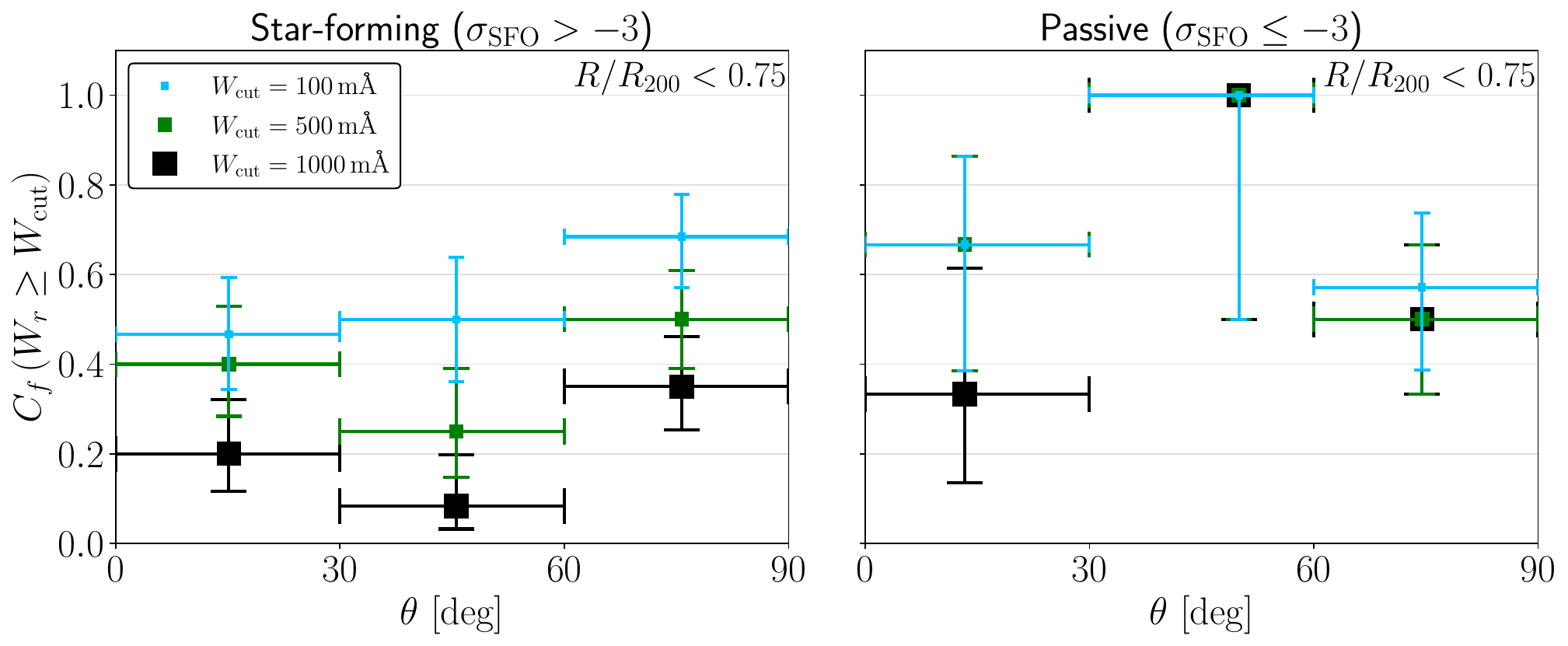}
    
\caption{
\textit{Top row:} 2D spatial distribution of \mgii\ detections (circles) and 
non-detections (inverted triangles) for star-forming (left) and passive (right) 
galaxies. Detection absorption strength is indicated by the color bar at right. 
Azimuthal angle $\theta$ is defined such that $0^\circ$ corresponds to alignment 
with the galaxy disk and $90^\circ$ to alignment with the pole. \textit{Bottom 
row:} covering fraction versus azimuthal angle for star-forming (left) and 
passive (right) galaxies, restricted to $R/R_{\rm 200} < 0.75$, at three 
equivalent width cutoffs. The sampling of passive 
disk-oriented sightlines is sparse, producing the larger error bars seen in 
the passive panel.
}
\label{fig:az_angle_combo}
\end{figure*}

Our results fit naturally into a feedback-recycling framework for the cool 
CGM post-cosmic noon along the quenching sequence. The enhanced \mgii\ covering fraction and equivalent 
width we find in the inner halos of star-forming galaxies 
($R/R_{\rm 200} \le 0.25$) are consistent with cosmological simulations in 
which the inner CGM is continually replenished by metal-rich outflows that 
cool and condense \citep{Appleby2021,Ford2013,Hafen2019,Muratov2017}. In 
these models, winds preferentially emerge along the minor axis while cool 
gas condenses and accretes along the major axis. This picture is consistent 
with the significant polar- and tentative disk-aligned cool CGM enhancement in our 
star-forming sample  (Figure~\ref{fig:az_angle_combo}).

In contrast, passive galaxies often occupy higher-mass hot halos and have higher virial-temperature gas, lengthening cooling times and suppressing the formation or survival of 
cool \mgii\ in the inner CGM \citep{Appleby2021}. We find that the inner-CGM 
covering fraction of strong \mgii\ ($W_c = 1000$~m\AA) is suppressed by a 
factor of 2.3 
in passive galaxies relative to star-forming galaxies 
(Figure~\ref{fig:coveringfraction_inner}), broadly consistent with the hot-mode 
accretion picture, in which massive halos retain less cool CGM gas than halos still fed by cold-mode accretion 
\citep{Dekel2005,Keres2006}. A global transition in halo properties likely 
occurs between these two regimes during quenching \citep{Appleby2023}, 
matching the decline in \mgii\ absorption we observe for green valley 
galaxies.

Although passive galaxies have decreased $R_f$ and inner-CGM $C_f$ values, our results show that passive galaxies are not fully depleted of cool gas, 
consistent with mechanisms that inhibit further ionization, such as 
self-shielding or turbulent entrainment \citep{Oppenheimer2018}. Luminous 
red galaxies are similarly known to host cool \mgii\ clumps thought to be 
preserved by the same mechanisms, but prevented from accreting \citep{Berg_2019,Afruni_2019,Huang2021}.

A subset of our passive sample may also be post-starburst: galaxies with a recent 
episode of elevated star formation followed by rapid quenching ($<1$~Gyr), 
whose \mgii\ has not yet had time to deplete. We therefore search our 
isolated quiescent sample ($\sigma_{\rm SFO} < -4.5$) for post-starburst 
signatures among \mgii\ detections.

As a case study, we highlight one galaxy with high-resolution spectroscopy that appears to be a textbook post-starburst system caught in the act of CGM transition. Using zCOSMOS spectra from the ESO Archive \citep{Lilly2009} and 
the K+A selection of \citet{Vergani_2009} ($D4000 > 1.3$, H$\delta > 3$~\AA, 
[\ion{O}{3}] $< -3$~\AA), we identify one post-starburst galaxy in our 
sample, ID~820041, shown in the bottom panel of Figure~\ref{fig:MgII_UVES}. 
Despite being classified as passive, it shows strong \mgii\ absorption 
($\log N > 14$) and complex, multi-component kinematics. With absorption spanning $\sim500$ \kms, it is the most kinematically extended absorber in our sample. HST imaging reveals a likely companion at $\approx 1.2''$ separation and a consistent photometric redshift ($z_{\rm phot} = 0.6866$ and $0.6777$), suggesting an ongoing interaction or merger. We interpret ID~820041 as direct evidence that the decline of cool CGM gas lags behind the cessation of star 
formation: merger-driven feedback can temporarily sustain \mgii\ absorption 
even after quenching begins, offering a concrete, high-resolution illustration 
of the same transitional behavior inferred statistically from the 
green valley population above.

The halo potential provides a key physical framework for interpreting the distribution and survival of cool CGM gas. The extended \mgii\ radial profile we measure around group 
environment galaxies beyond $\sim$200~kpc may trace gas displaced by tidal 
interactions, or more plausibly, diffuse cool gas expelled from 
individual halos and now bound to the group potential rather than to any 
single galaxy \citep{Bordoloi2011,Nielsen_2018,Dutta2020}. More broadly, our finding that inner-halo covering fractions scale with $\sigma_{\rm SFO}$ 
and $\log_{10}(\mathrm{sSFR})$ agrees closely with \citet{Huang2021} and 
\citet{Chaudary_2025}. Normalizing impact parameter by halo potential 
sharpens this contrast between star-forming and passive systems, both in 
covering fraction and in radial profile residuals. This normalization 
removes the leading-order dependence on halo size and isolates the link 
between \mgii\ enhancement and star-formation activity.

As halos grow over cosmic time,
simulations predict a shift from efficient gas escape at high redshift toward greater CGM retention and recycling at later times. FIRE simulations show that the mass outflow rates of metals decrease by a factor of 
2--5 between $0.25\,R_{\rm 200}$ and $R_{\rm 200}$ \citep{Muratov2017}, and 
that the CGM retains $>50\%$ of its metals at $z < 2$ \citep{Hafen2019}. In 
contrast, at $z > 2$--3 only $\sim$30\% of \mgii\ absorbers are fully bound 
to galaxy halos \citep{Bordoloi2024,Rudie_2019}. Our sample, with a median 
redshift of $z = 0.51$, shows \mgii\ absorption that is almost entirely 
bound. This is consistent with a picture in which gas is more readily 
ejected from galaxies at early times ($z > 2$--3). Post-cosmic-noon 
hierarchical halo growth instead transforms the CGM into a recycling 
reservoir, in which cool material circulates between the ISM and inner CGM 
on Gyr timescales without large-scale expulsion into the IGM.

We note that even in this large survey several limitations remain. Our sample contains substantially fewer green valley, starburst, and quiescent galaxies than star-forming galaxies, owing to the need for spectroscopically confirmed redshifts, which 
preferentially limits the sampling of passive systems. Our passive-disk subsample is particularly 
small; enlarging this sample is necessary to robustly compare the azimuthal 
angle dependence between star-forming and passive disk galaxies. Stellar 
masses and star formation rates are drawn from heterogeneous catalogs with 
different estimation methodologies, introducing additional uncertainty and 
potential systematic offsets. In future work, expanding the high-redshift ($z \ge 1$) 
inner-CGM sample would test whether these trends persist closer to cosmic 
noon, where cold-mode accretion is more prevalent, enabling a more complete 
picture of halo evolution over time. Comparing observations of the full multiphase CGM with 
forward-modeled simulations that track both metal recycling and 
environment-specific processes will be essential for understanding the role 
of the cool CGM in galaxy quenching.

\subsection{Summary}

From a combined dataset of 716 galaxies: 169 galaxies from this work and 
547 galaxies from six archival datasets spanning $0.07 < z < 2.7$, we find:

\begin{enumerate}

\item \textbf{Radial profile residuals trace quenching.} After normalizing by halo potential, the scatter in $\log_{10} W_r$ around the mean radial profile correlates significantly with $\sigma_{\rm SFO}$ and $\log_{10}(\mathrm{sSFR})$, and anticorrelate with 
$\log_{10}(M_*/M_\odot)$, likely due to mass--sSFR covariance. The residual fraction $R_f$ decreases as $\sigma_{\rm SFO}$ 
drops, tracking the quenching sequence from star-forming through the green 
valley to passive galaxies. This indicates that the scatter in \mgii\ absorption around the mean radial profile is not random, but traces the star-formation activity in the host 
galaxy with CGM \mgii\ absorption strength reducing as galaxies quench.

\item \textbf{Inner-CGM \mgii\ trends with quenching.} At 
$R/R_{\rm 200} < 0.25$, the covering fraction of strong \mgii\ 
($W_c = 1000$~m\AA) is enhanced by a factor of 2.3 in star-forming 
galaxies relative to passive galaxies, with the enhancement increasing 
toward higher $W_c$. Separating the sample into four $\sigma_{\rm SFO}$ 
categories reveals intermediate $C_f$ for green valley galaxies, indicating 
a continuous decline in \mgii\ absorption during quenching rather than a 
sharp transition.

\item \textbf{Environmental dependence.} When associating each absorber with 
its closest galaxy, group-environment galaxies show a slightly extended, 
low-$W_r$ \mgii\ profile, with $C_f$ flattening beyond $\sim$200~kpc. This is 
consistent with weaker \mgii\ clouds becoming bound to the group potential 
and redistributed outward through group-scale kinematic processes.

\item \textbf{Active kinematics in star-forming galaxies.} The large 
majority of clouds are gravitationally bound ($|v|/v_{\rm esc} \le 1$), with 
column density declining away from the galaxy systemic velocity. Clouds 
around passive galaxies show higher mean absolute velocities than those around 
star-forming galaxies, but once normalized by escape velocity, the two 
populations show comparable kinematics. Normalized velocity dispersions, 
however, are larger for star-forming galaxies, indicating a more 
dynamically active CGM than around passive galaxies, consistent with a 
picture of virialized passive halos and kinematically disturbed 
star-forming halos.

\item \textbf{Azimuthal dependence.}  Strong ($W_c=1000~\mathrm{m\AA}$) \mgii\ absorption around star-forming disk galaxies shows a clear polar enhancement within
$R/R_{\rm 200}<0.75$. Strong absorption for disk-oriented covering fraction is also
higher than at intermediate angles, but the two are consistent within their
uncertainties, suggesting a possible moderate disk enhancement. 
Detections extend to larger $R/R_{\rm 200}$ along polar angles than along 
disk angles, consistent with bipolar inflow--outflow models in which 
outflows oriented perpendicular to the disk plane drive cool gas to large 
galactocentric distances.

\end{enumerate}

\begin{acknowledgments}
This paper includes data gathered with the 6.5 meter Magellan Telescopes located at Las Campanas Observatory, Chile. This research has made use of the NASA/IPAC Infrared Science Archive, which is funded by the National Aeronautics and Space Administration and operated by the California Institute of Technology. This work is based on observations collected at the European Southern Observatory under ESO programmes 083.A-0401(A) and 086.A-0974(A). This work is based on data obtained from the ESO Science Archive Facility with DOI's: https://doi.eso.org/10.18727/archive/50, https://doi.org/10.18727/archive/71.
Some of the data presented herein were obtained at Keck Observatory, which is a private 501(c)3 non-profit organization operated as a scientific partnership among the California Institute of Technology, the University of California, and the National Aeronautics and Space Administration. The Observatory was made possible by the generous financial support of the W. M. Keck Foundation. This research has made use of the Keck Observatory Archive (KOA), which is operated by the W. M. Keck Observatory and the NASA Exoplanet Science Institute (NExScI), under contract with the National Aeronautics and Space Administration. JKW gratefully acknowledges support from NSF-AST 1812521 and NSF-CAREER 2044303.

\end{acknowledgments}
\facilities{Magellan:MagE, VLT:UVES, VLT:X-shooter, Keck:I (HIRES), IRSA, HST (WFC3)}

\software{astropy \citep{2013A&A...558A..33A,2018AJ....156..123A}, 
          rbcodes \citep{rbcodes},
          rbvfit \citep{rbvfit}.
          }

\bibliography{bibliography}

\begin{thebibliography}{}
\expandafter\ifx\csname natexlab\endcsname\relax\def\natexlab#1{#1}\fi
\providecommand{\url}[1]{\href{#1}{#1}}
\providecommand{\dodoi}[1]{doi:~\href{http://doi.org/#1}{\nolinkurl{#1}}}
\providecommand{\doeprint}[1]{\href{http://ascl.net/#1}{\nolinkurl{http://ascl.net/#1}}}
\providecommand{\doarXiv}[1]{\href{https://arxiv.org/abs/#1}{\nolinkurl{https://arxiv.org/abs/#1}}}

\bibitem[{A. {Afruni} {et~al.}(2019){Afruni}, {Fraternali}, \&
  {Pezzulli}}]{Afruni_2019}
{Afruni}, A., {Fraternali}, F., \& {Pezzulli}, G. 2019, \bibinfo{title}{{Cool
  circumgalactic gas of passive galaxies from cosmological inflow},} \aap, 625,
  A11, \dodoi{10.1051/0004-6361/201835002}

\bibitem[{D. {Angl{\'e}s-Alc{\'a}zar} {et~al.}(2017){Angl{\'e}s-Alc{\'a}zar},
  {Faucher-Gigu{\`e}re}, {Kere{\v{s}}}, {Hopkins}, {Quataert}, \&
  {Murray}}]{Alcazar2017}
{Angl{\'e}s-Alc{\'a}zar}, D., {Faucher-Gigu{\`e}re}, C.-A., {Kere{\v{s}}}, D.,
  {et~al.} 2017, \bibinfo{title}{{The cosmic baryon cycle and galaxy mass
  assembly in the FIRE simulations},} \mnras, 470, 4698,
  \dodoi{10.1093/mnras/stx1517}

\bibitem[{J. {Angthopo} {et~al.}(2020){Angthopo}, {Ferreras}, \&
  {Silk}}]{Angthopo2020}
{Angthopo}, J., {Ferreras}, I., \& {Silk}, J. 2020, \bibinfo{title}{{A detailed
  look at the stellar populations in green valley galaxies},} \mnras, 495,
  2720, \dodoi{10.1093/mnras/staa1276}

\bibitem[{S. {Appleby} {et~al.}(2023){Appleby}, {Dav{\'e}}, {Sorini}, {Cui}, \&
  {Christiansen}}]{Appleby2023}
{Appleby}, S., {Dav{\'e}}, R., {Sorini}, D., {Cui}, W., \& {Christiansen}, J.
  2023, \bibinfo{title}{{The physical nature of circumgalactic medium absorbers
  in SIMBA},} \mnras, 519, 5514, \dodoi{10.1093/mnras/stad025}

\bibitem[{S. {Appleby} {et~al.}(2021){Appleby}, {Dav{\'e}}, {Sorini},
  {Storey-Fisher}, \& {Smith}}]{Appleby2021}
{Appleby}, S., {Dav{\'e}}, R., {Sorini}, D., {Storey-Fisher}, K., \& {Smith},
  B. 2021, \bibinfo{title}{{The low-redshift circumgalactic medium in SIMBA},}
  \mnras, 507, 2383, \dodoi{10.1093/mnras/stab2310}

\bibitem[{ {Astropy Collaboration} {et~al.}(2013){Astropy Collaboration},
  {Robitaille}, {Tollerud}, {Greenfield}, {Droettboom}, {Bray}, {Aldcroft},
  {Davis}, {Ginsburg}, {Price-Whelan}, {Kerzendorf}, {Conley}, {Crighton},
  {Barbary}, {Muna}, {Ferguson}, {Grollier}, {Parikh}, {Nair}, {Unther},
  {Deil}, {Woillez}, {Conseil}, {Kramer}, {Turner}, {Singer}, {Fox}, {Weaver},
  {Zabalza}, {Edwards}, {Azalee Bostroem}, {Burke}, {Casey}, {Crawford},
  {Dencheva}, {Ely}, {Jenness}, {Labrie}, {Lim}, {Pierfederici}, {Pontzen},
  {Ptak}, {Refsdal}, {Servillat}, \& {Streicher}}]{2013A&A...558A..33A}
{Astropy Collaboration}, {Robitaille}, T.~P., {Tollerud}, E.~J., {et~al.} 2013,
  \bibinfo{title}{{Astropy: A community Python package for astronomy},} \aap,
  558, A33, \dodoi{10.1051/0004-6361/201322068}

\bibitem[{ {Astropy Collaboration} {et~al.}(2018){Astropy Collaboration},
  {Price-Whelan}, {Sip{\H{o}}cz}, {G{\"u}nther}, {Lim}, {Crawford}, {Conseil},
  {Shupe}, {Craig}, {Dencheva}, {Ginsburg}, {VanderPlas}, {Bradley},
  {P{\'e}rez-Su{\'a}rez}, {de Val-Borro}, {Aldcroft}, {Cruz}, {Robitaille},
  {Tollerud}, {Ardelean}, {Babej}, {Bach}, {Bachetti}, {Bakanov}, {Bamford},
  {Barentsen}, {Barmby}, {Baumbach}, {Berry}, {Biscani}, {Boquien}, {Bostroem},
  {Bouma}, {Brammer}, {Bray}, {Breytenbach}, {Buddelmeijer}, {Burke},
  {Calderone}, {Cano Rodr{\'\i}guez}, {Cara}, {Cardoso}, {Cheedella}, {Copin},
  {Corrales}, {Crichton}, {D'Avella}, {Deil}, {Depagne}, {Dietrich}, {Donath},
  {Droettboom}, {Earl}, {Erben}, {Fabbro}, {Ferreira}, {Finethy}, {Fox},
  {Garrison}, {Gibbons}, {Goldstein}, {Gommers}, {Greco}, {Greenfield},
  {Groener}, {Grollier}, {Hagen}, {Hirst}, {Homeier}, {Horton}, {Hosseinzadeh},
  {Hu}, {Hunkeler}, {Ivezi{\'c}}, {Jain}, {Jenness}, {Kanarek}, {Kendrew},
  {Kern}, {Kerzendorf}, {Khvalko}, {King}, {Kirkby}, {Kulkarni}, {Kumar},
  {Lee}, {Lenz}, {Littlefair}, {Ma}, {Macleod}, {Mastropietro}, {McCully},
  {Montagnac}, {Morris}, {Mueller}, {Mumford}, {Muna}, {Murphy}, {Nelson},
  {Nguyen}, {Ninan}, {N{\"o}the}, {Ogaz}, {Oh}, {Parejko}, {Parley}, {Pascual},
  {Patil}, {Patil}, {Plunkett}, {Prochaska}, {Rastogi}, {Reddy Janga},
  {Sabater}, {Sakurikar}, {Seifert}, {Sherbert}, {Sherwood-Taylor}, {Shih},
  {Sick}, {Silbiger}, {Singanamalla}, {Singer}, {Sladen}, {Sooley},
  {Sornarajah}, {Streicher}, {Teuben}, {Thomas}, {Tremblay}, {Turner},
  {Terr{\'o}n}, {van Kerkwijk}, {de la Vega}, {Watkins}, {Weaver}, {Whitmore},
  {Woillez}, {Zabalza}, \& {Astropy Contributors}}]{2018AJ....156..123A}
{Astropy Collaboration}, {Price-Whelan}, A.~M., {Sip{\H{o}}cz}, B.~M., {et~al.}
  2018, \bibinfo{title}{{The Astropy Project: Building an Open-science Project
  and Status of the v2.0 Core Package},} \aj, 156, 123,
  \dodoi{10.3847/1538-3881/aabc4f}

\bibitem[{P. {Behroozi} {et~al.}(2019){Behroozi}, {Wechsler}, {Hearin}, \&
  {Conroy}}]{Behroozi2019}
{Behroozi}, P., {Wechsler}, R.~H., {Hearin}, A.~P., \& {Conroy}, C. 2019,
  \bibinfo{title}{{UNIVERSEMACHINE: The correlation between galaxy growth and
  dark matter halo assembly from z = 0-10},} \mnras, 488, 3143,
  \dodoi{10.1093/mnras/stz1182}

\bibitem[{M.~A. {Berg} {et~al.}(2019){Berg}, {Howk}, {Lehner}, {Wotta},
  {O'Meara}, {Bowen}, {Burchett}, {Peeples}, \& {Tejos}}]{Berg_2019}
{Berg}, M.~A., {Howk}, J.~C., {Lehner}, N., {et~al.} 2019, \bibinfo{title}{{The
  Red Dead Redemption Survey of Circumgalactic Gas about Massive Galaxies. I.
  Mass and Metallicity of the Cool Phase},} \apj, 883, 5,
  \dodoi{10.3847/1538-4357/ab378e}

\bibitem[{J. {Bergeron}(1986){Bergeron}}]{Bergeron_1986}
{Bergeron}, J. 1986, \bibinfo{title}{{The MG II absorption system in the QSO
  PKS 2128-12 : a galaxy disc/halo with a radius of 65 kpc.},} \aap, 155, L8

\bibitem[{J. {Bergeron} \& P. {Boiss{\'e}}(1991){Bergeron} \&
  {Boiss{\'e}}}]{Bergeron_1991}
{Bergeron}, J., \& {Boiss{\'e}}, P. 1991, \bibinfo{title}{{A sample of galaxies
  giving rise to Mg II quasar absorption systems.},} \aap, 243, 344

\bibitem[{J.~J. {Bochanski} {et~al.}(2009){Bochanski}, {Hennawi}, {Simcoe},
  {Prochaska}, {West}, {Burgasser}, {Burles}, {Bernstein}, {Williams}, \&
  {Murphy}}]{Bochanski2009}
{Bochanski}, J.~J., {Hennawi}, J.~F., {Simcoe}, R.~A., {et~al.} 2009,
  \bibinfo{title}{{MASE: A New Data-Reduction Pipeline for the Magellan
  Echellette Spectrograph},} \pasp, 121, 1409, \dodoi{10.1086/648597}

\bibitem[{R. Bordoloi \& J. Higginson(2025)Bordoloi \& Higginson}]{rbvfit}
Bordoloi, R., \& Higginson, J. 2025, \bibinfo{title}{rbvfit: A Voigt profile
  fitting tool,}, 2.0.0 Zenodo, \dodoi{10.5281/zenodo.16318060}

\bibitem[{R. {Bordoloi} {et~al.}(2014{\natexlab{a}}){Bordoloi}, {Lilly},
  {Kacprzak}, \& {Churchill}}]{Bordoloi2014_mgii}
{Bordoloi}, R., {Lilly}, S.~J., {Kacprzak}, G.~G., \& {Churchill}, C.~W.
  2014{\natexlab{a}}, \bibinfo{title}{{Modeling the Distribution of Mg II
  Absorbers around Galaxies Using Background Galaxies and Quasars},} \apj, 784,
  108, \dodoi{10.1088/0004-637X/784/2/108}

\bibitem[{R. Bordoloi {et~al.}(2025)Bordoloi, Liu, Clark, Higginson, \&
  Flores}]{rbcodes}
Bordoloi, R., Liu, B., Clark, S., Higginson, J., \& Flores, D. 2025,
  \bibinfo{title}{rongmon/rbcodes: rbcodes v2.0.0,}, v2.0.0 Zenodo,
  \dodoi{10.5281/zenodo.6079263}

\bibitem[{R. {Bordoloi} {et~al.}(2018){Bordoloi}, {Prochaska}, {Tumlinson},
  {Werk}, {Tripp}, \& {Burchett}}]{Bordoloi2018}
{Bordoloi}, R., {Prochaska}, J.~X., {Tumlinson}, J., {et~al.} 2018,
  \bibinfo{title}{{On the CGM Fundamental Plane: The Halo Mass Dependency of
  Circumgalactic H I},} \apj, 864, 132, \dodoi{10.3847/1538-4357/aad8ac}

\bibitem[{R. {Bordoloi} {et~al.}(2011){Bordoloi}, {Lilly}, {Knobel},
  {Bolzonella}, {Kampczyk}, {Carollo}, {Iovino}, {Zucca}, {Contini}, {Kneib},
  {Le Fevre}, {Mainieri}, {Renzini}, {Scodeggio}, {Zamorani}, {Balestra},
  {Bardelli}, {Bongiorno}, {Caputi}, {Cucciati}, {de la Torre}, {de Ravel},
  {Garilli}, {Kova{\v{c}}}, {Lamareille}, {Le Borgne}, {Le Brun}, {Maier},
  {Mignoli}, {Pello}, {Peng}, {Perez Montero}, {Presotto}, {Scarlata},
  {Silverman}, {Tanaka}, {Tasca}, {Tresse}, {Vergani}, {Barnes}, {Cappi},
  {Cimatti}, {Coppa}, {Diener}, {Franzetti}, {Koekemoer}, {L{\'o}pez-Sanjuan},
  {McCracken}, {Moresco}, {Nair}, {Oesch}, {Pozzetti}, \&
  {Welikala}}]{Bordoloi2011}
{Bordoloi}, R., {Lilly}, S.~J., {Knobel}, C., {et~al.} 2011,
  \bibinfo{title}{{The Radial and Azimuthal Profiles of Mg II Absorption around
  0.5 < z < 0.9 zCOSMOS Galaxies of Different Colors, Masses, and
  Environments},} \apj, 743, 10, \dodoi{10.1088/0004-637X/743/1/10}

\bibitem[{R. {Bordoloi} {et~al.}(2014{\natexlab{b}}){Bordoloi}, {Lilly},
  {Hardmeier}, {Contini}, {Kneib}, {Le Fevre}, {Mainieri}, {Renzini},
  {Scodeggio}, {Zamorani}, {Bardelli}, {Bolzonella}, {Bongiorno}, {Caputi},
  {Carollo}, {Cucciati}, {de la Torre}, {de Ravel}, {Garilli}, {Iovino},
  {Kampczyk}, {Kova{\v{c}}}, {Knobel}, {Lamareille}, {Le Borgne}, {Le Brun},
  {Maier}, {Mignoli}, {Oesch}, {Pello}, {Peng}, {Perez Montero}, {Presotto},
  {Silverman}, {Tanaka}, {Tasca}, {Tresse}, {Vergani}, {Zucca}, {Cappi},
  {Cimatti}, {Coppa}, {Franzetti}, {Koekemoer}, {Moresco}, {Nair}, \&
  {Pozzetti}}]{Bordoloi2014outflow}
{Bordoloi}, R., {Lilly}, S.~J., {Hardmeier}, E., {et~al.} 2014{\natexlab{b}},
  \bibinfo{title}{{The Dependence of Galactic Outflows on the Properties and
  Orientation of zCOSMOS Galaxies at z \raisebox{-0.5ex}\textasciitilde 1},}
  \apj, 794, 130, \dodoi{10.1088/0004-637X/794/2/130}

\bibitem[{R. {Bordoloi} {et~al.}(2014{\natexlab{c}}){Bordoloi}, {Tumlinson},
  {Werk}, {Oppenheimer}, {Peeples}, {Prochaska}, {Tripp}, {Katz}, {Dav{\'e}},
  {Fox}, {Thom}, {Ford}, {Weinberg}, {Burchett}, \& {Kollmeier}}]{Bordoloi2014}
{Bordoloi}, R., {Tumlinson}, J., {Werk}, J.~K., {et~al.} 2014{\natexlab{c}},
  \bibinfo{title}{{The COS-Dwarfs Survey: The Carbon Reservoir around Sub-L*
  Galaxies},} \apj, 796, 136, \dodoi{10.1088/0004-637X/796/2/136}

\bibitem[{R. {Bordoloi} {et~al.}(2024){Bordoloi}, {Simcoe}, {Matthee},
  {Kashino}, {Mackenzie}, {Lilly}, {Eilers}, {Liu}, {DePalma}, {Yue}, \& {P.
  Naidu}}]{Bordoloi2024}
{Bordoloi}, R., {Simcoe}, R.~A., {Matthee}, J., {et~al.} 2024,
  \bibinfo{title}{{EIGER IV. The Cool {}10$^{4}$ K Circumgalactic Environment
  of High-redshift Galaxies Reveals Remarkably Efficient Intergalactic Medium
  Enrichment},} \apj, 963, 28, \dodoi{10.3847/1538-4357/ad1b63}

\bibitem[{S. {Borthakur} {et~al.}(2015){Borthakur}, {Heckman}, {Tumlinson},
  {Bordoloi}, {Thom}, {Catinella}, {Schiminovich}, {Dav{\'e}}, {Kauffmann},
  {Moran}, \& {Saintonge}}]{Borthakur_2015}
{Borthakur}, S., {Heckman}, T., {Tumlinson}, J., {et~al.} 2015,
  \bibinfo{title}{{Connection between the Circumgalactic Medium and the
  Interstellar Medium of Galaxies: Results from the COS-GASS Survey},} \apj,
  813, 46, \dodoi{10.1088/0004-637X/813/1/46}

\bibitem[{N. {Bouch{\'e}} {et~al.}(2012){Bouch{\'e}}, {Hohensee}, {Vargas},
  {Kacprzak}, {Martin}, {Cooke}, \& {Churchill}}]{Bouche2012}
{Bouch{\'e}}, N., {Hohensee}, W., {Vargas}, R., {et~al.} 2012,
  \bibinfo{title}{{Physical properties of galactic winds using background
  quasars},} \mnras, 426, 801, \dodoi{10.1111/j.1365-2966.2012.21114.x}

\bibitem[{J.~N. {Burchett} {et~al.}(2013){Burchett}, {Tripp}, {Werk}, {Howk},
  {Prochaska}, {Ford}, \& {Dav{\'e}}}]{Burchett_2013}
{Burchett}, J.~N., {Tripp}, T.~M., {Werk}, J.~K., {et~al.} 2013,
  \bibinfo{title}{{A Deep Search for Faint Galaxies Associated with Very
  Low-redshift C IV Absorbers: A Case with Cold-accretion Characteristics},}
  \apjl, 779, L17, \dodoi{10.1088/2041-8205/779/2/L17}

\bibitem[{C.~M. {Casey} {et~al.}(2023){Casey}, {Kartaltepe}, {Drakos},
  {Franco}, {Harish}, {Paquereau}, {Ilbert}, {Rose}, {Cox}, {Nightingale},
  {Robertson}, {Silverman}, {Koekemoer}, {Massey}, {McCracken}, {Rhodes},
  {Akins}, {Allen}, {Amvrosiadis}, {Arango-Toro}, {Bagley}, {Bongiorno},
  {Capak}, {Champagne}, {Chartab}, {Ch{\'a}vez Ortiz}, {Chworowsky}, {Cooke},
  {Cooper}, {Darvish}, {Ding}, {Faisst}, {Finkelstein}, {Fujimoto}, {Gentile},
  {Gillman}, {Gould}, {Gozaliasl}, {Hayward}, {He}, {Hemmati}, {Hirschmann},
  {Jahnke}, {Jin}, {Khostovan}, {Kokorev}, {Lambrides}, {Laigle}, {Larson},
  {Leung}, {Liu}, {Liaudat}, {Long}, {Magdis}, {Mahler}, {Mainieri}, {Manning},
  {Maraston}, {Martin}, {McCleary}, {McKinney}, {McPartland}, {Mobasher},
  {Pattnaik}, {Renzini}, {Rich}, {Sanders}, {Sattari}, {Scognamiglio},
  {Scoville}, {Sheth}, {Shuntov}, {Sparre}, {Suzuki}, {Talia}, {Toft},
  {Trakhtenbrot}, {Urry}, {Valentino}, {Vanderhoof}, {Vardoulaki}, {Weaver},
  {Whitaker}, {Wilkins}, {Yang}, \& {Zavala}}]{Casey_2023}
{Casey}, C.~M., {Kartaltepe}, J.~S., {Drakos}, N.~E., {et~al.} 2023,
  \bibinfo{title}{{COSMOS-Web: An Overview of the JWST Cosmic Origins Survey},}
  \apj, 954, 31, \dodoi{10.3847/1538-4357/acc2bc}

\bibitem[{R. {Chaudhary} {et~al.}(2026){Chaudhary}, {Joshi}, {Das},
  {Fumagalli}, {Kacprzak}, {Fossati}, {P{\'e}roux}, \& {Ho}}]{Chaudary_2025}
{Chaudhary}, R., {Joshi}, R., {Das}, S., {et~al.} 2026,
  \bibinfo{title}{{Baryonic Ecosystem IN Galaxies (BEINGMgII): III. Cool gas
  reservoirs at 0.3 {\ensuremath{\leq}} z {\ensuremath{\leq}} 1.6 in the Dark
  Energy Survey},} \aap, 709, A238, \dodoi{10.1051/0004-6361/202557080}

\bibitem[{P. {Chauke} {et~al.}(2019){Chauke}, {van der Wel}, {Pacifici},
  {Bezanson}, {Wu}, {Gallazzi}, {Straatman}, {Franx}, {Bari{\v{s}}i{\'c}},
  {Bell}, {van Houdt}, {Maseda}, {Muzzin}, {Sobral}, \& {Spilker}}]{Chauke2019}
{Chauke}, P., {van der Wel}, A., {Pacifici}, C., {et~al.} 2019,
  \bibinfo{title}{{Rejuvenation in z {\ensuremath{\sim}} 0.8 Quiescent Galaxies
  in LEGA-C},} \apj, 877, 48, \dodoi{10.3847/1538-4357/ab164d}

\bibitem[{H.-W. {Chen} {et~al.}(2010){Chen}, {Helsby}, {Gauthier}, {Shectman},
  {Thompson}, \& {Tinker}}]{Chen2010}
{Chen}, H.-W., {Helsby}, J.~E., {Gauthier}, J.-R., {et~al.} 2010,
  \bibinfo{title}{{An Empirical Characterization of Extended Cool Gas Around
  Galaxies Using Mg II Absorption Features},} \apj, 714, 1521,
  \dodoi{10.1088/0004-637X/714/2/1521}

\bibitem[{H.-W. {Chen} \& F.~S. {Zahedy}(2026){Chen} \& {Zahedy}}]{Chen_2026}
{Chen}, H.-W., \& {Zahedy}, F.~S. 2026, in Encyclopedia of Astrophysics, Volume
  4, Vol.~4, 370--400, \dodoi{10.1016/B978-0-443-21439-4.00059-6}

\bibitem[{S.-F.~S. {Chen} {et~al.}(2017){Chen}, {Simcoe}, {Torrey},
  {Ba{\~n}ados}, {Cooksey}, {Cooper}, {Furesz}, {Matejek}, {Miller}, {Turner},
  {Venemans}, {Decarli}, {Farina}, {Mazzucchelli}, \& {Walter}}]{Chen2017}
{Chen}, S.-F.~S., {Simcoe}, R.~A., {Torrey}, P., {et~al.} 2017,
  \bibinfo{title}{{Mg II Absorption at 2 < Z < 7 with Magellan/Fire. III. Full
  Statistics of Absorption toward 100 High-redshift QSOs},} \apj, 850, 188,
  \dodoi{10.3847/1538-4357/aa9707}

\bibitem[{M. {Cherrey} {et~al.}(2025){Cherrey}, {Bouch{\'e}}, {Zabl},
  {Schroetter}, {Wendt}, {Langan}, {Schaye}, {Wisotzki}, {Guo}, \&
  {Pessa}}]{Cherrey_2025}
{Cherrey}, M., {Bouch{\'e}}, N.~F., {Zabl}, J., {et~al.} 2025,
  \bibinfo{title}{{MusE GAs FLOw and Wind (MEGAFLOW): XIII. Cool gas traced by
  Mg II around isolated galaxies},} \aap, 694, A117,
  \dodoi{10.1051/0004-6361/202451165}

\bibitem[{C.~W. {Churchill} {et~al.}(2025){Churchill}, {Abbas}, {Kacprzak}, \&
  {Nielsen}}]{Churchill2025}
{Churchill}, C.~W., {Abbas}, A., {Kacprzak}, G.~G., \& {Nielsen}, N.~M. 2025,
  \bibinfo{title}{{13 Billion Years of MgII Absorber Evolution},} arXiv
  e-prints, arXiv:2510.01430, \dodoi{10.48550/arXiv.2510.01430}

\bibitem[{C.~W. {Churchill} {et~al.}(2013){Churchill}, {Trujillo-Gomez},
  {Nielsen}, \& {Kacprzak}}]{Churchill2013}
{Churchill}, C.~W., {Trujillo-Gomez}, S., {Nielsen}, N.~M., \& {Kacprzak},
  G.~G. 2013, \bibinfo{title}{{MAGIICAT III. Interpreting Self-similarity of
  the Circumgalactic Medium with Virial Mass Using Mg II Absorption},} \apj,
  779, 87, \dodoi{10.1088/0004-637X/779/1/87}

\bibitem[{C. {Conroy}(2013){Conroy}}]{Conroy2013}
{Conroy}, C. 2013, \bibinfo{title}{{Modeling the Panchromatic Spectral Energy
  Distributions of Galaxies},} \araa, 51, 393,
  \dodoi{10.1146/annurev-astro-082812-141017}

\bibitem[{ {COSMOS Project}(2024){COSMOS Project}}]{IRSA_COSMOS_DOI}
{COSMOS Project}. 2024, \bibinfo{title}{{Cosmic Evolution Survey with HST
  (COSMOS)},}, IRSA COSMOS Archive IPAC, \dodoi{10.26131/IRSA178}

\bibitem[{M. {Damen} {et~al.}(2009){Damen}, {Labb{\'e}}, {Franx}, {van Dokkum},
  {Taylor}, \& {Gawiser}}]{Damen_2009}
{Damen}, M., {Labb{\'e}}, I., {Franx}, M., {et~al.} 2009, \bibinfo{title}{{The
  Evolution of the Specific Star Formation Rate of Massive Galaxies to z
  \raisebox{-0.5ex}\textasciitilde 1.8 in the Extended Chandra Deep Field
  South},} \apj, 690, 937, \dodoi{10.1088/0004-637X/690/1/937}

\bibitem[{B. {Darvish} {et~al.}(2017){Darvish}, {Mobasher}, {Martin}, {Sobral},
  {Scoville}, {Stroe}, {Hemmati}, \& {Kartaltepe}}]{Darvish_2017}
{Darvish}, B., {Mobasher}, B., {Martin}, D.~C., {et~al.} 2017,
  \bibinfo{title}{{VizieR Online Data Catalog: Cosmic web of galaxies in the
  COSMOS field (Darvish+, 2017)},}, VizieR On-line Data Catalog: J/ApJ/837/16.
  Originally published in: 2017ApJ...837...16D
  \dodoi{10.26093/cds/vizier.18370016}

\bibitem[{S. {Das} {et~al.}(2025){Das}, {Joshi}, {Chaudhary}, {Fumagalli},
  {Fossati}, {P{\'e}roux}, \& {Ho}}]{Das_2025}
{Das}, S., {Joshi}, R., {Chaudhary}, R., {et~al.} 2025,
  \bibinfo{title}{{Baryonic Ecosystem IN Galaxies (BEINGMgII): II. Unveiling
  the nature of galaxies harbouring cool gas reservoirs},} \aap, 695, A207,
  \dodoi{10.1051/0004-6361/202452494}

\bibitem[{A. {Dekel} \& Y. {Birnboim}(2006){Dekel} \& {Birnboim}}]{Dekel2005}
{Dekel}, A., \& {Birnboim}, Y. 2006, \bibinfo{title}{{Galaxy bimodality due to
  cold flows and shock heating},} \mnras, 368, 2,
  \dodoi{10.1111/j.1365-2966.2006.10145.x}

\bibitem[{C. {Diener} {et~al.}(2013){Diener}, {Lilly}, {Knobel}, {Zamorani},
  {Lemson}, {Kampczyk}, {Scoville}, {Carollo}, {Contini}, {Kneib}, {Le Fevre},
  {Mainieri}, {Renzini}, {Scodeggio}, {Bardelli}, {Bolzonella}, {Bongiorno},
  {Caputi}, {Cucciati}, {de la Torre}, {de Ravel}, {Franzetti}, {Garilli},
  {Iovino}, {Kova{\v{c}}}, {Lamareille}, {Le Borgne}, {Le Brun}, {Maier},
  {Mignoli}, {Pello}, {Peng}, {Perez Montero}, {Presotto}, {Silverman},
  {Tanaka}, {Tasca}, {Tresse}, {Vergani}, {Zucca}, {Bordoloi}, {Cappi},
  {Cimatti}, {Coppa}, {Koekemoer}, {L{\'o}pez-Sanjuan}, {McCracken}, {Moresco},
  {Nair}, {Pozzetti}, \& {Welikala}}]{Diener_2013}
{Diener}, C., {Lilly}, S.~J., {Knobel}, C., {et~al.} 2013,
  \bibinfo{title}{{Proto-groups at 1.8 < z < 3 in the zCOSMOS-deep Sample},}
  \apj, 765, 109, \dodoi{10.1088/0004-637X/765/2/109}

\bibitem[{R. {Dutta} {et~al.}(2020){Dutta}, {Fumagalli}, {Fossati},
  {Lofthouse}, {Prochaska}, {Arrigoni Battaia}, {Bielby}, {Cantalupo}, {Cooke},
  {Murphy}, \& {O'Meara}}]{Dutta2020}
{Dutta}, R., {Fumagalli}, M., {Fossati}, M., {et~al.} 2020,
  \bibinfo{title}{{MUSE Analysis of Gas around Galaxies (MAGG) - II:
  metal-enriched halo gas around z {\ensuremath{\sim}} 1 galaxies},} \mnras,
  499, 5022, \dodoi{10.1093/mnras/staa3147}

\bibitem[{A.~A. {Dutton} \& A.~V. {Macci{\`o}}(2014){Dutton} \&
  {Macci{\`o}}}]{Dutton_2014}
{Dutton}, A.~A., \& {Macci{\`o}}, A.~V. 2014, \bibinfo{title}{{Cold dark matter
  haloes in the Planck era: evolution of structural parameters for Einasto and
  NFW profiles},} \mnras, 441, 3359, \dodoi{10.1093/mnras/stu742}

\bibitem[{C.-A. {Faucher-Gigu{\`e}re} \& S.~P. {Oh}(2023){Faucher-Gigu{\`e}re}
  \& {Oh}}]{Faucher_2023}
{Faucher-Gigu{\`e}re}, C.-A., \& {Oh}, S.~P. 2023, \bibinfo{title}{{Key
  Physical Processes in the Circumgalactic Medium},} \araa, 61, 131,
  \dodoi{10.1146/annurev-astro-052920-125203}

\bibitem[{A.~B. {Ford} {et~al.}(2013){Ford}, {Oppenheimer}, {Dav{\'e}}, {Katz},
  {Kollmeier}, \& {Weinberg}}]{Ford2013}
{Ford}, A.~B., {Oppenheimer}, B.~D., {Dav{\'e}}, R., {et~al.} 2013,
  \bibinfo{title}{{Hydrogen and metal line absorption around low-redshift
  galaxies in cosmological hydrodynamic simulations},} \mnras, 432, 89,
  \dodoi{10.1093/mnras/stt393}

\bibitem[{L.~K. {Guha} {et~al.}(2022){Guha}, {Srianand}, {Dutta}, {Joshi},
  {Noterdaeme}, \& {Petitjean}}]{guha2022}
{Guha}, L.~K., {Srianand}, R., {Dutta}, R., {et~al.} 2022,
  \bibinfo{title}{{Host galaxies of ultrastrong Mg II absorbers at z 0.5},}
  \mnras, 513, 3836, \dodoi{10.1093/mnras/stac1106}

\bibitem[{Z. {Hafen} {et~al.}(2019){Hafen}, {Faucher-Gigu{\`e}re},
  {Angl{\'e}s-Alc{\'a}zar}, {Stern}, {Kere{\v{s}}}, {Hummels}, {Esmerian},
  {Garrison-Kimmel}, {El-Badry}, {Wetzel}, {Chan}, {Hopkins}, \&
  {Murray}}]{Hafen2019}
{Hafen}, Z., {Faucher-Gigu{\`e}re}, C.-A., {Angl{\'e}s-Alc{\'a}zar}, D.,
  {et~al.} 2019, \bibinfo{title}{{The origins of the circumgalactic medium in
  the FIRE simulations},} \mnras, 488, 1248, \dodoi{10.1093/mnras/stz1773}

\bibitem[{S.~H. {Ho} {et~al.}(2017){Ho}, {Martin}, {Kacprzak}, \&
  {Churchill}}]{Ho2017}
{Ho}, S.~H., {Martin}, C.~L., {Kacprzak}, G.~G., \& {Churchill}, C.~W. 2017,
  \bibinfo{title}{{Quasars Probing Galaxies. I. Signatures of Gas Accretion at
  Redshift Approximately 0.2},} \apj, 835, 267,
  \dodoi{10.3847/1538-4357/835/2/267}

\bibitem[{S.~H. {Ho} {et~al.}(2020){Ho}, {Martin}, \& {Schaye}}]{Ho_2020}
{Ho}, S.~H., {Martin}, C.~L., \& {Schaye}, J. 2020,
  \bibinfo{title}{{Morphological and Rotation Structures of Circumgalactic Mg
  II Gas in the EAGLE Simulation and the Dependence on Galaxy Properties},}
  \apj, 904, 76, \dodoi{10.3847/1538-4357/abbe88}

\bibitem[{Y.-H. {Huang} {et~al.}(2021){Huang}, {Chen}, {Shectman}, {Johnson},
  {Zahedy}, {Helsby}, {Gauthier}, \& {Thompson}}]{Huang2021}
{Huang}, Y.-H., {Chen}, H.-W., {Shectman}, S.~A., {et~al.} 2021,
  \bibinfo{title}{{A complete census of circumgalactic Mg II at redshift z
  {\ensuremath{\lesssim}} 0.5},} \mnras, 502, 4743,
  \dodoi{10.1093/mnras/stab360}

\bibitem[{O. {Ilbert} {et~al.}(2013){Ilbert}, {McCracken}, {Le F{\`e}vre},
  {Capak}, {Dunlop}, {Karim}, {Renzini}, {Caputi}, {Boissier}, {Arnouts},
  {Aussel}, {Comparat}, {Guo}, {Hudelot}, {Kartaltepe}, {Kneib}, {Krogager},
  {Le Floc'h}, {Lilly}, {Mellier}, {Milvang-Jensen}, {Moutard}, {Onodera},
  {Richard}, {Salvato}, {Sanders}, {Scoville}, {Silverman}, {Taniguchi},
  {Tasca}, {Thomas}, {Toft}, {Tresse}, {Vergani}, {Wolk}, \&
  {Zirm}}]{Ilbert_2013}
{Ilbert}, O., {McCracken}, H.~J., {Le F{\`e}vre}, O., {et~al.} 2013,
  \bibinfo{title}{{Mass assembly in quiescent and star-forming galaxies since z
  ≃ 4 from UltraVISTA},} \aap, 556, A55, \dodoi{10.1051/0004-6361/201321100}

\bibitem[{O. {Ilbert} {et~al.}(2015){Ilbert}, {Arnouts}, {Le Floc'h}, {Aussel},
  {Bethermin}, {Capak}, {Hsieh}, {Kajisawa}, {Karim}, {Le F{\`e}vre}, {Lee},
  {Lilly}, {McCracken}, {Michel-Dansac}, {Moutard}, {Renzini}, {Salvato},
  {Sanders}, {Scoville}, {Sheth}, {Silverman}, {Smol{\v{c}}i{\'c}},
  {Taniguchi}, \& {Tresse}}]{Ilbert}
{Ilbert}, O., {Arnouts}, S., {Le Floc'h}, E., {et~al.} 2015,
  \bibinfo{title}{{Evolution of the specific star formation rate function at z<
  1.4 Dissecting the mass-SFR plane in COSMOS and GOODS},} \aap, 579, A2,
  \dodoi{10.1051/0004-6361/201425176}

\bibitem[{G.~G. {Kacprzak} {et~al.}(2013){Kacprzak}, {Cooke}, {Churchill},
  {Ryan-Weber}, \& {Nielsen}}]{Kacprzak2013}
{Kacprzak}, G.~G., {Cooke}, J., {Churchill}, C.~W., {Ryan-Weber}, E.~V., \&
  {Nielsen}, N.~M. 2013, \bibinfo{title}{{The Smooth Mg II Gas Distribution
  through the Interstellar/Extra-planar/Halo Interface},} \apjl, 777, L11,
  \dodoi{10.1088/2041-8205/777/1/L11}

\bibitem[{R.~C. {Kennicutt} \& N.~J. {Evans}(2012){Kennicutt} \&
  {Evans}}]{Kennicutt_2012}
{Kennicutt}, R.~C., \& {Evans}, N.~J. 2012, \bibinfo{title}{{Star Formation in
  the Milky Way and Nearby Galaxies},} \araa, 50, 531,
  \dodoi{10.1146/annurev-astro-081811-125610}

\bibitem[{D. {Kere{\v{s}}} {et~al.}(2005){Kere{\v{s}}}, {Katz}, {Weinberg}, \&
  {Dav{\'e}}}]{Keres2006}
{Kere{\v{s}}}, D., {Katz}, N., {Weinberg}, D.~H., \& {Dav{\'e}}, R. 2005,
  \bibinfo{title}{{How do galaxies get their gas?},} \mnras, 363, 2,
  \dodoi{10.1111/j.1365-2966.2005.09451.x}

\bibitem[{C. {Knobel} {et~al.}(2012){Knobel}, {Lilly}, {Iovino}, {Kova{\v{c}}},
  {Bschorr}, {Presotto}, {Oesch}, {Kampczyk}, {Carollo}, {Contini}, {Kneib},
  {Le Fevre}, {Mainieri}, {Renzini}, {Scodeggio}, {Zamorani}, {Bardelli},
  {Bolzonella}, {Bongiorno}, {Caputi}, {Cucciati}, {de la Torre}, {de Ravel},
  {Franzetti}, {Garilli}, {Lamareille}, {Le Borgne}, {Le Brun}, {Maier},
  {Mignoli}, {Pello}, {Peng}, {Perez Montero}, {Silverman}, {Tanaka}, {Tasca},
  {Tresse}, {Vergani}, {Zucca}, {Barnes}, {Bordoloi}, {Cappi}, {Cimatti},
  {Coppa}, {Koekemoer}, {L{\'o}pez-Sanjuan}, {McCracken}, {Moresco}, {Nair},
  {Pozzetti}, \& {Welikala}}]{Knobel_2012}
{Knobel}, C., {Lilly}, S.~J., {Iovino}, A., {et~al.} 2012, \bibinfo{title}{{The
  zCOSMOS 20k Group Catalog},} \apj, 753, 121,
  \dodoi{10.1088/0004-637X/753/2/121}

\bibitem[{A.~M. {Koekemoer} {et~al.}(2007){Koekemoer}, {Aussel}, {Calzetti},
  {Capak}, {Giavalisco}, {Kneib}, {Leauthaud}, {Le F{\`e}vre}, {McCracken},
  {Massey}, {Mobasher}, {Rhodes}, {Scoville}, \& {Shopbell}}]{Koekemoer_2007}
{Koekemoer}, A.~M., {Aussel}, H., {Calzetti}, D., {et~al.} 2007,
  \bibinfo{title}{{The COSMOS Survey: Hubble Space Telescope Advanced Camera
  for Surveys Observations and Data Processing},} \apjs, 172, 196,
  \dodoi{10.1086/520086}

\bibitem[{K. {Kova{\v{c}}} {et~al.}(2014){Kova{\v{c}}}, {Lilly}, {Knobel},
  {Bschorr}, {Peng}, {Carollo}, {Contini}, {Kneib}, {Le F{\'e}vre}, {Mainieri},
  {Renzini}, {Scodeggio}, {Zamorani}, {Bardelli}, {Bolzonella}, {Bongiorno},
  {Caputi}, {Cucciati}, {de la Torre}, {de Ravel}, {Franzetti}, {Garilli},
  {Iovino}, {Kampczyk}, {Lamareille}, {Le Borgne}, {Le Brun}, {Maier},
  {Mignoli}, {Oesch}, {Pello}, {Montero}, {Presotto}, {Silverman}, {Tanaka},
  {Tasca}, {Tresse}, {Vergani}, {Zucca}, {Aussel}, {Koekemoer}, {Le Floc'h},
  {Moresco}, \& {Pozzetti}}]{Kovac2014}
{Kova{\v{c}}}, K., {Lilly}, S.~J., {Knobel}, C., {et~al.} 2014,
  \bibinfo{title}{{zCOSMOS 20k: satellite galaxies are the main drivers of
  environmental effects in the galaxy population at least to z
  {\ensuremath{\sim}} 0.7},} \mnras, 438, 717, \dodoi{10.1093/mnras/stt2241}

\bibitem[{T.-W. {Lan} \& H. {Mo}(2018){Lan} \& {Mo}}]{Lan2018}
{Lan}, T.-W., \& {Mo}, H. 2018, \bibinfo{title}{{The Circumgalactic Medium of
  eBOSS Emission Line Galaxies: Signatures of Galactic Outflows in Gas
  Distribution and Kinematics},} \apj, 866, 36,
  \dodoi{10.3847/1538-4357/aadc08}

\bibitem[{S.~J. {Lilly} {et~al.}(2007){Lilly}, {Le F{\`e}vre}, {Renzini},
  {Zamorani}, {Scodeggio}, {Contini}, {Carollo}, {Hasinger}, {Kneib}, {Iovino},
  {Le Brun}, {Maier}, {Mainieri}, {Mignoli}, {Silverman}, {Tasca},
  {Bolzonella}, {Bongiorno}, {Bottini}, {Capak}, {Caputi}, {Cimatti},
  {Cucciati}, {Daddi}, {Feldmann}, {Franzetti}, {Garilli}, {Guzzo}, {Ilbert},
  {Kampczyk}, {Kovac}, {Lamareille}, {Leauthaud}, {Le Borgne}, {McCracken},
  {Marinoni}, {Pello}, {Ricciardelli}, {Scarlata}, {Vergani}, {Sanders},
  {Schinnerer}, {Scoville}, {Taniguchi}, {Arnouts}, {Aussel}, {Bardelli},
  {Brusa}, {Cappi}, {Ciliegi}, {Finoguenov}, {Foucaud}, {Franceschini},
  {Halliday}, {Impey}, {Knobel}, {Koekemoer}, {Kurk}, {Maccagni}, {Maddox},
  {Marano}, {Marconi}, {Meneux}, {Mobasher}, {Moreau}, {Peacock}, {Porciani},
  {Pozzetti}, {Scaramella}, {Schiminovich}, {Shopbell}, {Smail}, {Thompson},
  {Tresse}, {Vettolani}, {Zanichelli}, \& {Zucca}}]{Lilly_2007}
{Lilly}, S.~J., {Le F{\`e}vre}, O., {Renzini}, A., {et~al.} 2007,
  \bibinfo{title}{{zCOSMOS: A Large VLT/VIMOS Redshift Survey Covering 0 < z <
  3 in the COSMOS Field},} \apjs, 172, 70, \dodoi{10.1086/516589}

\bibitem[{S.~J. {Lilly} {et~al.}(2009){Lilly}, {Le Brun}, {Maier}, {Mainieri},
  {Mignoli}, {Scodeggio}, {Zamorani}, {Carollo}, {Contini}, {Kneib}, {Le
  F{\`e}vre}, {Renzini}, {Bardelli}, {Bolzonella}, {Bongiorno}, {Caputi},
  {Coppa}, {Cucciati}, {de la Torre}, {de Ravel}, {Franzetti}, {Garilli},
  {Iovino}, {Kampczyk}, {Kovac}, {Knobel}, {Lamareille}, {Le Borgne}, {Pello},
  {Peng}, {P{\'e}rez-Montero}, {Ricciardelli}, {Silverman}, {Tanaka}, {Tasca},
  {Tresse}, {Vergani}, {Zucca}, {Ilbert}, {Salvato}, {Oesch}, {Abbas},
  {Bottini}, {Capak}, {Cappi}, {Cassata}, {Cimatti}, {Elvis}, {Fumana},
  {Guzzo}, {Hasinger}, {Koekemoer}, {Leauthaud}, {Maccagni}, {Marinoni},
  {McCracken}, {Memeo}, {Meneux}, {Porciani}, {Pozzetti}, {Sanders},
  {Scaramella}, {Scarlata}, {Scoville}, {Shopbell}, \& {Taniguchi}}]{Lilly2009}
{Lilly}, S.~J., {Le Brun}, V., {Maier}, C., {et~al.} 2009, \bibinfo{title}{{The
  zCOSMOS 10k-Bright Spectroscopic Sample},} \apjs, 184, 218,
  \dodoi{10.1088/0067-0049/184/2/218}

\bibitem[{S. {Lopez} {et~al.}(2018){Lopez}, {Tejos}, {Ledoux}, {Barrientos},
  {Sharon}, {Rigby}, {Gladders}, {Bayliss}, \& {Pessa}}]{Lopez2018}
{Lopez}, S., {Tejos}, N., {Ledoux}, C., {et~al.} 2018, \bibinfo{title}{{A
  clumpy and anisotropic galaxy halo at redshift 1 from gravitational-arc
  tomography},} \nat, 554, 493, \dodoi{10.1038/nature25436}

\bibitem[{C.~L. {Martin} {et~al.}(2019){Martin}, {Ho}, {Kacprzak}, \&
  {Churchill}}]{Martin2019}
{Martin}, C.~L., {Ho}, S.~H., {Kacprzak}, G.~G., \& {Churchill}, C.~W. 2019,
  \bibinfo{title}{{Kinematics of Circumgalactic Gas: Feeding Galaxies and
  Feedback},} \apj, 878, 84, \dodoi{10.3847/1538-4357/ab18ac}

\bibitem[{R. {Massey}(2010){Massey}}]{Massey_2010}
{Massey}, R. 2010, \bibinfo{title}{{Charge transfer inefficiency in the Hubble
  Space Telescope since Servicing Mission 4},} \mnras, 409, L109,
  \dodoi{10.1111/j.1745-3933.2010.00959.x}

\bibitem[{A.~L. {Muratov} {et~al.}(2017){Muratov}, {Kere{\v{s}}},
  {Faucher-Gigu{\`e}re}, {Hopkins}, {Ma}, {Angl{\'e}s-Alc{\'a}zar}, {Chan},
  {Torrey}, {Hafen}, {Quataert}, \& {Murray}}]{Muratov2017}
{Muratov}, A.~L., {Kere{\v{s}}}, D., {Faucher-Gigu{\`e}re}, C.-A., {et~al.}
  2017, \bibinfo{title}{{Metal flows of the circumgalactic medium, and the
  metal budget in galactic haloes},} \mnras, 468, 4170,
  \dodoi{10.1093/mnras/stx667}

\bibitem[{A. {Muzzin} {et~al.}(2013){Muzzin}, {Marchesini}, {Stefanon},
  {Franx}, {Milvang-Jensen}, {Dunlop}, {Fynbo}, {Brammer}, {Labb{\'e}}, \& {van
  Dokkum}}]{Muzzin_2013}
{Muzzin}, A., {Marchesini}, D., {Stefanon}, M., {et~al.} 2013,
  \bibinfo{title}{{A Public K$_{s}$ -selected Catalog in the COSMOS/ULTRAVISTA
  Field: Photometry, Photometric Redshifts, and Stellar Population
  Parameters},} \apjs, 206, 8, \dodoi{10.1088/0067-0049/206/1/8}

\bibitem[{D. {Nelson} {et~al.}(2019){Nelson}, {Pillepich}, {Springel},
  {Pakmor}, {Weinberger}, {Genel}, {Torrey}, {Vogelsberger}, {Marinacci}, \&
  {Hernquist}}]{Nelson_2019}
{Nelson}, D., {Pillepich}, A., {Springel}, V., {et~al.} 2019,
  \bibinfo{title}{{First results from the TNG50 simulation: galactic outflows
  driven by supernovae and black hole feedback},} \mnras, 490, 3234,
  \dodoi{10.1093/mnras/stz2306}

\bibitem[{N.~M. {Nielsen} {et~al.}(2013){Nielsen}, {Churchill}, \&
  {Kacprzak}}]{Nielsen_2013}
{Nielsen}, N.~M., {Churchill}, C.~W., \& {Kacprzak}, G.~G. 2013,
  \bibinfo{title}{{MAGIICAT II. General Characteristics of the Mg II Absorbing
  Circumgalactic Medium},} \apj, 776, 115, \dodoi{10.1088/0004-637X/776/2/115}

\bibitem[{N.~M. {Nielsen} {et~al.}(2018){Nielsen}, {Kacprzak}, {Pointon},
  {Churchill}, \& {Murphy}}]{Nielsen_2018}
{Nielsen}, N.~M., {Kacprzak}, G.~G., {Pointon}, S.~K., {Churchill}, C.~W., \&
  {Murphy}, M.~T. 2018, \bibinfo{title}{{MAGIICAT VI. The Mg II Intragroup
  Medium Is Kinematically Complex},} \apj, 869, 153,
  \dodoi{10.3847/1538-4357/aaedbd}

\bibitem[{B.~D. {Oppenheimer} {et~al.}(2018){Oppenheimer}, {Schaye}, {Crain},
  {Werk}, \& {Richings}}]{Oppenheimer2018}
{Oppenheimer}, B.~D., {Schaye}, J., {Crain}, R.~A., {Werk}, J.~K., \&
  {Richings}, A.~J. 2018, \bibinfo{title}{{The multiphase circumgalactic medium
  traced by low metal ions in EAGLE zoom simulations},} \mnras, 481, 835,
  \dodoi{10.1093/mnras/sty2281}

\bibitem[{W.~J. {Pearson} {et~al.}(2023){Pearson}, {Pistis}, {Figueira},
  {Ma{\l}ek}, {Moutard}, {Vergani}, \& {Pollo}}]{Pearson2023}
{Pearson}, W.~J., {Pistis}, F., {Figueira}, M., {et~al.} 2023,
  \bibinfo{title}{{Influence of star-forming galaxy selection on the galaxy
  main sequence},} \aap, 679, A35, \dodoi{10.1051/0004-6361/202346396}

\bibitem[{C. {P{\'e}roux} {et~al.}(2020){P{\'e}roux}, {Nelson}, {van de Voort},
  {Pillepich}, {Marinacci}, {Vogelsberger}, \& {Hernquist}}]{Peroux_2020}
{P{\'e}roux}, C., {Nelson}, D., {van de Voort}, F., {et~al.} 2020,
  \bibinfo{title}{{Predictions for the angular dependence of gas mass flow rate
  and metallicity in the circumgalactic medium},} \mnras, 499, 2462,
  \dodoi{10.1093/mnras/staa2888}

\bibitem[{ {Planck Collaboration} {et~al.}(2020){Planck Collaboration},
  {Aghanim}, {Akrami}, {Ashdown}, {Aumont}, {Baccigalupi}, {Ballardini},
  {Banday}, {Barreiro}, {Bartolo}, {Basak}, {Battye}, {Benabed}, {Bernard},
  {Bersanelli}, {Bielewicz}, {Bock}, {Bond}, {Borrill}, {Bouchet}, {Boulanger},
  {Bucher}, {Burigana}, {Butler}, {Calabrese}, {Cardoso}, {Carron},
  {Challinor}, {Chiang}, {Chluba}, {Colombo}, {Combet}, {Contreras}, {Crill},
  {Cuttaia}, {de Bernardis}, {de Zotti}, {Delabrouille}, {Delouis}, {Di
  Valentino}, {Diego}, {Dor{\'e}}, {Douspis}, {Ducout}, {Dupac}, {Dusini},
  {Efstathiou}, {Elsner}, {En{\ss}lin}, {Eriksen}, {Fantaye}, {Farhang},
  {Fergusson}, {Fernandez-Cobos}, {Finelli}, {Forastieri}, {Frailis},
  {Fraisse}, {Franceschi}, {Frolov}, {Galeotta}, {Galli}, {Ganga},
  {G{\'e}nova-Santos}, {Gerbino}, {Ghosh}, {Gonz{\'a}lez-Nuevo}, {G{\'o}rski},
  {Gratton}, {Gruppuso}, {Gudmundsson}, {Hamann}, {Handley}, {Hansen},
  {Herranz}, {Hildebrandt}, {Hivon}, {Huang}, {Jaffe}, {Jones}, {Karakci},
  {Keih{\"a}nen}, {Keskitalo}, {Kiiveri}, {Kim}, {Kisner}, {Knox},
  {Krachmalnicoff}, {Kunz}, {Kurki-Suonio}, {Lagache}, {Lamarre}, {Lasenby},
  {Lattanzi}, {Lawrence}, {Le Jeune}, {Lemos}, {Lesgourgues}, {Levrier},
  {Lewis}, {Liguori}, {Lilje}, {Lilley}, {Lindholm}, {L{\'o}pez-Caniego},
  {Lubin}, {Ma}, {Mac{\'\i}as-P{\'e}rez}, {Maggio}, {Maino}, {Mandolesi},
  {Mangilli}, {Marcos-Caballero}, {Maris}, {Martin}, {Martinelli},
  {Mart{\'\i}nez-Gonz{\'a}lez}, {Matarrese}, {Mauri}, {McEwen}, {Meinhold},
  {Melchiorri}, {Mennella}, {Migliaccio}, {Millea}, {Mitra},
  {Miville-Desch{\^e}nes}, {Molinari}, {Montier}, {Morgante}, {Moss}, {Natoli},
  {N{\o}rgaard-Nielsen}, {Pagano}, {Paoletti}, {Partridge}, {Patanchon},
  {Peiris}, {Perrotta}, {Pettorino}, {Piacentini}, {Polastri}, {Polenta},
  {Puget}, {Rachen}, {Reinecke}, {Remazeilles}, {Renzi}, {Rocha}, {Rosset},
  {Roudier}, {Rubi{\~n}o-Mart{\'\i}n}, {Ruiz-Granados}, {Salvati}, {Sandri},
  {Savelainen}, {Scott}, {Shellard}, {Sirignano}, {Sirri}, {Spencer},
  {Sunyaev}, {Suur-Uski}, {Tauber}, {Tavagnacco}, {Tenti}, {Toffolatti},
  {Tomasi}, {Trombetti}, {Valenziano}, {Valiviita}, {Van Tent}, {Vibert},
  {Vielva}, {Villa}, {Vittorio}, {Wandelt}, {Wehus}, {White}, {White},
  {Zacchei}, \& {Zonca}}]{Planck2020}
{Planck Collaboration}, {Aghanim}, N., {Akrami}, Y., {et~al.} 2020,
  \bibinfo{title}{{Planck 2018 results. VI. Cosmological parameters},} \aap,
  641, A6, \dodoi{10.1051/0004-6361/201833910}

\bibitem[{A. {Renzini} \& Y.-j. {Peng}(2015){Renzini} \& {Peng}}]{Renzini_2015}
{Renzini}, A., \& {Peng}, Y.-j. 2015, \bibinfo{title}{{An Objective Definition
  for the Main Sequence of Star-forming Galaxies},} \apjl, 801, L29,
  \dodoi{10.1088/2041-8205/801/2/L29}

\bibitem[{P. {Richter} {et~al.}(2016){Richter}, {Wakker}, {Fechner}, {Herenz},
  {Tepper-Garc{\'\i}a}, \& {Fox}}]{Richter_2016}
{Richter}, P., {Wakker}, B.~P., {Fechner}, C., {et~al.} 2016,
  \bibinfo{title}{{An HST/COS legacy survey of intervening Si III absorption in
  the extended gaseous halos of low-redshift galaxies},} \aap, 590, A68,
  \dodoi{10.1051/0004-6361/201527038}

\bibitem[{K.~H.~R. {Rubin} {et~al.}(2018){Rubin}, {Diamond-Stanic}, {Coil},
  {Crighton}, \& {Moustakas}}]{Rubin_2018}
{Rubin}, K. H.~R., {Diamond-Stanic}, A.~M., {Coil}, A.~L., {Crighton}, N.
  H.~M., \& {Moustakas}, J. 2018, \bibinfo{title}{{Galaxies Probing Galaxies in
  PRIMUS. I. Sample, Spectroscopy, and Characteristics of the
  z\textbackslashsim 0.5 Mg II-absorbing Circumgalactic Medium},} \apj, 853,
  95, \dodoi{10.3847/1538-4357/aa9792}

\bibitem[{K.~H.~R. {Rubin} {et~al.}(2012){Rubin}, {Prochaska}, {Koo}, \&
  {Phillips}}]{Rubin2012}
{Rubin}, K. H.~R., {Prochaska}, J.~X., {Koo}, D.~C., \& {Phillips}, A.~C. 2012,
  \bibinfo{title}{{The Direct Detection of Cool, Metal-enriched Gas Accretion
  onto Galaxies at z \raisebox{-0.5ex}\textasciitilde 0.5},} \apjl, 747, L26,
  \dodoi{10.1088/2041-8205/747/2/L26}

\bibitem[{G.~C. {Rudie} {et~al.}(2019){Rudie}, {Steidel}, {Pettini}, {Trainor},
  {Strom}, {Hummels}, {Reddy}, \& {Shapley}}]{Rudie_2019}
{Rudie}, G.~C., {Steidel}, C.~C., {Pettini}, M., {et~al.} 2019,
  \bibinfo{title}{{Column Density, Kinematics, and Thermal State of
  Metal-bearing Gas within the Virial Radius of z {\ensuremath{\sim}} 2
  Star-forming Galaxies in the Keck Baryonic Structure Survey},} \apj, 885, 61,
  \dodoi{10.3847/1538-4357/ab4255}

\bibitem[{S. {Salim}(2014){Salim}}]{Salim_2014}
{Salim}, S. 2014, \bibinfo{title}{{Green Valley Galaxies},} Serbian
  Astronomical Journal, 189, 1, \dodoi{10.2298/SAJ1489001S}

\bibitem[{S. {Salim} {et~al.}(2014){Salim}, {Lee}, {Ly}, {Brinchmann},
  {Dav{\'e}}, {Dickinson}, {Salzer}, \& {Charlot}}]{Salim2014}
{Salim}, S., {Lee}, J.~C., {Ly}, C., {et~al.} 2014, \bibinfo{title}{{A Critical
  Look at the Mass-Metallicity-Star Formation Rate Relation in the Local
  Universe. I. An Improved Analysis Framework and Confounding Systematics},}
  \apj, 797, 126, \dodoi{10.1088/0004-637X/797/2/126}

\bibitem[{K. {Schawinski} {et~al.}(2014){Schawinski}, {Urry}, {Simmons},
  {Fortson}, {Kaviraj}, {Keel}, {Lintott}, {Masters}, {Nichol}, {Sarzi},
  {Skibba}, {Treister}, {Willett}, {Wong}, \& {Yi}}]{Schawinski2014}
{Schawinski}, K., {Urry}, C.~M., {Simmons}, B.~D., {et~al.} 2014,
  \bibinfo{title}{{The green valley is a red herring: Galaxy Zoo reveals two
  evolutionary pathways towards quenching of star formation in early- and
  late-type galaxies},} \mnras, 440, 889, \dodoi{10.1093/mnras/stu327}

\bibitem[{I. {Schroetter} {et~al.}(2016){Schroetter}, {Bouch{\'e}}, {Wendt},
  {Contini}, {Finley}, {Pell{\'o}}, {Bacon}, {Cantalupo}, {Marino}, {Richard},
  {Lilly}, {Schaye}, {Soto}, {Steinmetz}, {Straka}, \&
  {Wisotzki}}]{Schroetter_2016}
{Schroetter}, I., {Bouch{\'e}}, N., {Wendt}, M., {et~al.} 2016,
  \bibinfo{title}{{Muse Gas Flow and Wind (MEGAFLOW). I. First MUSE Results on
  Background Quasars},} \apj, 833, 39, \dodoi{10.3847/1538-4357/833/1/39}

\bibitem[{A. {Shaban} {et~al.}(2025){Shaban}, {Bordoloi}, {O'Meara}, {Sharon},
  {Tejos}, {Lopez}, {Ledoux}, {Barrientos}, \& {Rigby}}]{Shaban2025}
{Shaban}, A., {Bordoloi}, R., {O'Meara}, J.~M., {et~al.} 2025,
  \bibinfo{title}{{Spatially Resolved Circumgalactic Medium around a
  Star-forming Galaxy Driving a Galactic Outflow at z ≍ 0.8},} \apj, 986,
  190, \dodoi{10.3847/1538-4357/add0b9}

\bibitem[{J.~S. {Speagle} {et~al.}(2014){Speagle}, {Steinhardt}, {Capak}, \&
  {Silverman}}]{Speagle2014}
{Speagle}, J.~S., {Steinhardt}, C.~L., {Capak}, P.~L., \& {Silverman}, J.~D.
  2014, \bibinfo{title}{{A Highly Consistent Framework for the Evolution of the
  Star-Forming ``Main Sequence'' from z \raisebox{-0.5ex}\textasciitilde 0-6},}
  \apjs, 214, 15, \dodoi{10.1088/0067-0049/214/2/15}

\bibitem[{K. {Tchernyshyov} {et~al.}(2022){Tchernyshyov}, {Werk}, {Wilde},
  {Prochaska}, {Tripp}, {Burchett}, {Bordoloi}, {Howk}, {Lehner}, {O'Meara},
  {Tejos}, \& {Tumlinson}}]{Tchernyshyov_2022}
{Tchernyshyov}, K., {Werk}, J.~K., {Wilde}, M.~C., {et~al.} 2022,
  \bibinfo{title}{{The CGM$^{2}$ Survey: Circumgalactic O VI from Dwarf to
  Massive Star-forming Galaxies},} \apj, 927, 147,
  \dodoi{10.3847/1538-4357/ac450c}

\bibitem[{N. {Tejos} {et~al.}(2021){Tejos}, {L{\'o}pez}, {Ledoux},
  {Fern{\'a}ndez-Figueroa}, {Rivas}, {Sharon}, {Johnston}, {Florian}, {D'Ago},
  {Katsianis}, {Barrientos}, {Berg}, {Corro-Guerra}, {Hamel}, {Moya-Sierralta},
  {Poudel}, {Rigby}, \& {Solimano}}]{Tejos2021}
{Tejos}, N., {L{\'o}pez}, S., {Ledoux}, C., {et~al.} 2021,
  \bibinfo{title}{{Telltale signs of metal recycling in the circumgalactic
  medium of a z 0.77 galaxy},} \mnras, 507, 663, \dodoi{10.1093/mnras/stab2147}

\bibitem[{J. {Tumlinson} {et~al.}(2017){Tumlinson}, {Peeples}, \&
  {Werk}}]{Tumlinson_2017}
{Tumlinson}, J., {Peeples}, M.~S., \& {Werk}, J.~K. 2017, \bibinfo{title}{{The
  Circumgalactic Medium},} \araa, 55, 389,
  \dodoi{10.1146/annurev-astro-091916-055240}

\bibitem[{J. {Tumlinson} {et~al.}(2013){Tumlinson}, {Thom}, {Werk},
  {Prochaska}, {Tripp}, {Katz}, {Dav{\'e}}, {Oppenheimer}, {Meiring}, {Ford},
  {O'Meara}, {Peeples}, {Sembach}, \& {Weinberg}}]{Tumlinson2013}
{Tumlinson}, J., {Thom}, C., {Werk}, J.~K., {et~al.} 2013, \bibinfo{title}{{The
  COS-Halos Survey: Rationale, Design, and a Census of Circumgalactic Neutral
  Hydrogen},} \apj, 777, 59, \dodoi{10.1088/0004-637X/777/1/59}

\bibitem[{A. {van der Wel} {et~al.}(2021){van der Wel}, {Bezanson},
  {D'Eugenio}, {Straatman}, {Franx}, {van Houdt}, {Maseda}, {Gallazzi}, {Wu},
  {Pacifici}, {Barisic}, {Brammer}, {Munoz-Mateos}, {Vervalcke}, {Zibetti},
  {Sobral}, {de Graaff}, {Calhau}, {Kaushal}, {Muzzin}, {Bell}, \& {van
  Dokkum}}]{Van_der_Wel_2021}
{van der Wel}, A., {Bezanson}, R., {D'Eugenio}, F., {et~al.} 2021,
  \bibinfo{title}{{The Large Early Galaxy Astrophysics Census (LEGA-C) Data
  Release 3: 3000 High-quality Spectra of K$_{s}$-selected Galaxies at z >
  0.6},} \apjs, 256, 44, \dodoi{10.3847/1538-4365/ac1356}

\bibitem[{D. {Vergani} {et~al.}(2010){Vergani}, {Zamorani}, {Lilly},
  {Lamareille}, {Halliday}, {Scodeggio}, {Vignali}, {Ciliegi}, {Bolzonella},
  {Bondi}, {Kova{\v{c}}}, {Knobel}, {Zucca}, {Caputi}, {Pozzetti}, {Bardelli},
  {Mignoli}, {Iovino}, {Carollo}, {Contini}, {Kneib}, {Le F{\`e}vre},
  {Mainieri}, {Renzini}, {Bongiorno}, {Coppa}, {Cucciati}, {de la Torre}, {de
  Ravel}, {Franzetti}, {Garilli}, {Kampczyk}, {Le Borgne}, {Le Brun}, {Maier},
  {Pello}, {Peng}, {Perez Montero}, {Ricciardelli}, {Silverman}, {Tanaka},
  {Tasca}, {Tresse}, {Abbas}, {Bottini}, {Cappi}, {Cassata}, {Cimatti},
  {Guzzo}, {Koekemoer}, {Leauthaud}, {Maccagni}, {Marinoni}, {McCracken},
  {Memeo}, {Meneux}, {Oesch}, {Porciani}, {Scaramella}, {Capak}, {Sanders},
  {Scoville}, \& {Taniguchi}}]{Vergani_2009}
{Vergani}, D., {Zamorani}, G., {Lilly}, S., {et~al.} 2010, \bibinfo{title}{{K+a
  galaxies in the zCOSMOS survey . Physical properties of systems in their
  post-starburst phase},} \aap, 509, A42, \dodoi{10.1051/0004-6361/200912802}

\bibitem[{J.~K. {Werk} {et~al.}(2012){Werk}, {Prochaska}, {Thom}, {Tumlinson},
  {Tripp}, {O'Meara}, \& {Meiring}}]{Werk2012}
{Werk}, J.~K., {Prochaska}, J.~X., {Thom}, C., {et~al.} 2012,
  \bibinfo{title}{{The COS-Halos Survey: Keck LRIS and Magellan MagE Optical
  Spectroscopy},} \apjs, 198, 3, \dodoi{10.1088/0067-0049/198/1/3}

\bibitem[{J.~K. {Werk} {et~al.}(2013){Werk}, {Prochaska}, {Thom}, {Tumlinson},
  {Tripp}, {O'Meara}, \& {Peeples}}]{Werk2013}
{Werk}, J.~K., {Prochaska}, J.~X., {Thom}, C., {et~al.} 2013,
  \bibinfo{title}{{The COS-Halos Survey: An Empirical Description of Metal-line
  Absorption in the Low-redshift Circumgalactic Medium},} \apjs, 204, 17,
  \dodoi{10.1088/0067-0049/204/2/17}

\bibitem[{J.~K. {Werk} {et~al.}(2014){Werk}, {Prochaska}, {Tumlinson},
  {Peeples}, {Tripp}, {Fox}, {Lehner}, {Thom}, {O'Meara}, {Ford}, {Bordoloi},
  {Katz}, {Tejos}, {Oppenheimer}, {Dav{\'e}}, \& {Weinberg}}]{Werk2014}
{Werk}, J.~K., {Prochaska}, J.~X., {Tumlinson}, J., {et~al.} 2014,
  \bibinfo{title}{{The COS-Halos Survey: Physical Conditions and Baryonic Mass
  in the Low-redshift Circumgalactic Medium},} \apj, 792, 8,
  \dodoi{10.1088/0004-637X/792/1/8}

\bibitem[{K.~E. {Whitaker} {et~al.}(2012){Whitaker}, {van Dokkum}, {Brammer},
  \& {Franx}}]{Whitaker2012}
{Whitaker}, K.~E., {van Dokkum}, P.~G., {Brammer}, G., \& {Franx}, M. 2012,
  \bibinfo{title}{{The Star Formation Mass Sequence Out to z = 2.5},} \apjl,
  754, L29, \dodoi{10.1088/2041-8205/754/2/L29}

\bibitem[{M.~C. {Wilde} {et~al.}(2021){Wilde}, {Werk}, {Burchett}, {Prochaska},
  {Tchernyshyov}, {Tripp}, {Tejos}, {Lehner}, {Bordoloi}, {O'Meara}, \&
  {Tumlinson}}]{Wilde_2021}
{Wilde}, M.~C., {Werk}, J.~K., {Burchett}, J.~N., {et~al.} 2021,
  \bibinfo{title}{{CGM$^{2}$ I: The Extent of the Circumgalactic Medium Traced
  by Neutral Hydrogen},} \apj, 912, 9, \dodoi{10.3847/1538-4357/abea14}

\bibitem[{F.~S. {Zahedy} {et~al.}(2019){Zahedy}, {Chen}, {Johnson}, {Pierce},
  {Rauch}, {Huang}, {Weiner}, \& {Gauthier}}]{Zahedy_2018}
{Zahedy}, F.~S., {Chen}, H.-W., {Johnson}, S.~D., {et~al.} 2019,
  \bibinfo{title}{{Characterizing circumgalactic gas around massive ellipticals
  at z {\ensuremath{\sim}} 0.4 - II. Physical properties and elemental
  abundances},} \mnras, 484, 2257, \dodoi{10.1093/mnras/sty3482}

\end{thebibliography}
\bibliographystyle{aasjournalv7}
\clearpage

\appendix
\restartappendixnumbering
\section{\mgii\ absorption profiles and SFMS fitting properties}

In the appendix we present the \mgii\ absorption profiles and fitted Voigt profiles to each of the detections and a table showing the properties of SFMS fits in each redshift bin.

\begin{figure*}[ht]
    \centering

    \includegraphics[width=.845
    \textwidth]{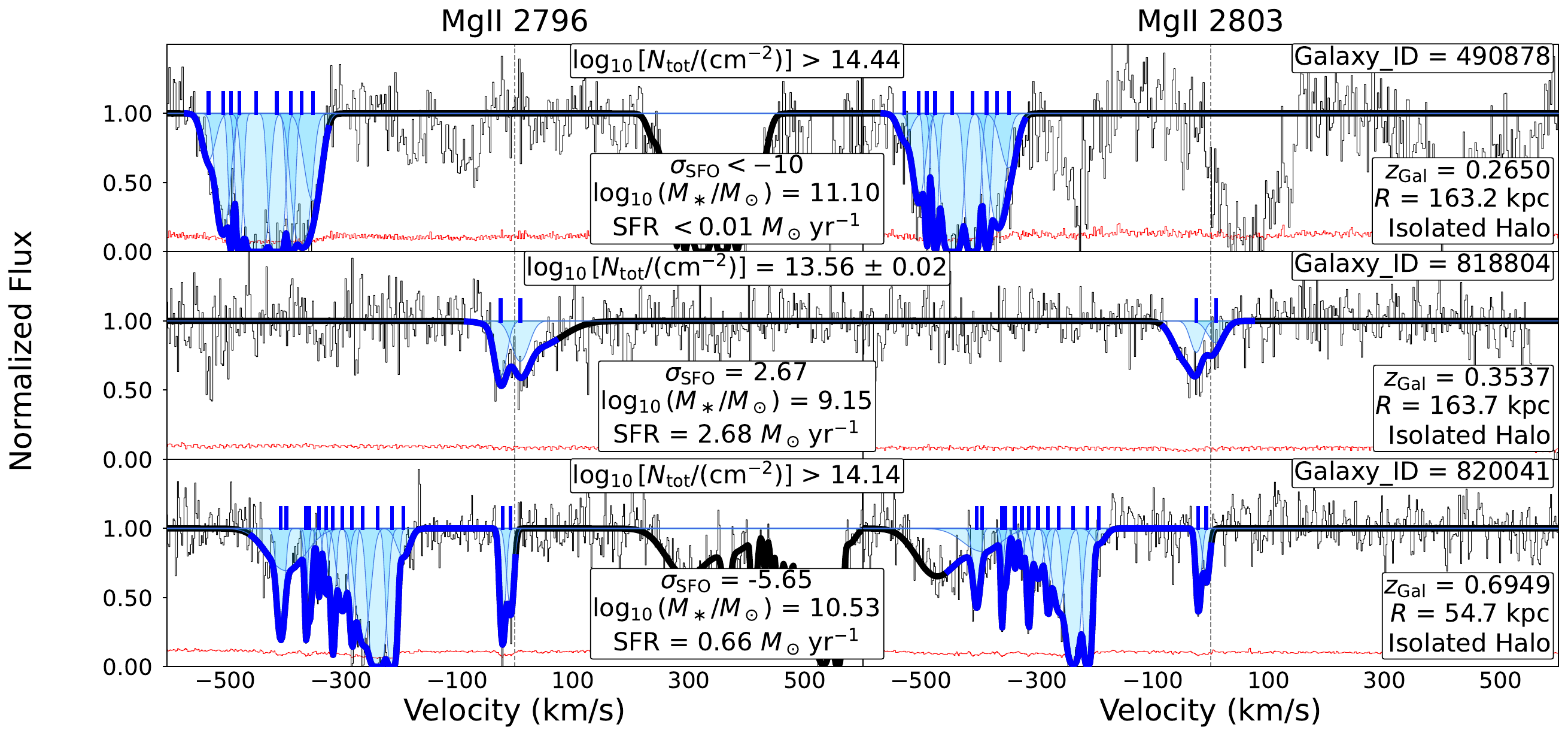}

    \caption{Normalized UVES spectra showing Voigt-profile fits for galaxy-associated \mgii\ detections. The absorption system associated with Galaxy 820041 contains 15 identifiable individual clouds.}
    \label{fig:MgII_UVES}
\end{figure*}
\begin{figure*}
    \centering

    \includegraphics[width=.665
    \textheight,keepaspectratio]{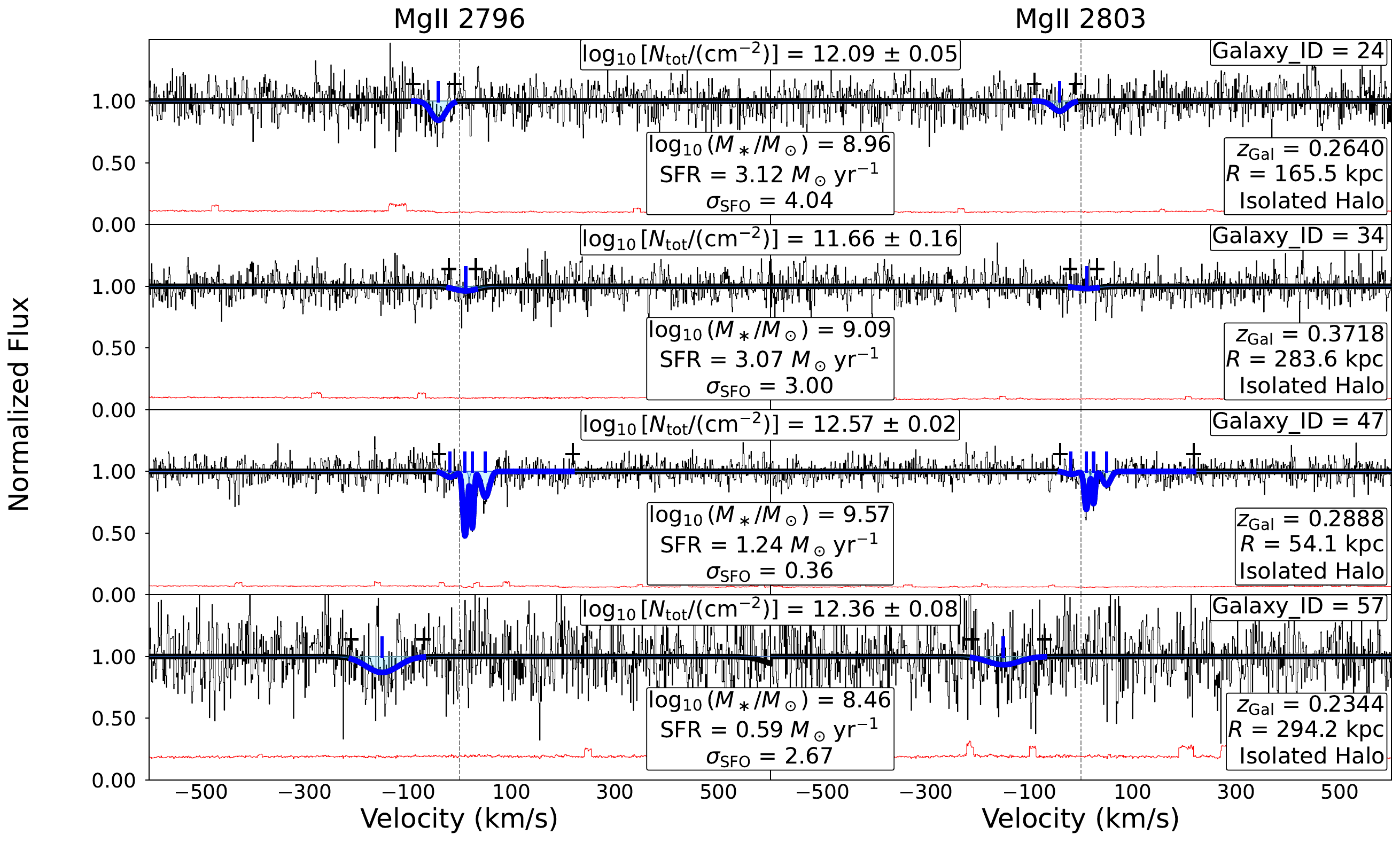}

    \caption{
    Normalized spectra and Voigt profiles of our HIRES absorber sample associated with the CGM$^2$ galaxy dataset.
    }
    \label{fig:MgII_HIRES}
\end{figure*}

\begin{figure*}[p]
\centering
\includegraphics[
height=1\textheight,
keepaspectratio
]{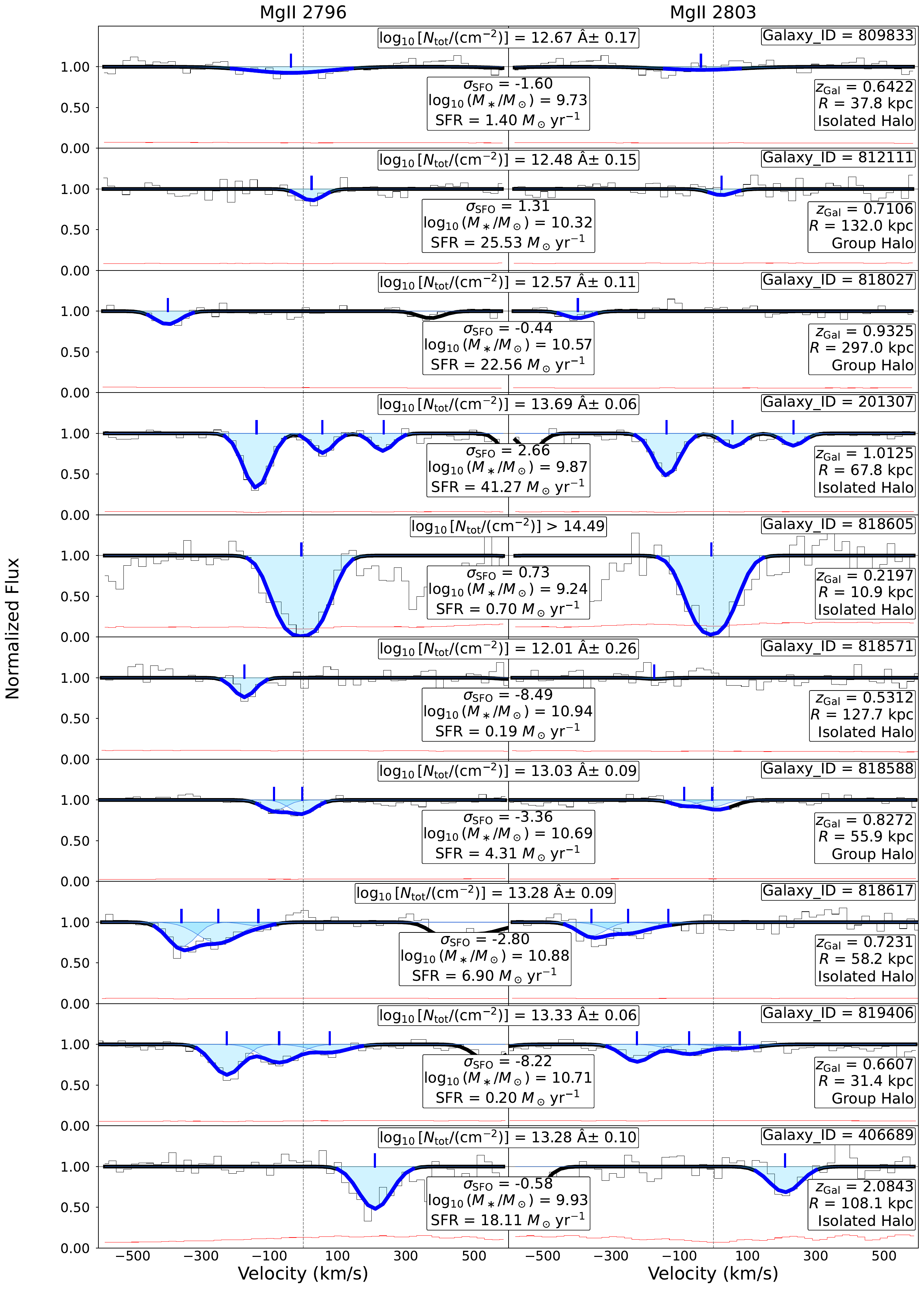}

\end{figure*}

\begin{figure*}[p]
\centering
\includegraphics[
height=1\textheight,
keepaspectratio
]{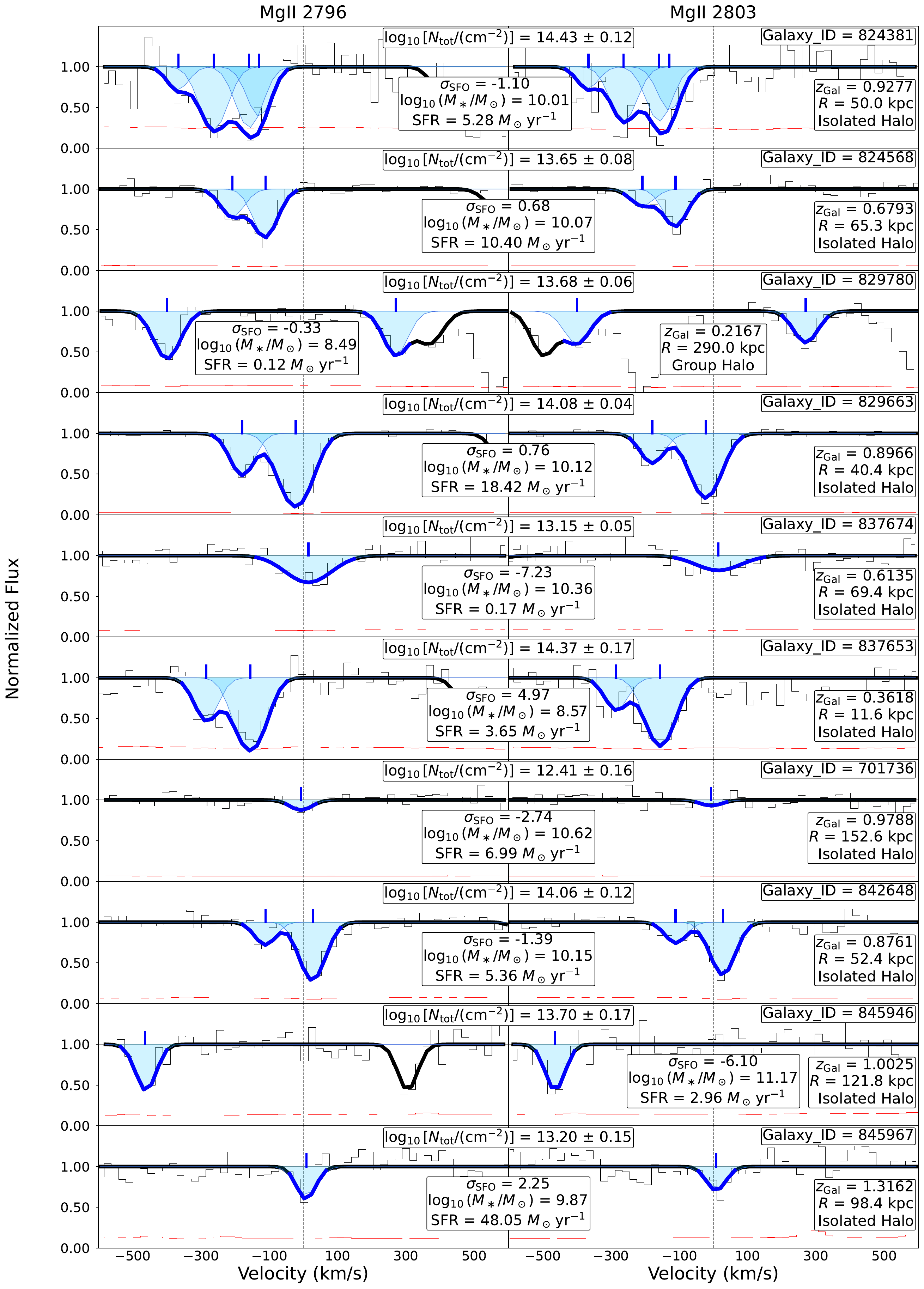}

\caption{Normalized MagE spectra centered on the rest frames of galaxies associated with detected \mgii\ $\lambda\lambda2796,2803$ absorption. The best-fitting Voigt profiles are shown in blue, and the spectral uncertainties are shown in red.}

\label{fig:MgII_MAGE_01}

\end{figure*}

\clearpage

\begin{figure*}[ht]
    \centering

    \includegraphics[width=.9
    \textwidth]{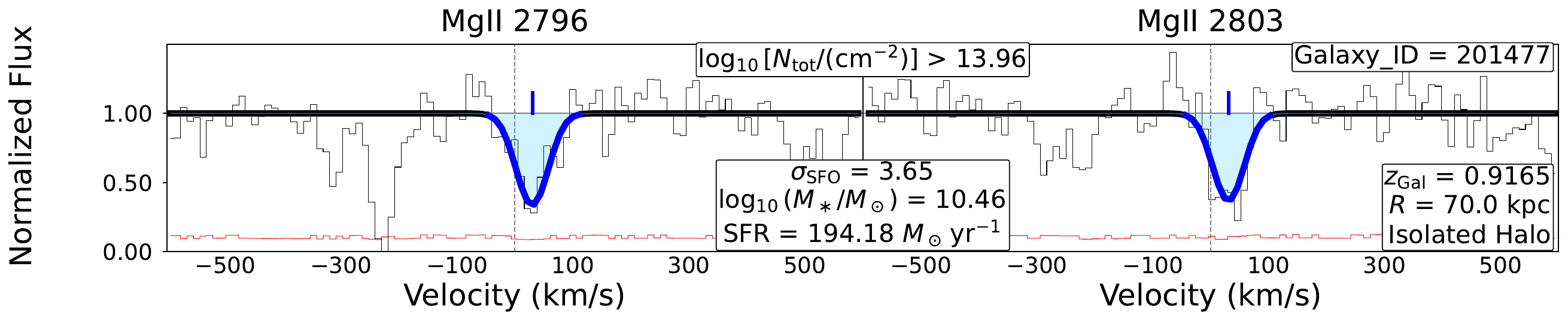}

    \caption{
    Normalized spectra of our single galaxy-associated X-shooter detection.
    }
    \label{fig:MgII_XSHOOTER}
\end{figure*}

\begin{table*}[h!]
\centering
\scriptsize
\setlength{\tabcolsep}{3pt}
\hspace*{-0.5in}\resizebox{\textwidth}{!}{
\begin{tabular}{cccccccccc}
\hline
\hline
Redshift Range & $z_{\rm center}$ & sSFR Cut & $N_{\rm gal}$ &
$\alpha_{\rm init}$ & $\gamma_{\rm init}$ & $\sigma_{\rm init}$ (dex) &
$\alpha_{\rm fit}$ & $\gamma_{\rm fit}$ & $\sigma_{\rm fit}$ (dex) \\
\hline
0.05--0.15 & 0.100 & $> -10.6$ & 256 & $-0.248 \pm 0.018$ & $-9.928 \pm 0.007$ & $0.339^{\dagger}$ & -0.331 & -9.912 & 0.230 \\
0.15--0.25 & 0.200 & $> -10.5$ & 684 & $-0.336 \pm 0.011$ & $-9.678 \pm 0.005$ & 0.269 & -0.301 & -9.735 & 0.235 \\
0.25--0.30 & 0.275 & $> -10.4$ & 435 & $-0.361 \pm 0.013$ & $-9.604 \pm 0.006$ & 0.227 & -0.281 & -9.617 & 0.237 \\
0.30--0.35 & 0.325 & $> -10.3$ & 703 & $-0.320 \pm 0.010$ & $-9.515 \pm 0.004$ & 0.234 & -0.269 & -9.545 & 0.239 \\
0.35--0.40 & 0.375 & $> -10.2$ & 1120 & $-0.293 \pm 0.008$ & $-9.488 \pm 0.003$ & 0.230 & -0.258 & -9.478 & 0.241 \\
0.40--0.45 & 0.425 & $> -10.1$ & 682 & $-0.276 \pm 0.010$ & $-9.436 \pm 0.004$ & 0.244 & -0.247 & -9.415 & 0.242 \\
0.45--0.50 & 0.475 & $> -10.0$ & 725 & $-0.234 \pm 0.009$ & $-9.400 \pm 0.005$ & 0.243 & -0.237 & -9.356 & 0.244 \\
0.50--0.55 & 0.525 & $> -9.9$ & 788 & $-0.251 \pm 0.010$ & $-9.300 \pm 0.004$ & 0.241 & -0.228 & -9.300 & 0.245 \\
0.55--0.60 & 0.575 & $> -9.8$ & 546 & $-0.150 \pm 0.011$ & $-9.255 \pm 0.005$ & 0.262 & -0.219 & -9.248 & 0.246 \\
0.60--0.65 & 0.625 & $> -9.8$ & 747 & $-0.205 \pm 0.009$ & $-9.252 \pm 0.004$ & 0.228 & -0.211 & -9.199 & 0.247 \\
0.65--0.70 & 0.675 & $> -9.7$ & 1154 & $-0.192 \pm 0.008$ & $-9.159 \pm 0.004$ & 0.227 & -0.203 & -9.153 & 0.249 \\
0.70--0.75 & 0.725 & $> -9.7$ & 1022 & $-0.223 \pm 0.008$ & $-9.092 \pm 0.004$ & 0.254 & -0.196 & -9.109 & 0.250 \\
0.75--0.85 & 0.800 & $> -9.6$ & 1601 & $-0.188 \pm 0.007$ & $-9.026 \pm 0.003$ & 0.256 & -0.185 & -9.049 & 0.251 \\
0.85--1.00 & 0.925 & $> -9.6$ & 4611 & $-0.102 \pm 0.004$ & $-8.976 \pm 0.002$ & 0.251 & -0.170 & -8.959 & 0.253 \\
1.00--1.20 & 1.100 & $> -9.5$ & 1680 & $-0.178 \pm 0.006$ & $-8.829 \pm 0.003$ & 0.273 & -0.152 & -8.853 & 0.256 \\
1.20--1.50 & 1.350 & $> -9.5$ & 759 & $-0.197 \pm 0.009$ & $-8.700 \pm 0.007$ & 0.252 & -0.131 & -8.731 & 0.259 \\
1.50--1.90 & 1.700 & $> -9.4$ & 286 & $-0.217 \pm 0.014$ & $-8.372^{\dagger} \pm 0.009$ & $0.558^{\dagger}$ & -0.110 & -8.602 & 0.262 \\
1.90--2.30 & 2.100 & $> -9.4$ & 179 & $0.023 \pm 0.025$ & $-8.649 \pm 0.009$ & $0.334^{\dagger}$ & -0.092 & -8.496 & 0.264 \\
2.30--2.60 & 2.450 & $> -9.3$ & 354 & $-0.083 \pm 0.019$ & $-8.483 \pm 0.008$ & $0.421^{\dagger}$ & -0.080 & -8.427 & 0.266 \\
2.60--2.90 & 2.750 & $> -9.3$ & 550 & $-0.205 \pm 0.014$ & $-8.299 \pm 0.005$ & $0.531^{\dagger}$ & -0.072 & -8.380 & 0.267 \\
2.90--4.00 & 3.450 & $> -9.2$ & 367 & $-0.084 \pm 0.015$ & $-8.353 \pm 0.007$ & $0.545^{\dagger}$ & -0.059 & -8.299 & 0.269 \\
\hline
\end{tabular}
}
\caption{
Star-forming main sequence fit parameters by redshift bin using the pivot-mass form
$\log_{10}(\mathrm{sSFR}_{\mathrm{SFMS}})=
\alpha(t)\left(\log_{10}(M_\star)-M_{\rm pivot}\right)+\gamma(t)$,
with $M_{\rm pivot}=9.7$. The columns $\alpha_{\rm init}$, $\gamma_{\rm init}$, and
$\sigma_{\rm init}$ give the initial per-bin sigma-clipped measurements, while
$\alpha_{\rm fit}$, $\gamma_{\rm fit}$, and $\sigma_{\rm fit}$ give example fitted
values evaluated at each $z_{\rm center}$ for comparison. Uncertainties on $\alpha_{\rm init}$ and $\gamma_{\rm init}$ are estimated by Monte Carlo propagation using adopted global uncertainties in $\log M_\star$ and $\log \mathrm{sSFR}$, applied uniformly to the galaxies in each redshift bin, with the SFMS refit after each perturbation. Dagger symbols mark initial
values excluded from their corresponding fit.
}
\label{tab:SFMS_parameters_pivot}
\end{table*}

\clearpage

\end{document}